\documentclass{jfm}

\usepackage{graphicx}
\usepackage{natbib}
\usepackage{caption}
\usepackage{subcaption}
\usepackage{epstopdf, epsfig}

\ifCUPmtlplainloaded \else
  \checkfont{eurm10}
  \iffontfound
    \IfFileExists{upmath.sty}
      {\typeout{^^JFound AMS Euler Roman fonts on the system,
                   using the 'upmath' package.^^J}%
       \usepackage{upmath}}
      {\typeout{^^JFound AMS Euler Roman fonts on the system, but you
                   dont seem to have the}%
       \typeout{'upmath' package installed. JFM.cls can take advantage
                 of these fonts,^^Jif you use 'upmath' package.^^J}%
      }
  \else
  \fi
\fi

\ifCUPmtlplainloaded \else
  \checkfont{msam10}
  \iffontfound
    \IfFileExists{amssymb.sty}
      {\typeout{^^JFound AMS Symbol fonts on the system, using the
                'amssymb' package.^^J}%
       \usepackage{amssymb}%
         
         \let\geq=\geqslant
      }{}
  \fi
\fi

\ifCUPmtlplainloaded \else
  \IfFileExists{amsbsy.sty}
    {\typeout{^^JFound the 'amsbsy' package on the system, using it.^^J}%
     \usepackage{amsbsy}}
    {}
\fi

\newsavebox{\astrutbox}
\sbox{\astrutbox}{\rule[-5pt]{0pt}{20pt}}

\title[Experimental investigation of flow behind a cube for moderate Reynolds numbers]{Experimental investigation of flow behind a cube for moderate Reynolds numbers}
\author[L.~Klotz, S.~Goujon-Durand, J.~Rokicki, J.E.~Wesfreid]%
{L. Klotz$^{1,2}$\thanks{Email address for correspondence: lukasz.klotz@espci.fr}, S. Goujon-Durand$^1$, J. Rokicki$^2$
\and J. E. Wesfreid$^1$\thanks{Email address for correspondence: wesfreid@pmmh.espci.fr},\ns,\ns}
\affiliation{$^1$Physique et M\'ecanique des Milieux H\'et\'erog\`enes, PMMH,
UMR 7636 \\ ESPCI - CNRS - UPMC - UPD - 75 005 Paris, France\\[\affilskip]
$^2$Institute of Aeronautics and Applied Mechanics, Warsaw University of Technology,\\ Nowowiejska 24, 00-665 Warsaw, Poland}
\pubyear{2010}
\volume{650}
\pagerange{119--126}
\date{6 May 2013; revised 4 March 2014; accepted 17 April 2017}
\begin{document}
\maketitle
\begin{abstract}
The wake behind a cube with a face normal to the flow was investigated experimentally in a water tunnel using laser induced fluorescence visualisation (LIF) and particle image velocimetry (PIV) techniques. Measurements were carried out for moderate Reynolds numbers between 100 and 400 and in this range a sequence of two flow bifurcations was confirmed. Values for both onsets were determined in the framework of Landau`s instability model. The measured longitudinal vorticity was separated into three components corresponding to each of the identified regimes. It was shown that the vorticity associated with a basic flow regime originates from corners of the bluff body, in contrast to the two other contributions which are related to instability effects. The present experimental results are compared with numerical simulation carried out earlier by Saha (\textit{Phys. Fluids}, vol. 16, 2004, pp. 1630-1646).
\end{abstract}
\section{INTRODUCTION}
Vortex shedding behind bluff bodies is a classic research area. From the first accurate measurements in the wake of a prismatic body by B\`{e}nard in 1905 and the first theoretical model by von K\'{a}rm\'{a}n in 1911 (\citealt{Provanzal2006}; \citealt{Wesfreid2006}), this subject has been one of the more active topics in fluid mechanics.

\indent The studies of the B\`{e}nard-von K\'{a}rm\'{a}n instability in two-dimensional systems characterise scenarios of transition with a well-defined Hopf bifurcation. The experimental study of \citet*{c_mathis}, concerning the wake behind a cylinder for Reynolds numbers near the threshold, proves the validity of Landau`s model for the description of the evolution of measured transversal velocity components. It is shown that the squared value of this quantity is proportional to $Re-Re_{cr}$, where $Re_{cr}$ is the Reynolds number of the onset of Hopf bifurcation. Complex Landau`s equations, including phase variables, allow the deduction of the linear dependence of frequency associated with vortex shedding on Reynolds number. Moreover, the vortex shedding process has been studied extensively using the concept of a global mode. Constant frequency of synchronised oscillations of shedding in the streamwise direction  is one of its major features. This frequency remains constant despite the fact that the flow profiles, which govern stability, vary with streamwise distance measured from the obstacle. Earlier experimental (\citealt*{c_global_mode_sophie}; \citealt*{c_global_mode_eduardo}) and numerical (\citealt*{c_global_mode_zielinska}) investigations have shown that the evolution of parameters describing the instability of the flow (e.g. transversal and longitudinal velocity components) has the character of a global mode and shows a scaling law for the spatial properties of the amplitude of the envelope of the perturbation as predicted by the 
Landau-Ginzburg model.

\indent The phenomenon of vortex shedding behind three dimensional bluff bodies has similarities with and differences from the classical two-dimensional case. For the simplest example of an axisymmetric body, a sphere, the following scenario is known. At low Reynolds numbers the separation bubble is axisymmetric and its length grows until a first regular transition occurs, after which this bubble loses axial symmetry. The value at the first onset is about Re=211 (\citealt*{s_patel}; \citealt*{s_tomboulides}; \citealt*{s_thompson}; \citealt*{s_bouchet}; \citealt{e_piotrek}). It leads to a three dimensional steady flow with planar symmetry, characterised by a double thread structure with two weak counter-rotating vortices. When the Reynolds number is increased, the flow becomes time-dependent and follows a Hopf transition. The existence of single-frequency hairpin vortex shedding is observed. Moreover, the symmetry plane from the previous regime is preserved. The Reynolds number of this second onset is reported to be about 275 (\citealt*{s_ormieres}; \citealt{s_thompson}; \citealt*{s_schouveiler}; \citealt{s_bouchet}; \citealt{e_konrad}; \citealt{e_adam}; \citealt{e_piotrek}). It should be noted that both experimental (\citealt*{s_schouveiler}; \citealt{e_konrad}) and numerical (\citealt{s_thompson}) investigations report the kinking of two trails of vorticity as a precursor to a second transition. Experiments with flow visualisations reveal that hairpin vortices are always shed on the same side (\citealt*{s_achenbach}; \citealt*{s_ormieres}; \citealt*{s_schouveiler};  \citealt{e_piotrek}). In contrast, numerical simulations of \citet{s_patel} and \citet*{d_fabre1} report two-sided hairpin vortex structures in the far wake. However, it seems that each side of a vortical structure has a different origin, and only one is caused by direct separation and shedding of part of the vortex ring formed immediately behind the bluff body in the recirculation zone (\citealt*{Brucker}).

\indent The flow instability behind a disk resembles topologically that observed for the sphere. Experimental investigations of the wake behind a disk (\citealt{e_piotrek}; \citealt*{e_tomek}) prove the existence of three subsequent regimes. In analogy to the sphere, regular bifurcation occurs with increasing Reynolds numbers, replacing axisymmetry by planar symmetry and preserving the time-invariance. It leads to the appearance of two counter-rotating filaments of vorticity. Finally, as the Reynolds number is increased, and after a Hopf transition, one-sided hairpin vortex shedding with single frequency is observed. In this regime, for higher Reynolds numbers, small irregularities between the heads of shed hairpins are present. In addition, similarly to the case of the sphere, the kinking of two trails of counter-rotating vortices is observed prior to a Hopf transition. For a disk of low aspect ratio $D/h$ (where $D$ denotes disk diameter and $h$ its thickness) onset values strongly depend on the aspect ratio, as verified experimentally in \citet{e_tomek}.

\indent In numerical simulations performed for the disk, the sequence of transitions is similar to the one observed in experiments. However, it is to be noted that one additional Hopf bifurcation between the two counter-rotating vortices and the hairpin vortex shedding regimes was reported. The description of this additional regime varies strongly between different studies (\citealt{d_fabre1}; \citealt*{d_shenoy}; \citealt*{d_meliga}). Their common feature is periodic rotation of shed structures. In particular, for a thick disk, \citet*{d_fabre2} obtained numerically a sequence of six bifurcations before the regime where regular hairpin vortex shedding is observed.

\indent Different physical situations occur when the flow behind a bluff body is not axisymmetric, as is the case for a prolate spheroid or a cube. The former case with the major axis oriented perpendicularly to the free-stream flow and for Re=50, 75, 100, 150, 200, 250 and 300 (Reynolds number is based on a equatorial diameter of the spheroid) has been analysed by \citet*{f_khoury}. The wake behind such an obstacle shares some of the features of the sphere as well as those of the planar wake behind a cylinder. For the lowest investigated Reynolds numbers (Re=50 and 75) the wake remains steady and strictly symmetric, about equatorial and meridional symmetry planes. At Re=100 the flow starts to be time-dependent, characterised by regular double-sided hairpin vortex shedding, with only one equatorial symmetry plane retained. Vortical structures formed behind the spheroid are similar to those observed for the sphere by \citealt*{s_patel}. At Re=200 the second frequency associated with hairpin vortex shedding is found. In addition no symmetry plane is present.

\indent In the case of a cube with a normal face to the flow, the basic flow exhibits orthogonal symmetry, namely four symmetry planes inclined at 45 degrees to each other. The wake of such an obstacle was investigated numerically in \citet*{k_raul}, \citet{k_saha1} and in \citet{k_saha2}. The studies of \citet{k_raul} concentrate on laminar flow at Reynolds numbers ranging from 10 to 100. The flow behind the obstacle forms a recirculation zone, whose size increases with the Reynolds number. It is also observed that corners of the bluff body contribute to a local increase of pressure in their vicinity. Moreover, no periodic vorticity shedding is observed for the investigated Reynolds number range. Their numerical results are also confirmed by experimental investigation of the drag coefficient of free-falling cubes (\citealt{k_raul}). 

\indent In Saha (\citeyear{k_saha1}, \citeyear{k_saha2}) the sequence of transitions is studied for a Reynolds number ranging from 50 to 400. The basic flow is characterised by the existence of four pairs of opposite-sign vortices. The length of the wake region behind the cube increases with increasing Reynolds number as is observed by \citet{k_raul} and the flow is found to be steady and orthogonally symmetric. Regular bifurcation from the basic state to the two counter-rotating vortices is reported at Reynolds number between 216 and 218, when the flow remains steady, but it breaks its orthogonal symmetry in such a manner that only one symmetry plane is preserved. The distribution of the longitudinal vorticity component (hereinafter denoted $\omega_x$) in this regime consists of a major bean-shaped pair of vortices and four minor vorticity filaments. A new Hopf bifurcation is found numerically to occur between the Reynolds numbers of 265 and 270. It leads to an unsteady regime where periodic hairpin vortex shedding occurs. The planar symmetry and its orientation are preserved from the previous regime. Based on streamlines, \citet{k_saha1} presents the temporal behaviour of shed hairpin vortices over a single period. The results are strikingly similar to those observed for the sphere in \citet*{s_patel}. The extraction of vortical structures described in \citet*{s_jeong} reveals the double-sided structure of the shed hairpins.

\indent The present study is concerned with the experimental investigation of the flow past a fixed cube for a range of Reynolds numbers of 100-400, which is based on the edge of the cube and on the average streamwise velocity in water channel. The front wall of the cube remains perpendicular to the inflow velocity. The main aim is to explore the sequence of transitions for a such a bluff body. As far as we know this is the first full experimental investigation of the wake behind a cube. The results obtained are compared with the numerical simulation carried out by \citet{k_saha1}. In Section~\ref{sec:SET} the experimental set-up used in our experiments is described. Results obtained, such as the evolution of the longitudinal vorticity component $\omega_x$ as a function of Reynolds number, as well as distance in the streamwise direction, the length of the recirculation zone, the frequency and the Strouhal number, are presented in Section~\ref{sec:RESULTS}. Section~\ref{sec:ANALYSIS} consists of two kinds of analysis of experimental data, namely an azimuthal Fourier decomposition of $\omega_x$ as a function of Reynolds number as well as the extraction of the components corresponding to each of the investigated regimes. Finally, Section~\ref{sec:CONC} is devoted to conclusions. 

\indent Unless otherwise stated, on all figures presented the direction of the flow is from the left to the right.

\section{EXPERIMENTAL SET-UP}\label{sec:SET}
\begin{figure}
	\begin{center}
	\begin{subfigure}[ht!]{0.8\textwidth}
	\centerline{\includegraphics[width=1\textwidth]{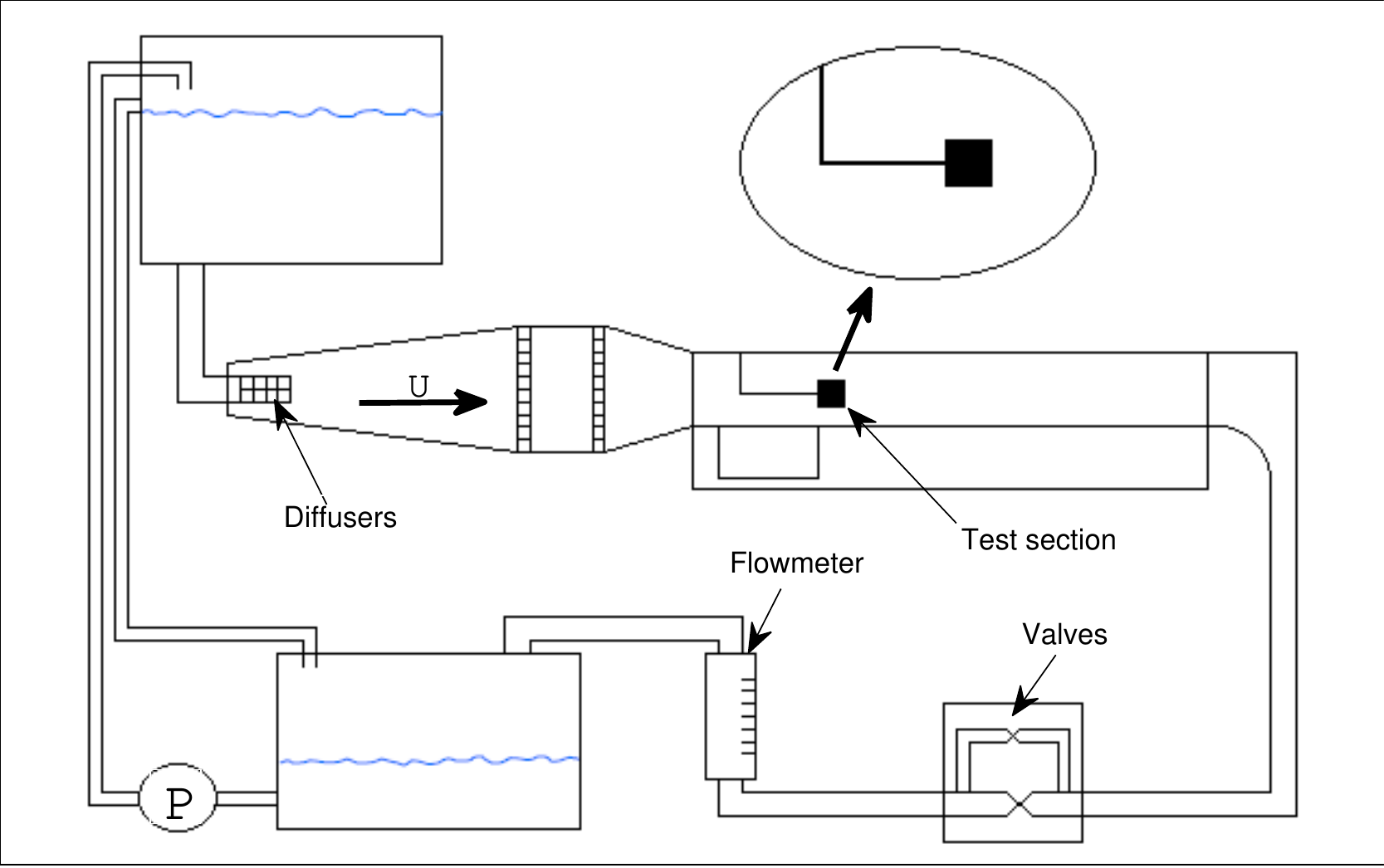}}
	\end{subfigure}
	\caption {(Colour online) Scheme of the water channel.}
	\label{fig:Figure1}
	\end{center}
\end{figure}

\indent The present experiments were carried out in a horizontal water channel adapted for low Reynolds numbers and used previously in similar experiments (see~\citealt*{c_thiria} and~\citealt*{phdMarais}). It is presented schematically in figure~\ref{fig:Figure1}. The internal dimensions of its cross-section and the length of the test section were 100 mm x 100 mm and 860 mm respectively. 

\indent The flow was induced by gravity, using a constant-level tank to provide a constant pressure gradient. Two honeycomb-type filters were present in the front of the test section to make the flow fully uniform. The turbulence level was lower than 2\%, obtained by considering the temporal standard deviations of measured velocity. It was demonstrated for the wake of a sphere that this level of free-stream turbulence has negligible influence on the vortex shedding (see~\citealt*{t_wu}, ~\citealt*{t_mittal} and~\citealt*{t_bagchi}). Velocity of the flow in the channel was controlled by means of two throttling valves and measured by a calibrated flow meter.
\\
\indent The cube with edge length of $d=12$ mm was mounted in the central part of the cross-section at the beginning of the test section. The leading face was perpendicular to the undisturbed flow, while its lateral faces were parallel to the internal walls of the water channel. The support system consisted of a rigid, nearly horizontal, bent tube of diameter $\phi=1.7$ mm with a threaded end to which the cube was screwed. The length of the horizontal part of the tube was equal to 100 mm to diminish the influence of its vertical part on the flow. This length was still sufficiently small to avoid elastic oscillations of the cube support. On the other hand the Reynolds number, based on the diameter of the support tube, was at maximum 50, limiting the existence of the B\`{e}nard-von K\'{a}rm\'{a}n vortex street. As a result it was possible to investigate the wake behind the cube as if it were an isolated body.

\begin{figure}
	\begin{center}
	\begin{minipage}[!hb]{0.27\linewidth}
		\begin{subfigure}[b]{\textwidth}
       			\centering
                \includegraphics[width=\textwidth]{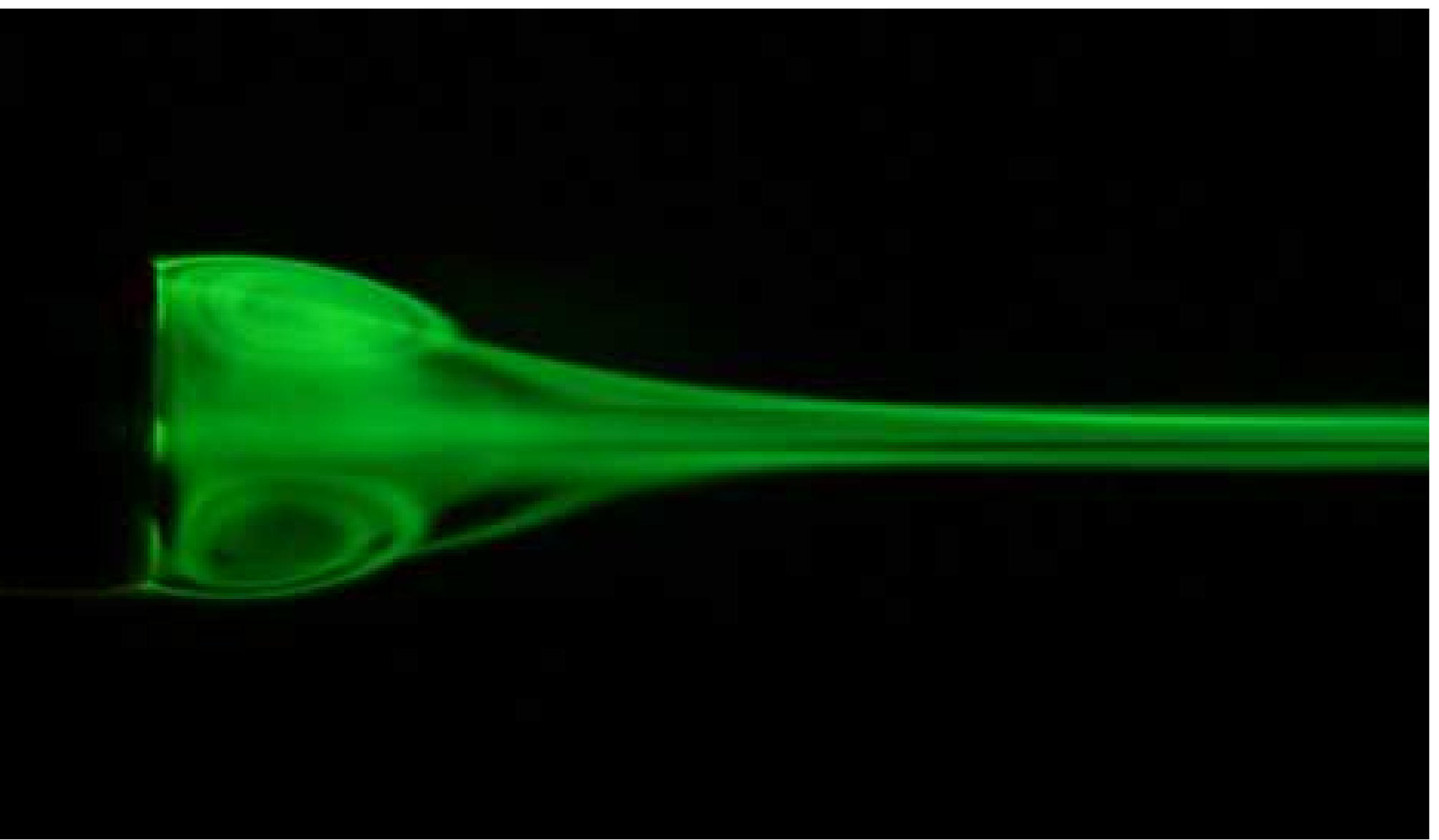}
                \caption{side view}
                \label{fig:Figure2a}
   
        \end{subfigure}
        \begin{subfigure}[b]{\textwidth}
                \centering
                \includegraphics[width=\textwidth]{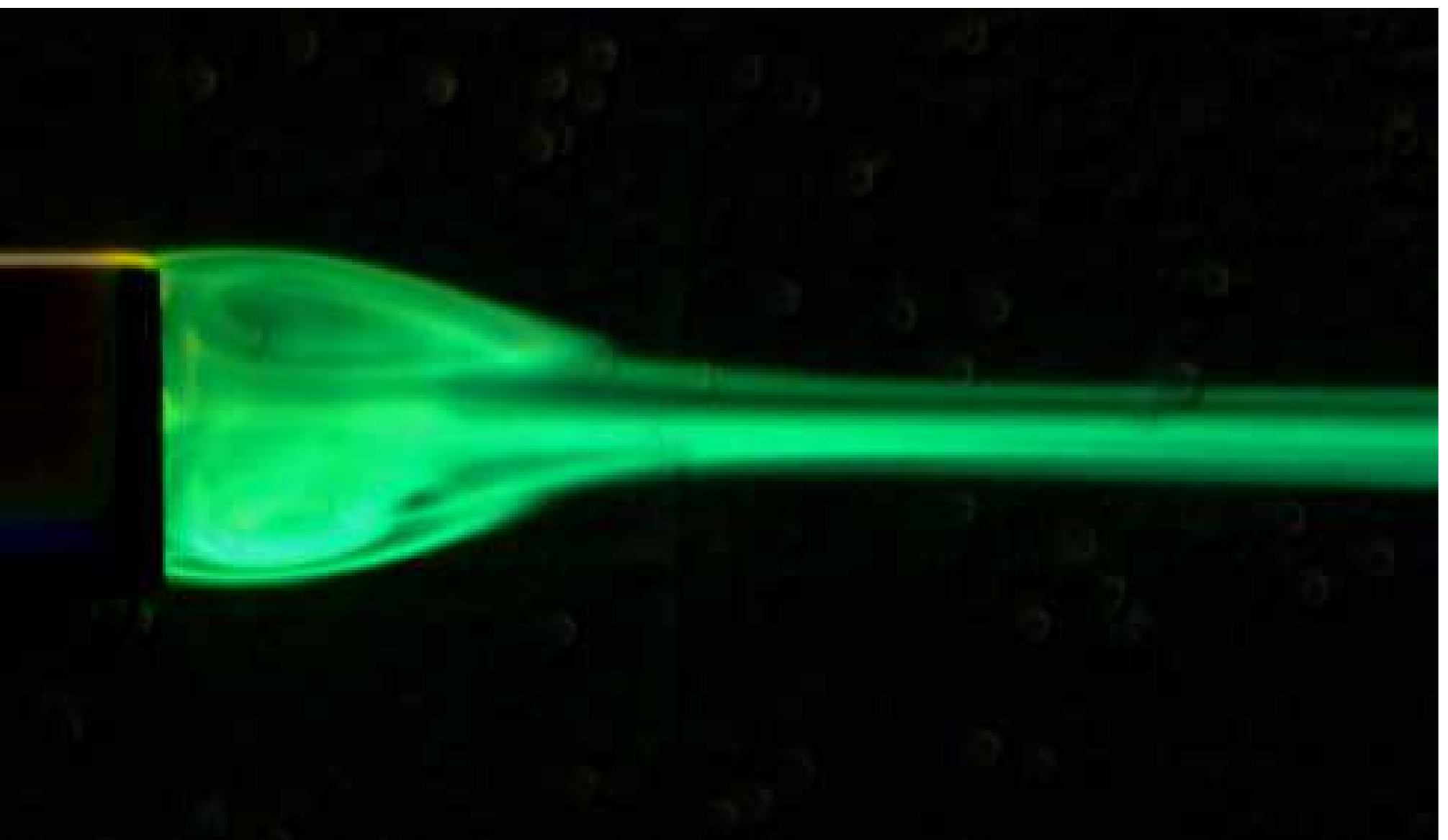}
                \caption{top view}
                \label{fig:Figure2b}
        \end{subfigure}
	\end{minipage}
	\begin{minipage}[!hb]{0.63\linewidth}
		\begin{subfigure}[b]{\textwidth}
       			\centering
                \includegraphics[width=\textwidth]{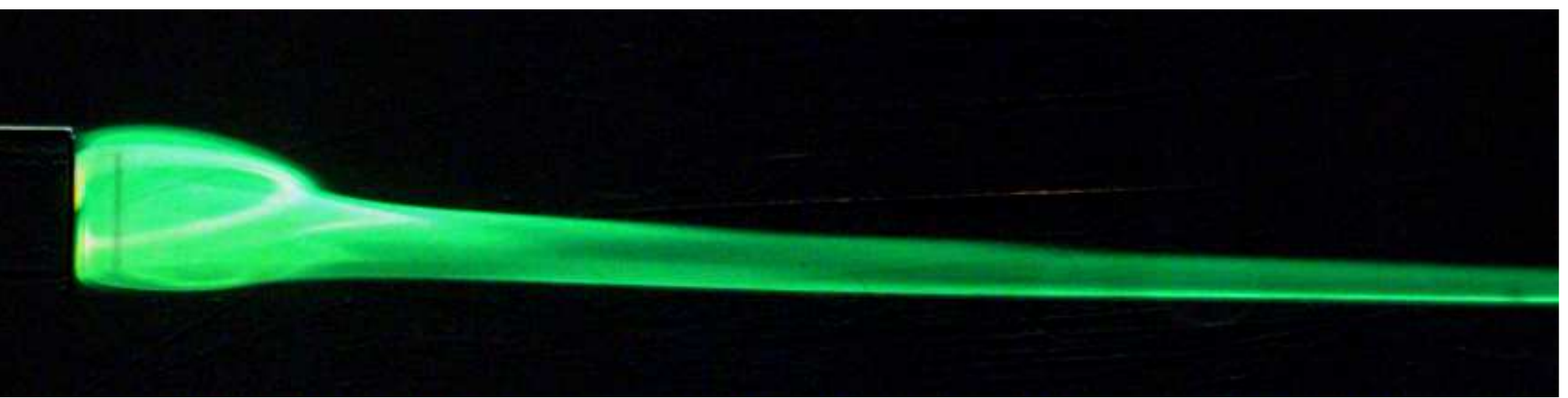}
                \caption{side view}
                \label{fig:Figure2c}
   
        \end{subfigure}
        \begin{subfigure}[b]{\textwidth}
                \centering
                \includegraphics[width=\textwidth]{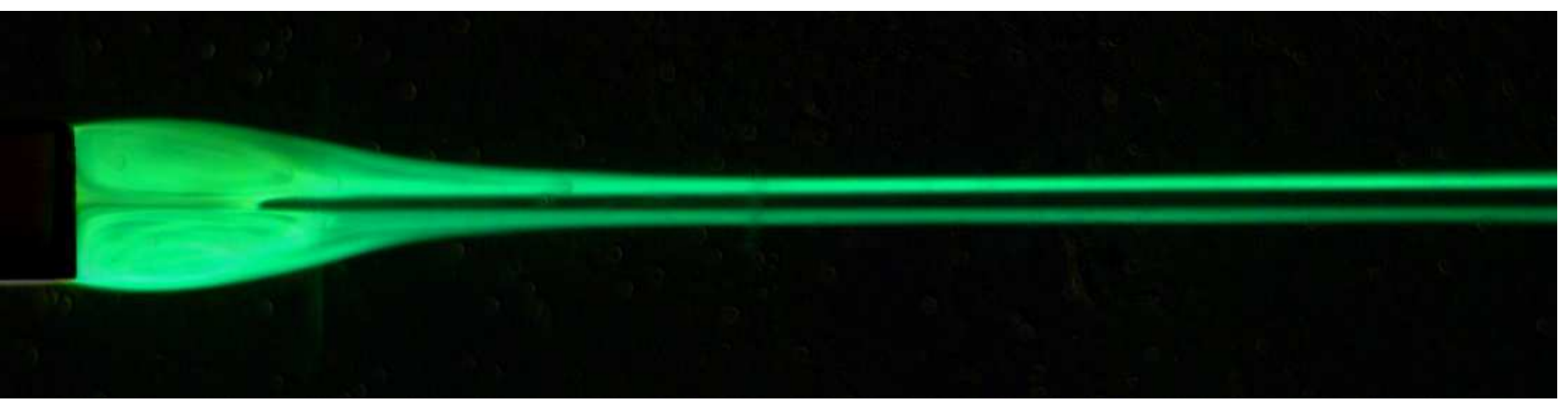}
                \caption{top view}
                \label{fig:Figure2d}
        \end{subfigure}
	\end{minipage}
	
	\begin{minipage}[!hb]{0.91\linewidth}
		\begin{subfigure}[b]{\textwidth}
       			\centering
                \includegraphics[width=\textwidth]{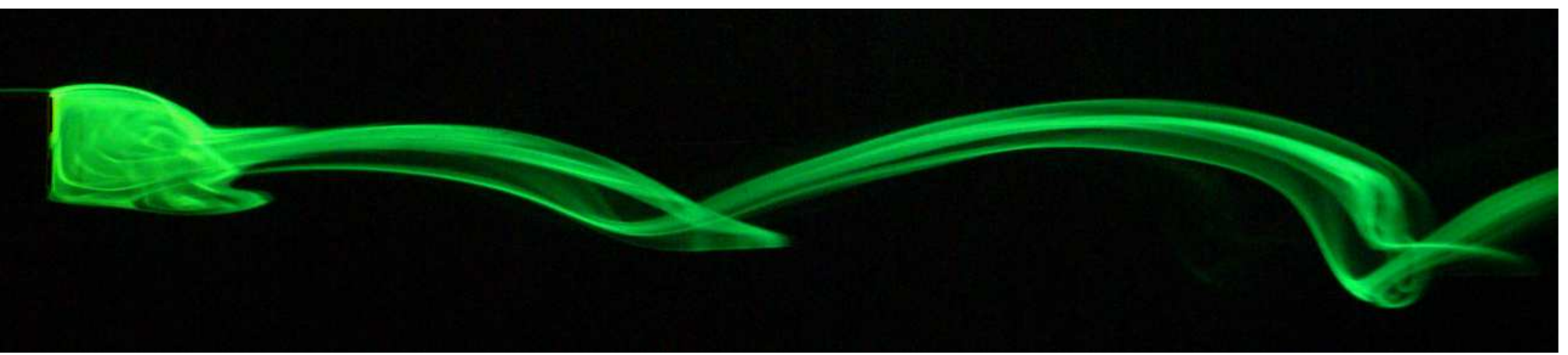}
                \caption{side view}
                \label{fig:Figure2e}
   
        \end{subfigure}
        \begin{subfigure}[b]{\textwidth}
                \centering
                \includegraphics[width=\textwidth]{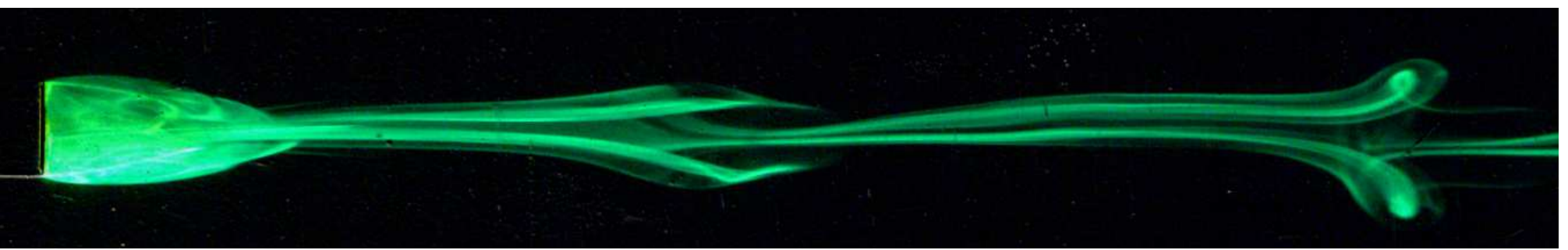}
                \caption{top view}
                \label{fig:Figure2f}
        \end{subfigure}
	\end{minipage}
	\captionsetup{format=default, justification=justified, width=12cm}
	\caption{Visualisations patterns of three consecutive regimes: \textit{(a,b)} the basic flow at $Re=100$, 
	\mbox{\textit{(c,d)} the two counter-rotating vortices regime at $Re=250$}, \textit{(e,f)} the hairpin vortex shedding regime at $Re=300$.}
	\label{fig:Figure2}
	\end{center}
\end{figure}

\indent For qualitative study of three-dimensional vortical structures behind the cube, laser induced fluorescence (LIF) visualisations were performed. The tube delivering the colourant into the recirculation zone was also used as the support of the cube. The fluorescein dye was ejected from holes placed on the rear wall of the cube and was forced to flow by a controlled syringe pump. Figure~\ref{fig:Figure2d} confirms that there was no significant perturbation of the main flow through the introduction of the dye. The configuration of nine holes placed evenly on the rear wall of the cube was chosen as providing the best results. 

\indent The fluorescein was excited by means of an appropriate light source (argon laser in the case of two-dimensional visualisations, and a UV-lamp for three-dimensional observations).

\indent A two-dimensional particle image velocimetry (PIV) method was used to obtain quantitative data about the flow field velocity. We have used a standard image PIV set-up, which consisted of an ImagePro 1600x1200 12 bit CCD camera with Nikkor f50mm, f85 mm and f70-180 mm lenses, a Minilite ND:YAG, double-pulsed laser, as well as software and hardware delivered by LaVision$^{\textregistered}$. The flow was seeded with spherical particles of a typical diameter of 11 $\mu$m. For all the PIV measurements, we used an interrogation window of 32x32 pixels with an overlap of 50\%. The distance between two adjacent vectors in the plane of measurements was equal to about 0.6 mm in the plane parallel to the rear wall of the cube and 0.8 mm in the measurements of the recirculation zone. The error in the measurements of the time-averaged vorticity fields was estimated to remain below 5\%. 

\indent Data acquisition was done with 15 Hz frequency. Unless otherwise stated, for each Reynolds number 1000 pairs of snapshots were recorded. From each pair of snapshots one instantaneous velocity field was determined. Subsequently a streamwise vorticity component was calculated from the measured velocity of the flow, by numerical derivation. 

\setcounter{bottomnumber}{6}
\begin{figure}
	\begin{center}
	\begin{minipage}[!hb]{\textwidth}
		\begin{center}
		\begin{subfigure}[b]{0.9\textwidth}
       			\centering
                \includegraphics[width=\textwidth]{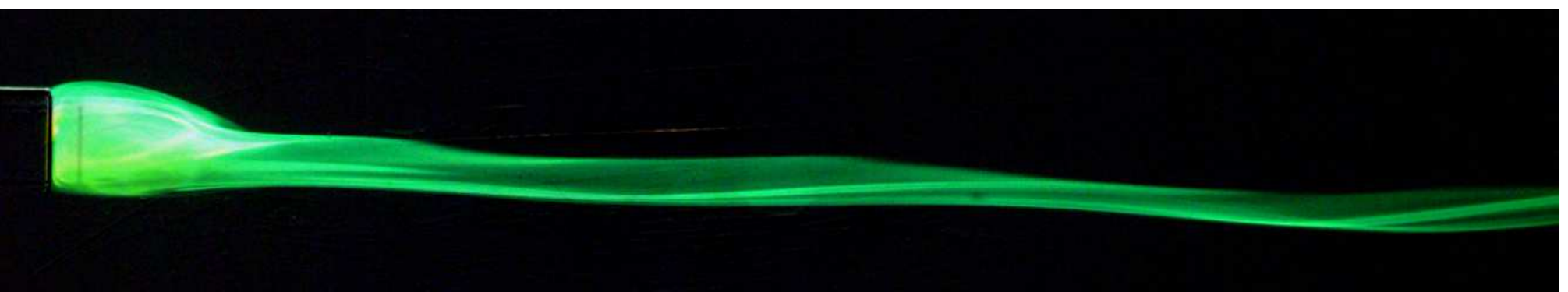}
                \caption{}
                \label{fig:Figure3a}   
        \end{subfigure}
        
        \begin{subfigure}[b]{0.9\textwidth}
                \centering
                \includegraphics[width=\textwidth]{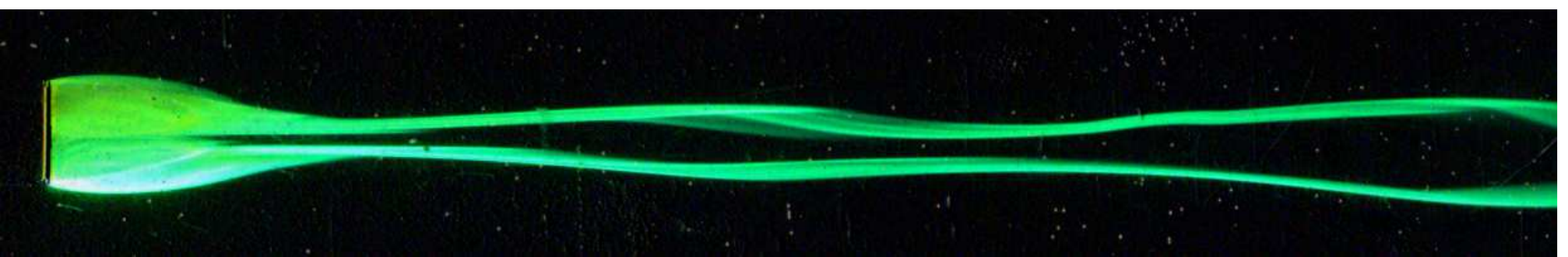}
                \caption{}
                \label{fig:Figure3b}
        \end{subfigure}
        \end{center}
	\end{minipage}	
	\caption{Peristaltic oscillation of the two trails of counter-rotating vortices prior to regular shedding of hairpin vortices at $Re=277$: \textit{(a)} side view, \textit{(b)} top view. }
	\label{fig:Figure3}
	\end{center}
\end{figure}
\section{EXPERIMENTAL RESULTS}\label{sec:RESULTS}
\subsection{Flow regimes}
\begin{figure}
	\begin{center}
	\begin{minipage}[!hb]{\linewidth}
		\begin{center}
		\begin{subfigure}[b]{0.37\textwidth}
       			\centering
                \includegraphics[width=\textwidth]{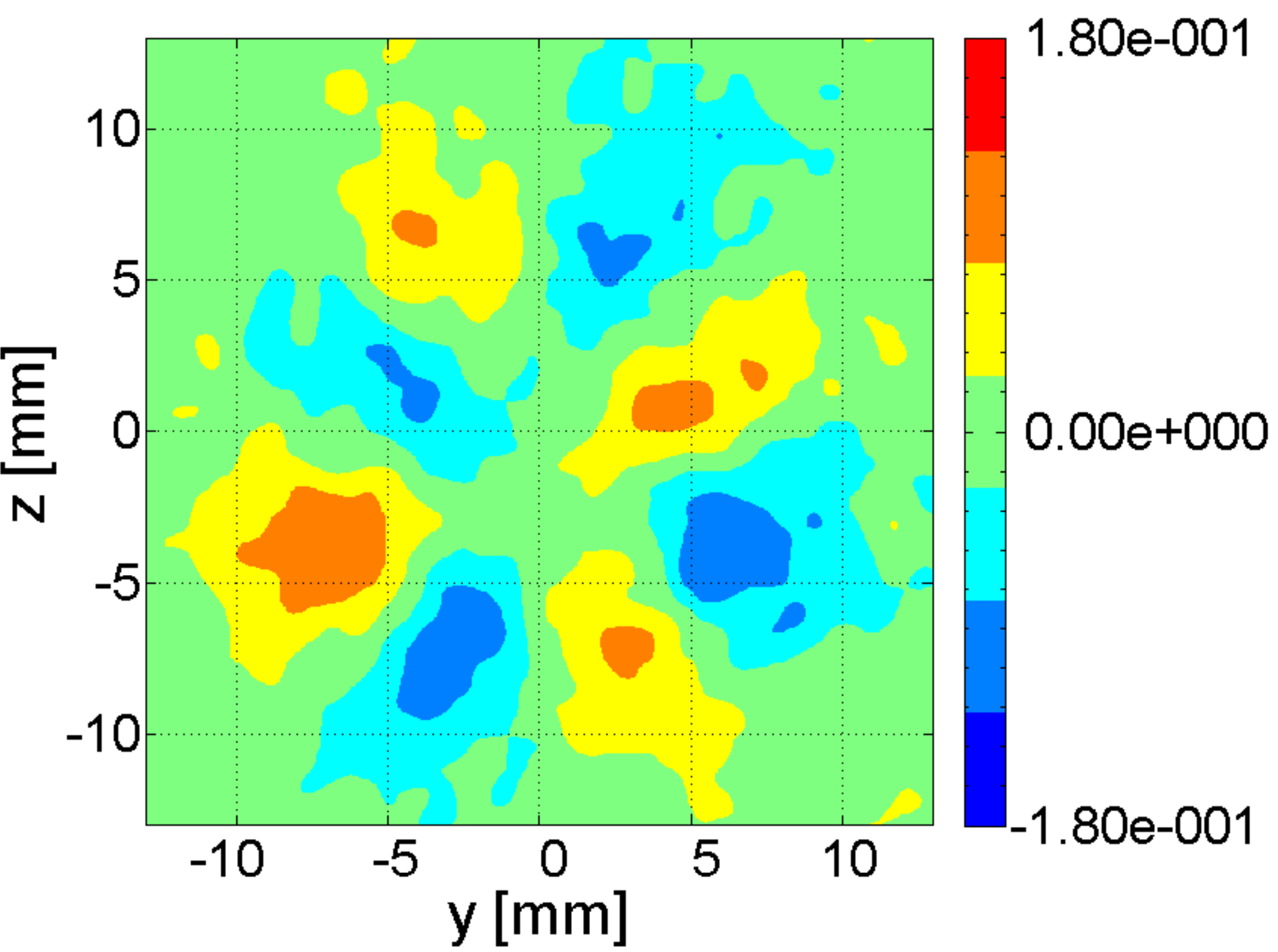}
                \caption{}
                \label{fig:Figure4a}   
        \end{subfigure}
        \begin{subfigure}[b]{0.37\textwidth}
                \centering
                \includegraphics[width=\textwidth]{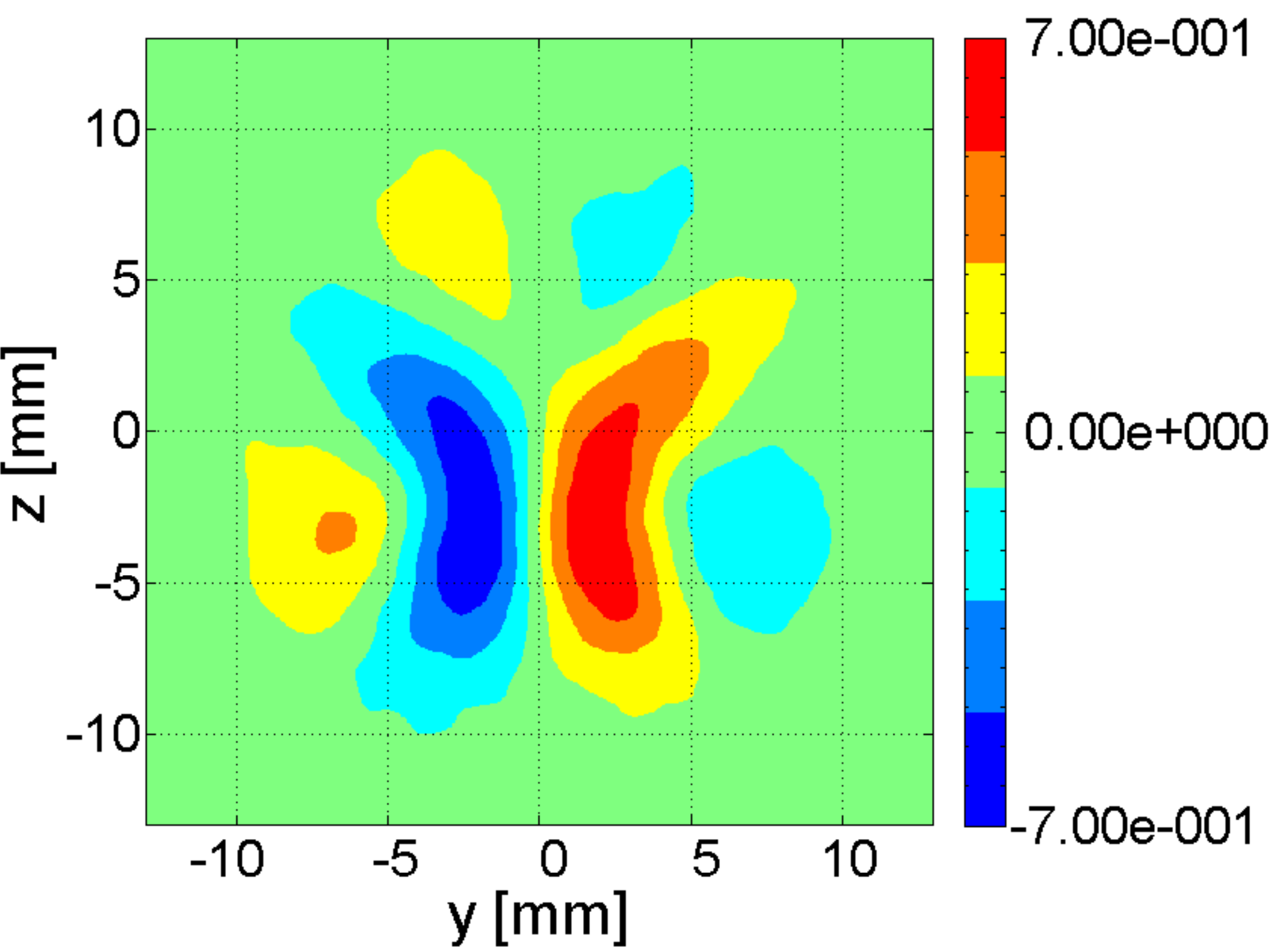}
                \caption{}
                \label{fig:Figure4b}
        \end{subfigure}
        \end{center}
	\end{minipage}	
	\begin{minipage}[!hb]{0.74\linewidth}
		\begin{subfigure}[b]{\textwidth}
       			\centering
                \includegraphics[width=\textwidth]{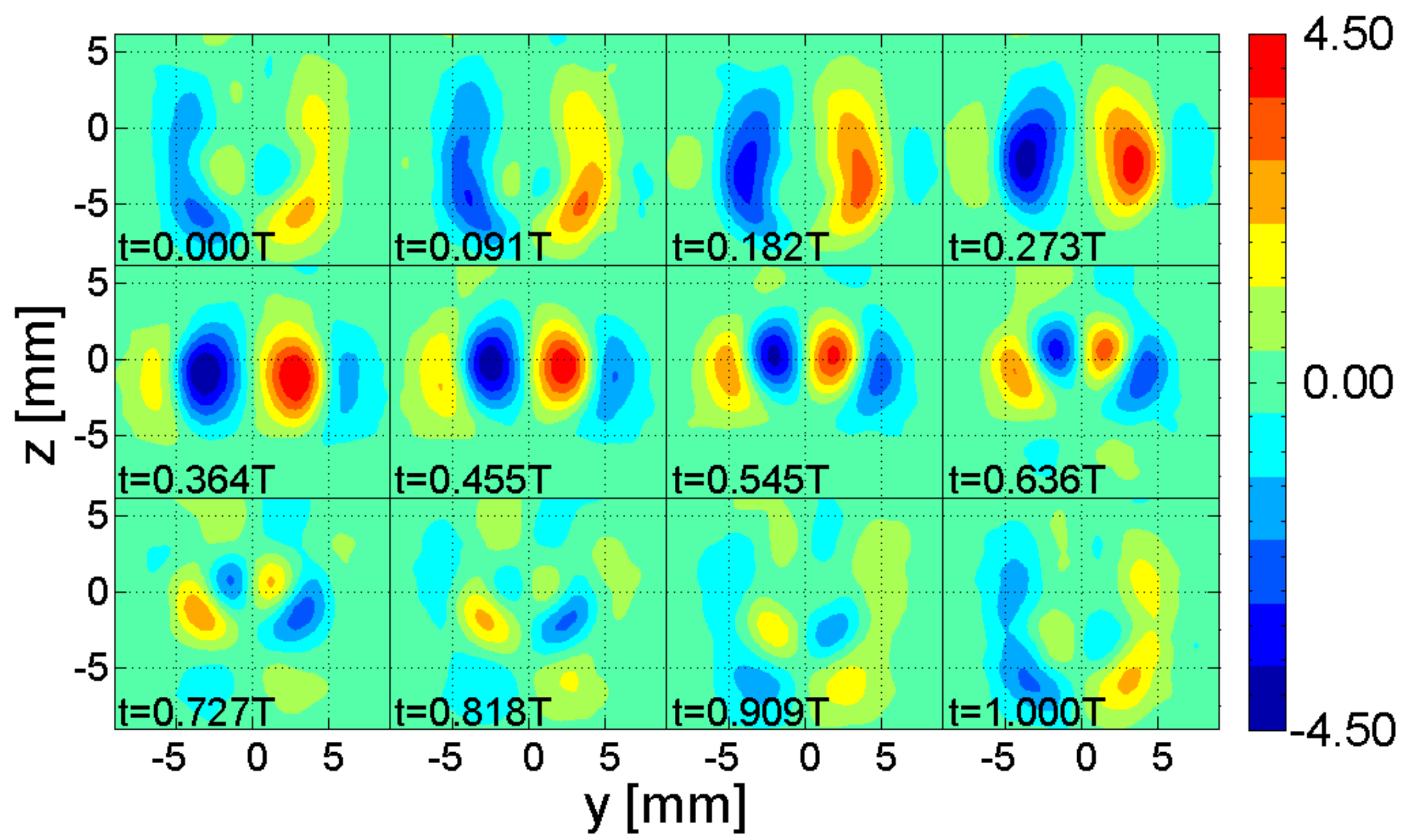}
                \caption{}
                \label{fig:Figure4c}
        \end{subfigure}
	\end{minipage}
	\caption{(Colour online) Longitudinal vorticity fields at a cross-section placed at $x/d=1.5$ from PIV measurements: \textit{(a)} the steady basic flow, averaged over all captured images at $Re=132$; \textit{(b)} the two steady counter-rotating vortices regime, averaged over all captured images at $Re=239$; \textit{(c)} time evolution of $\omega_x$ in the hairpin vortex shedding regime at $Re=316$ (T is a single cycle).}
	\label{fig:Figure4}
	\end{center}
\end{figure}
\begin{figure}
	\begin{center}

		\begin{subfigure}[b]{0.49\textwidth}
       			\centering
                \includegraphics[width=\textwidth]{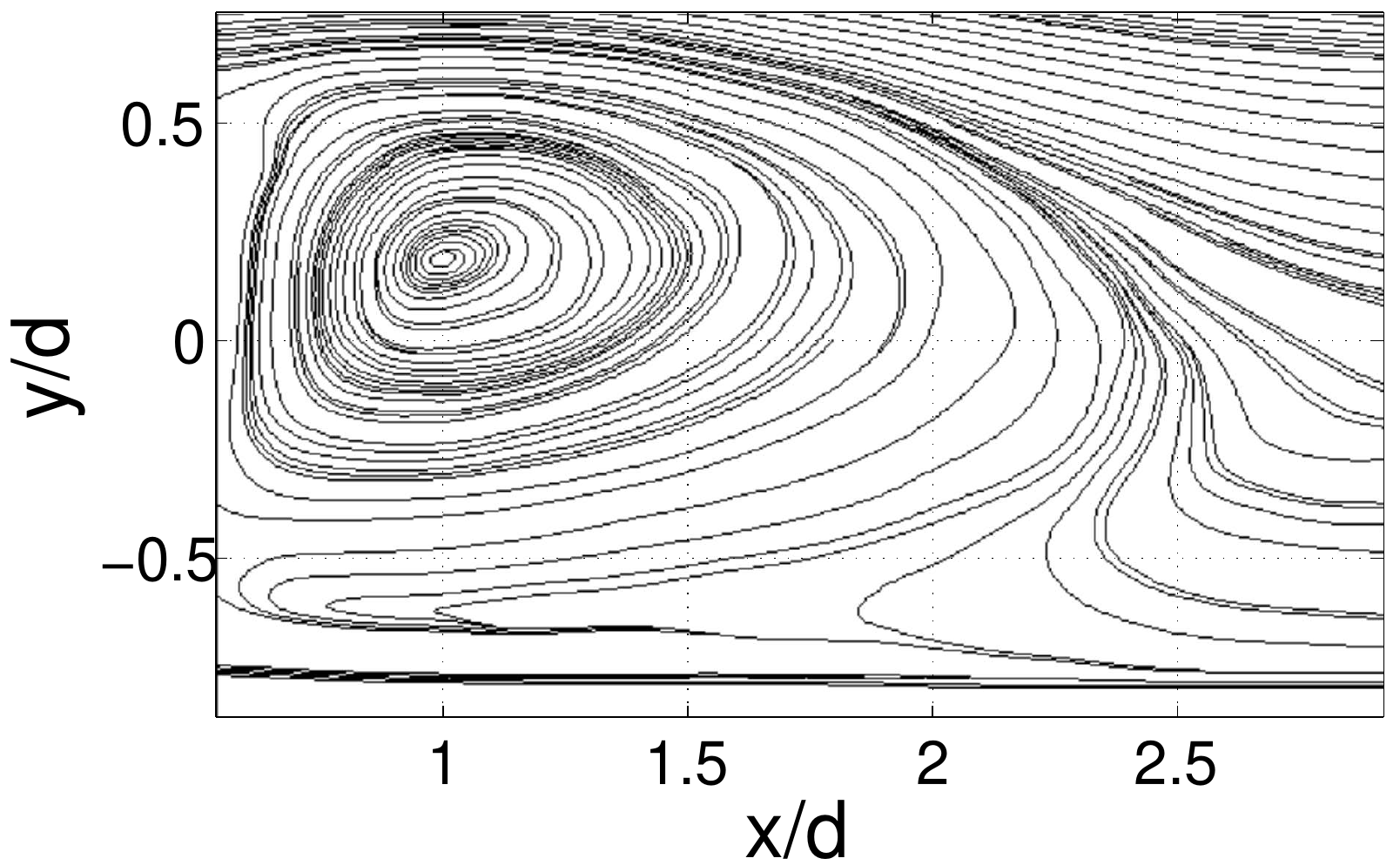}
                \caption{}
                \label{fig:Figure5a}
   
        \end{subfigure}
        \begin{subfigure}[b]{0.49\textwidth}
       			\centering
                \includegraphics[width=\textwidth]{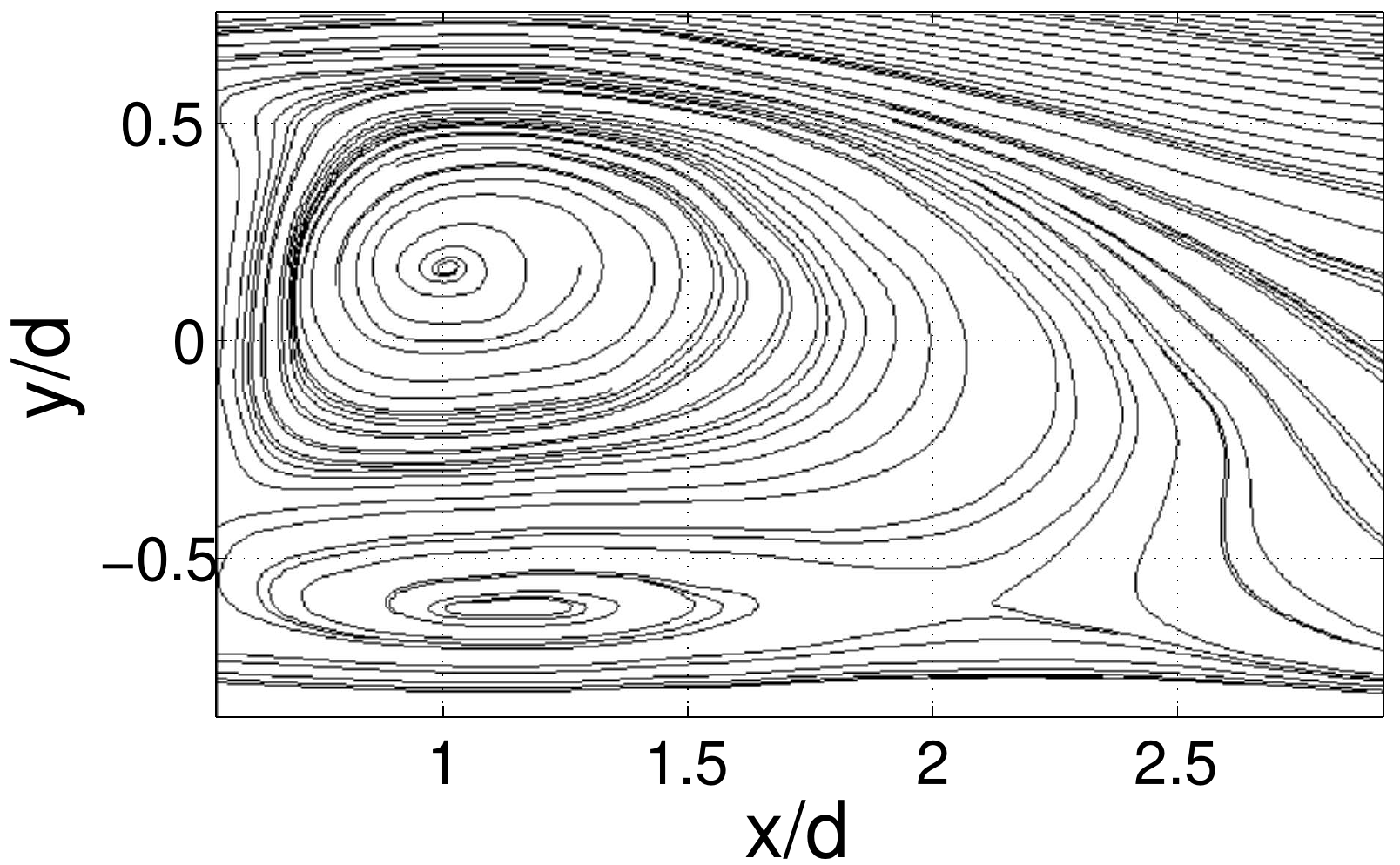}
                \caption{}
                \label{fig:Figure5b}
        \end{subfigure}

		\begin{subfigure}[b]{0.49\textwidth}
                \centering
                \includegraphics[width=\textwidth]{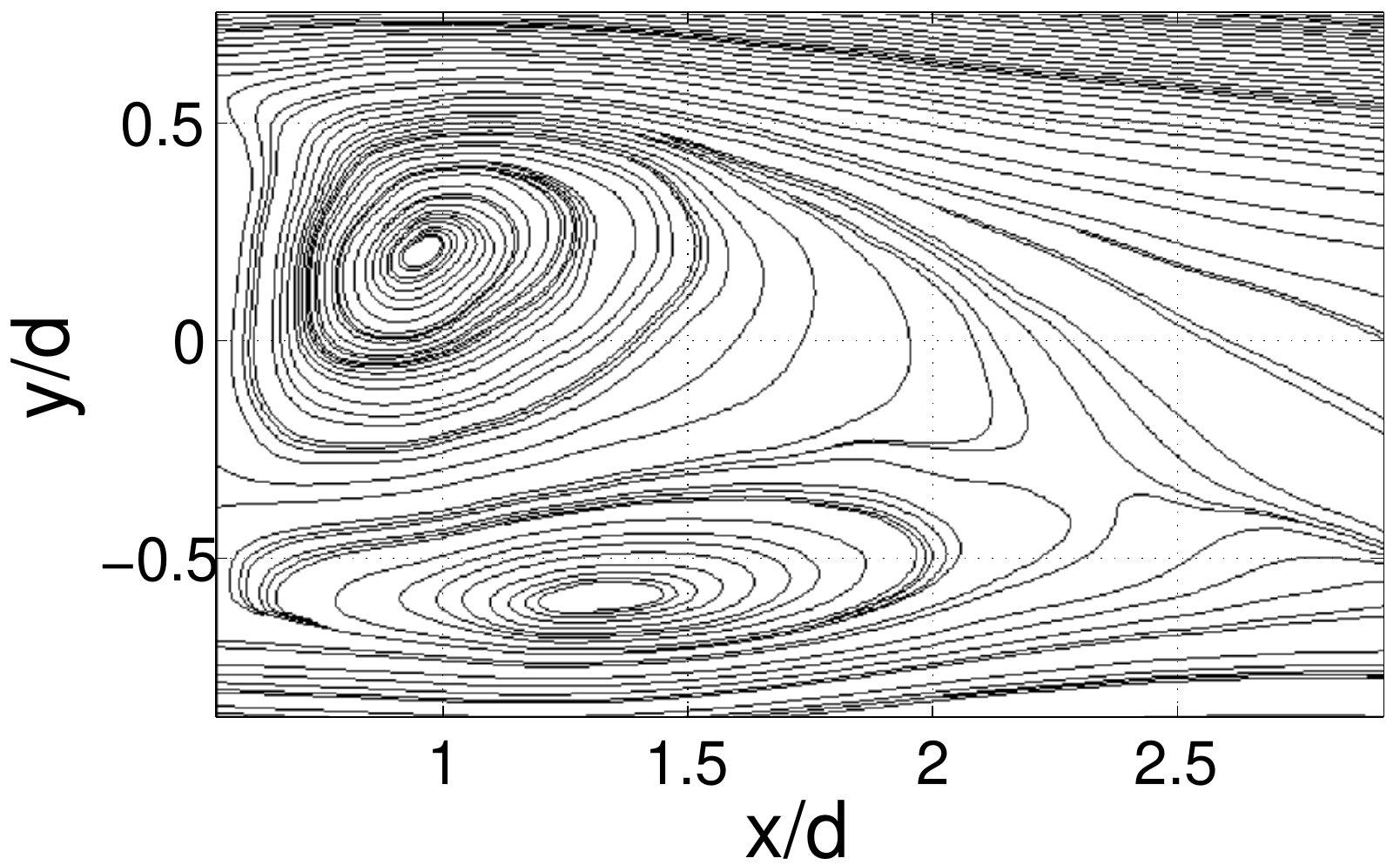}
                \caption{}
                \label{fig:Figure5c}
        \end{subfigure}
        \begin{subfigure}[b]{0.49\textwidth}
                \centering
                \includegraphics[width=\textwidth]{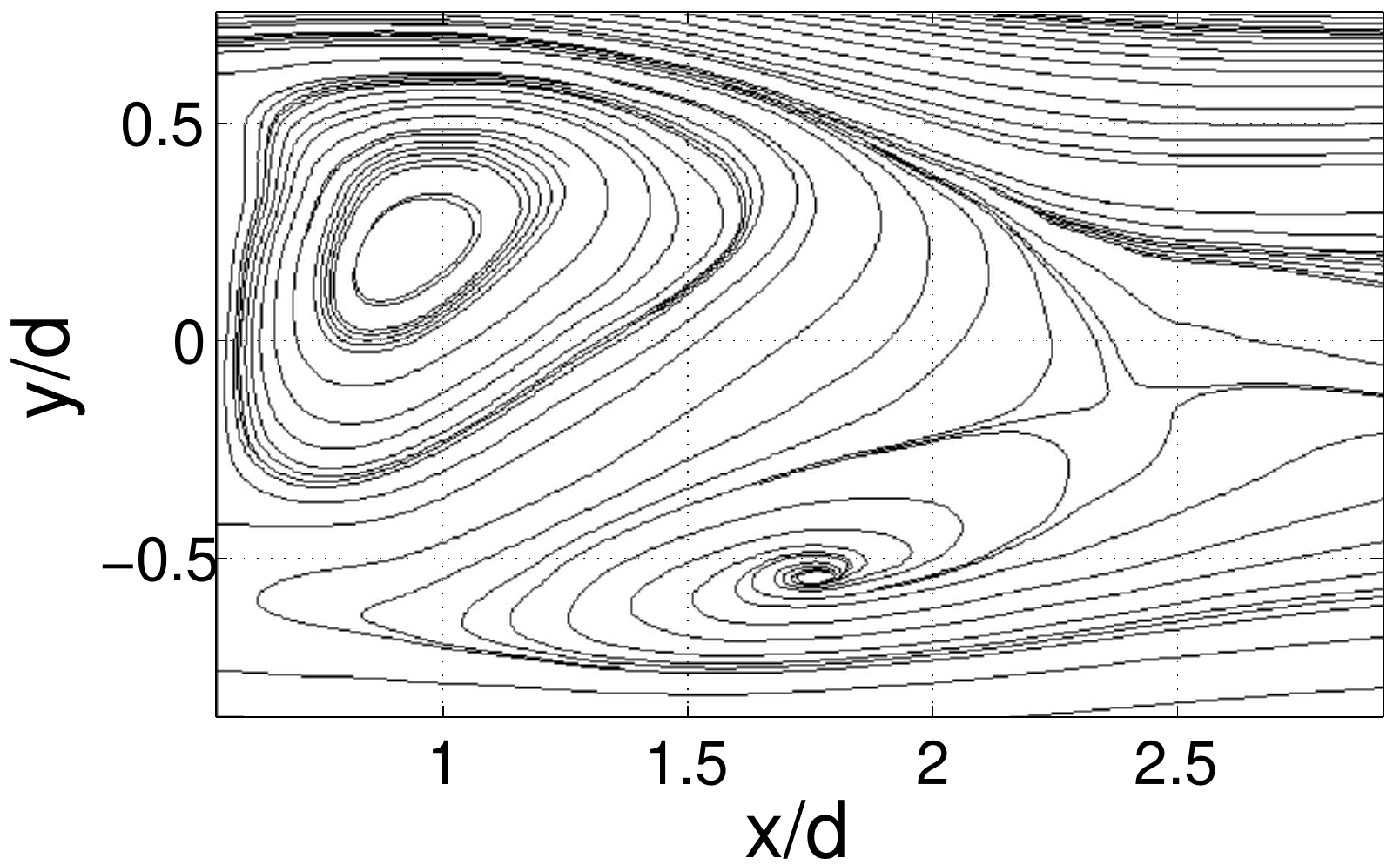}
                \caption{}
                \label{fig:Figure5d}
        \end{subfigure}

	\captionsetup{format=default, justification=justified, width=12cm}
	\caption{Streamline patterns of the lower one-sided vortex shed at $Re=331$, obtained through PIV measurements during a single cycle  $T$. \textit{(a)} $t=T/4$, \textit{(b)} $t=2T/4$, \textit{(c)} $t=3T/4$, \textit{(d)} $t=4T/4$.}
	\label{fig:Figure5}
	\end{center}
\end{figure}
In the range of Reynolds numbers from 100 to 400, three different well-defined flow regimes were confirmed in our experiments. The initial basic flow was found to be steady with orthogonal symmetry, characterised by the existence of four pairs of streamwise counter-rotating vortices originating at the lateral edges of the cube. It should be noted that it was not possible to visualise these basic flow vortices, using LIF, as they were apparently very weak. Figures~\ref{fig:Figure2a} and~\ref{fig:Figure2b} shows only large areas of diffused fluorescein. In contrast, it was possible to distinguish these four pairs of vortices using the PIV method (see figure~\ref{fig:Figure4a}) where red and blue correspond to anti-clockwise and clockwise longitudinal vorticity respectively. The presence of these vortices can be explained by analysis of the numerical results obtained by \citet{k_raul}. As mentioned in the introduction, they report the existence of local maxima of pressure in the vicinity of corners on the rear wall of the cube. This transversal pressure gradient induces motion of fluid toward the centre of this wall. Superposition of this motion with streamwise velocity results in the appearance of four pairs of weak counter-rotating vortices. One can consider this vorticity, associated with the basic flow, as extrinsic, not related to instability effects and without onset.

\indent Subsequently, after a regular transition from the basic flow to the two counter-rotating vortices regime, the observed flow remains steady and only one symmetry plane is preserved. In figure~\ref{fig:Figure2d} one can recognise two distinct filaments of fluorescein, which correspond to a major, bean-shaped pair of counter-rotating vortices visible in the cross-section displayed in figure~\ref{fig:Figure4b}. In addition, four minor filaments of vorticity are present, as a remaining component of the basic flow. In figure~\ref{fig:Figure2c} one can clearly observe the manifestation of the influence of the major, bean-shaped pair of vortices. In the vicinity of the symmetry plane this dipole of vorticity induces velocity oriented downward. Due to this, the trails of fluorescein move downwards while being convected downstream. 

\indent The presented flow structures of the basic flow (figure~\ref{fig:Figure4a}) and of the two counter-rotating vortices regime (figure~\ref{fig:Figure4b}) can be directly compared with numerical results obtained by Saha (see figures 4 and 11a in \citet{k_saha1}). It is useful to note that the spatial distribution of longitudinal vorticity is virtually the same for both stationary regimes. 

\indent Finally, a Hopf transition, namely from the two counter-rotating vortices to the hairpin vortex shedding regime, occurs for higher Reynolds numbers and leads to a time-periodic flow characterised by a regular one-sided hairpin vortex shedding process. This flow topology is confirmed using both LIF visualisations (figures~\ref{fig:Figure2e} and~\ref{fig:Figure2f}) and PIV measurements (figure~\ref{fig:Figure4c}). The symmetry plane of the previous regime, as well as its orientation, remain the same and are determined by the orientation of the vertical part of the support of the cube.

\indent We have also performed a preliminary study of the effect of cube rotation around the axis aligned with the undisturbed freestream for two Reynolds numbers $Re=250$ and $Re=330$, which are typical for the two counter-rotating vortices and hairpin vortex shedding regimes, respectively. In the first regime, the symmetry plane of the flow pattern seems to follow the rotation of the cube up to almost $\alpha$=30$^\circ$ (where $\alpha$ denotes an inclination of the side walls of the cube with respect to the vertical axis). For higher values of $\alpha$ the symmetry plane tends to be aligned with the vertical diagonal of the cube. Similar behaviour was also observed for the hairpin vortex shedding regime; however the switch of symmetry plane occurred around $\alpha$=20$^\circ$. The quantitative description of the observed effect requires further investigation. 

\indent In the present paper, however, the focus remains on the consecutive transition for the basic case of $\alpha$=0, where cube's side faces were parallel to the side walls of the water channel.

\indent For this case, figure~\ref{fig:Figure5} illustrates consecutive patterns of the flow as the vortex is shed from the recirculation zone. The hairpins appear on one side only, as a result of the deformation and tilting of the vortex ring formed immediately behind the obstacle. This process is analogous to those presented in \citet{k_saha1} for the cube and in \citet*{s_patel} for the sphere. 

\indent In the present study, peristaltic oscillation or kinking of two trails of counter-rotating vortices is observed prior to a Hopf bifurcation (figures~\ref{fig:Figure3a} and~\ref{fig:Figure3b}). Similar behaviour is reported for the case of a sphere (\citealt{s_thompson}; \citealt*{s_schouveiler}; \citealt{e_konrad}), for a disk of low aspect ratio (\citealt*{d_shenoy}; \citealt{e_piotrek}; \citealt{e_tomek}) and behind an afterbody (\citealt*{Bohorquez}). However, it should be noted that \citet*{d_shenoy} observed that the kinking is accompanied by periodic rotation in disks.

\indent Accurate values of $Re$ for transition onsets will be identified in Section~\ref{sec:ANALYSIS}.

\subsection{Length of recirculation zone}
In order to investigate the sequence of transitions, the length of the recirculation zone was determined using the PIV method for a sequence of Reynolds numbers. The laser sheet coincided with the symmetry plane during the whole investigated range of Reynolds numbers. For steady flow (the basic flow and the two counter-rotating vortices regime) the obtained velocity fields were averaged. In the non-stationary regime the recirculation region oscillates following the hairpin vortex shedding process. For this reason, at least 15 full periods were taken to obtain a time average. 

\begin{figure}
	\begin{center}
	\begin{subfigure}[ht!]{0.8\textwidth}
	\centerline{\includegraphics[width=1\textwidth]{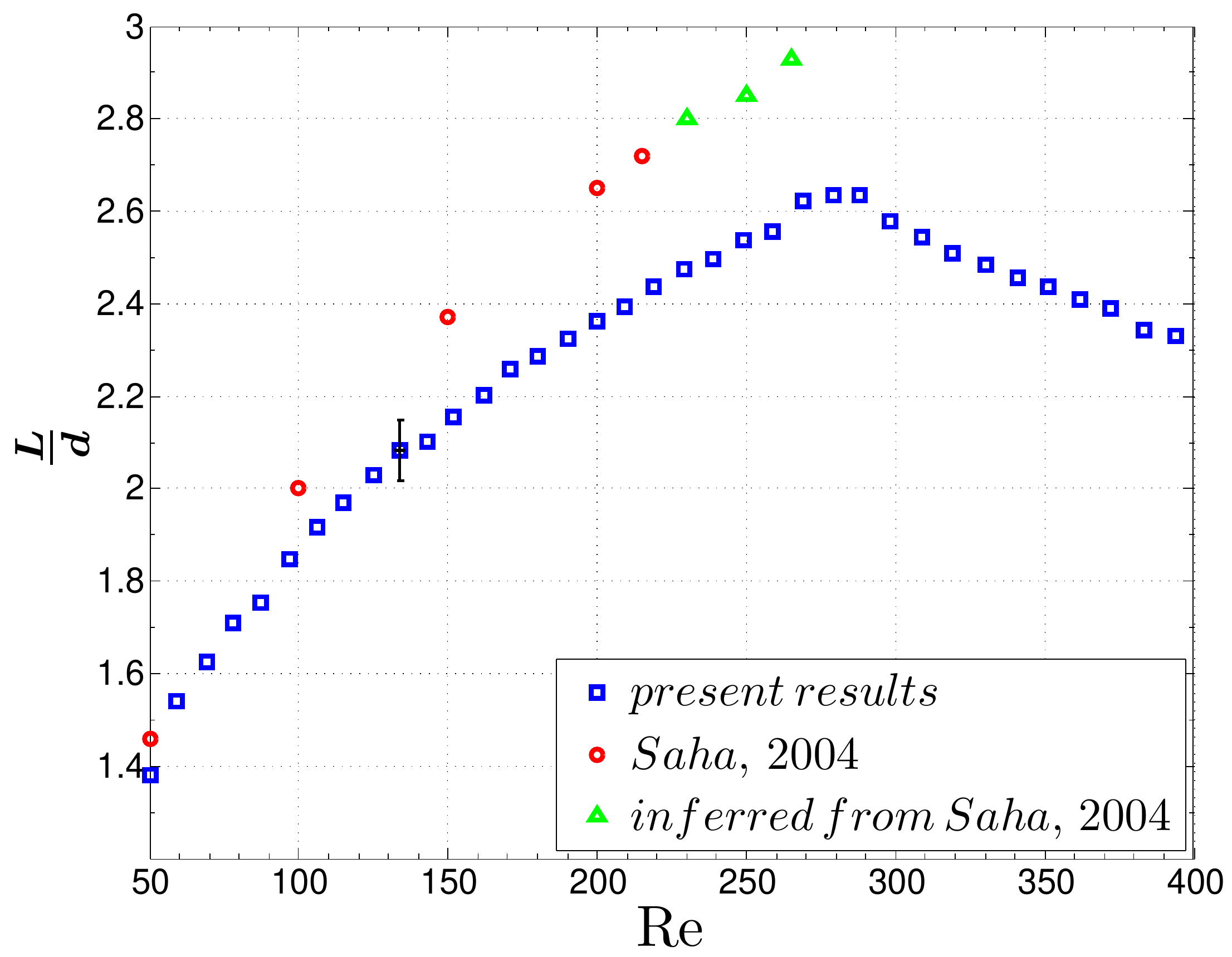}}
	\end{subfigure}
	\caption{(Colour online) Normalised length of recirculation zone as a function of Reynolds number.}
	\label{fig:Figure6}
	\end{center}
\end{figure}

\indent The length of a recirculation bubble was defined as the distance from the centre of the cube to the place where a streamwise component of velocity changes its sign. As it is difficult to identify this place directly, we have interpolated the positive and negative values in the vicinity of zero, which gave us the required length $L$ of the recirculation zone (normalised using a cube edge length $d$). Figure~\ref{fig:Figure6} presents the results obtained, which agree in trend with the numerical values reported by \citet{k_saha1} (see results in his table IV). Three additional values were inferred by the present authors from the streamlines pictures of \citet{k_saha1} (see his figures 10b, 14a and 14b) at the highest Re values. One can see that the recirculation zone grows with the Reynolds number up to a Hopf bifurcation from a stationary to an unsteady regime. For the basic flow, the evolution of $L/d$ deviates from a linear behaviour. The same effect was reported for a sphere by \citet*{s_tomboulides} and \citet{s_bouchet}. 

\indent Following the regular bifurcation, the length of the wake continues to grow with increasing Reynolds numbers. However the second bifurcation reverses this increasing trend. The analogous evolution of the recirculation zone, across the whole investigated Reynolds numbers range, was reported for the case of a disk by \citet{e_tomek}. For the case of the wake past a cylinder (\citet{c_nonlinear}) and a disk (\citet{d_meliga}), a similar decrease of the recirculation zone was observed in the unsteady regime. The nonlinear correction to the mean flow, generated by the Reynolds stress of the fluctuations of the unstable mode, induces an adverse pumping which stretches the recirculation loop. The fact that the two stationary counter-rotating vortices do not modify the recirculation length much implies that these structures do not contribute significantly to this mean flow correction.
\subsection{Frequency and Strouhal number}
\begin{figure}
	\begin{center}
	\begin{minipage}[!hb]{\textwidth}
		\begin{center}
		\begin{subfigure}[b]{0.8\textwidth}
       			\centering
                \includegraphics[width=\textwidth]{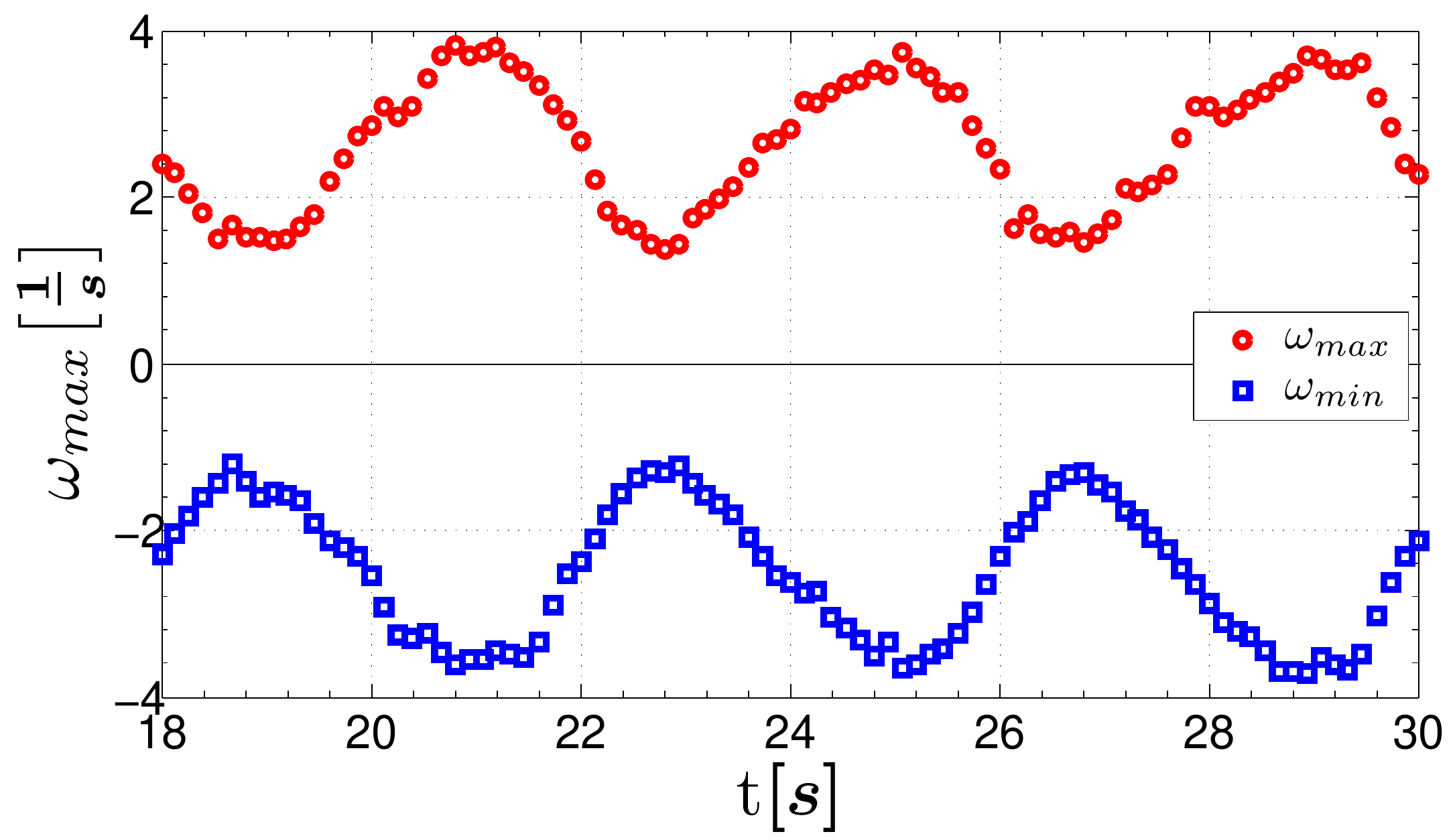}
                \caption{}
                \label{fig:Figure7a}   
        \end{subfigure}
		\begin{subfigure}[b]{0.8\textwidth}
       			\centering
                \includegraphics[width=\textwidth]{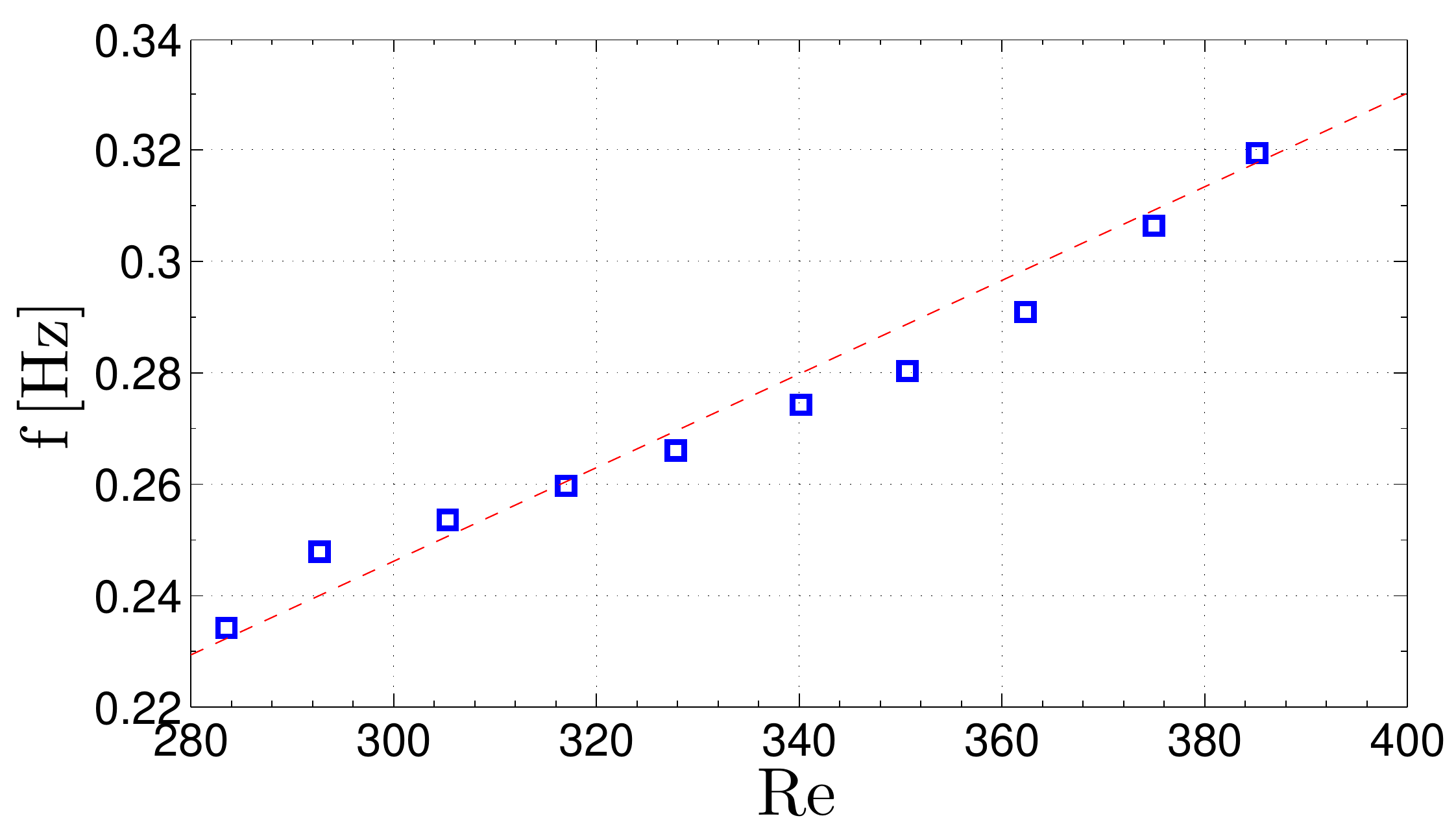}
                \caption{}
                \label{fig:Figure7b}   
        \end{subfigure}
        \begin{subfigure}[b]{0.8\textwidth}
                \centering
                \includegraphics[width=\textwidth]{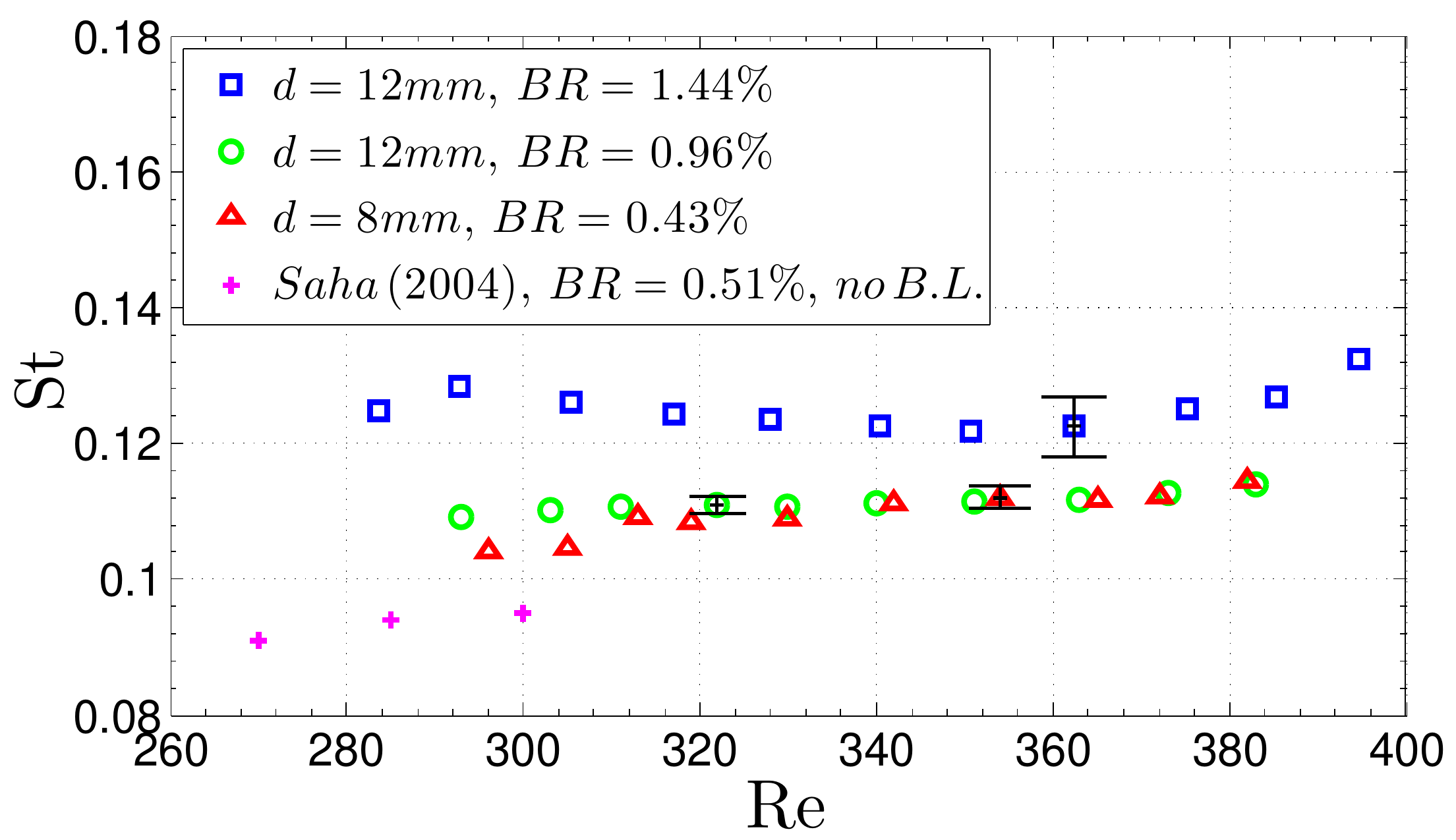}
                \caption{}
                \label{fig:Figure7c}
        \end{subfigure}
        \end{center}
	\end{minipage}	
	\caption{(Colour online) \textit{(a)} Evolution in time of the measured maximal (positive) and minimal (negative) values of longitudinal vorticity $\omega_x$, \textit{(b)} the observed frequency of hairpin vortex shedding as a function of Reynolds number and \textit{(c)} the corresponding Strouhal number for different cube size and tunnel blockage ratio.}
	\label{fig:Figure7}
	\end{center}
\end{figure}

To additionally characterise the unsteadiness of the flow, figure~\ref{fig:Figure7b} presents the evolution of frequency for a range of Reynolds numbers corresponding to the hairpin vortex shedding regime. The method to estimate the frequency is as follows: for each Reynolds number 800 pairs of PIV snapshots were recorded, corresponding to at least 12 periods of hairpin vortex oscillations. From the resulting vorticity fields, positive maxima and negative minima of vorticity were extracted. As the evolution of this quantity was periodic (see figure~\ref{fig:Figure7a}), we determined the time interval between several local maxima and, as a result, also the frequency of hairpin vortex shedding. It has nearly linear dependence on Reynolds number (with a small modulation observed). Similar behaviour was reported for the case of vortex shedding behind cylinders (\citealt{c_global_mode_sophie}), spheres (\citealt*{s_schouveiler}; \citealt{e_konrad}) and disks (\citealt{e_tomek}). Knowing the frequency, we have also determined the Strouhal number $St=(fd)/U$ as a function of the Reynolds number (figure~\ref{fig:Figure7c}). Up to Reynolds number Re=390, it remains almost constant and equal to 0.125 for these experimental conditions. 

\indent We performed two additional experiments in a similar water channel with larger cross-section (100 mm height and 150 mm width as compared to 100 mm by 100 mm in the original facility) and for the original and a smaller cube (edge length equal to 8 mm) to study the blockage effect. The frequency of hairpin vortex shedding was estimated by observation of 100 periods via LIF visualisations. As shown on figure~\ref{fig:Figure7c} the Strouhal number obtained was lower: 0.109 and 0.104 for cube edge length 12 mm and 8 mm, respectively. The corresponding blockage ($BR$) were 0.96\% and 0.43\%, in comparison with 1.44\% of the original channel ($BR$ is defined as a ratio of cube and channel cross-section). 

\indent The Strouhal number reported in \citet{k_saha1} is 0.094, but in his numerical channel the corresponding $BR$ was 0.51\%, with free-slip boundary conditions at the side walls, thus without the presence of a boundary layer.

\indent The difference of Strouhal number can be explained by the temporal theory of shear instabilities, according to which the initial frequency associated with the maximum growth of linear instability is proportional to the maximal vorticity in the shear layer. Higher $BR$ (or the presence of a boundary layer) implies the increase of the initial frequency and the Strouhal number as the maximal shear is higher (see \citet{st_monkewitz}).

\begin{figure}
	\begin{center}
	\begin{minipage}[!hb]{\textwidth}
		\begin{center}
		\begin{subfigure}[b]{0.8\textwidth}
       			\centering
                \includegraphics[width=\textwidth]{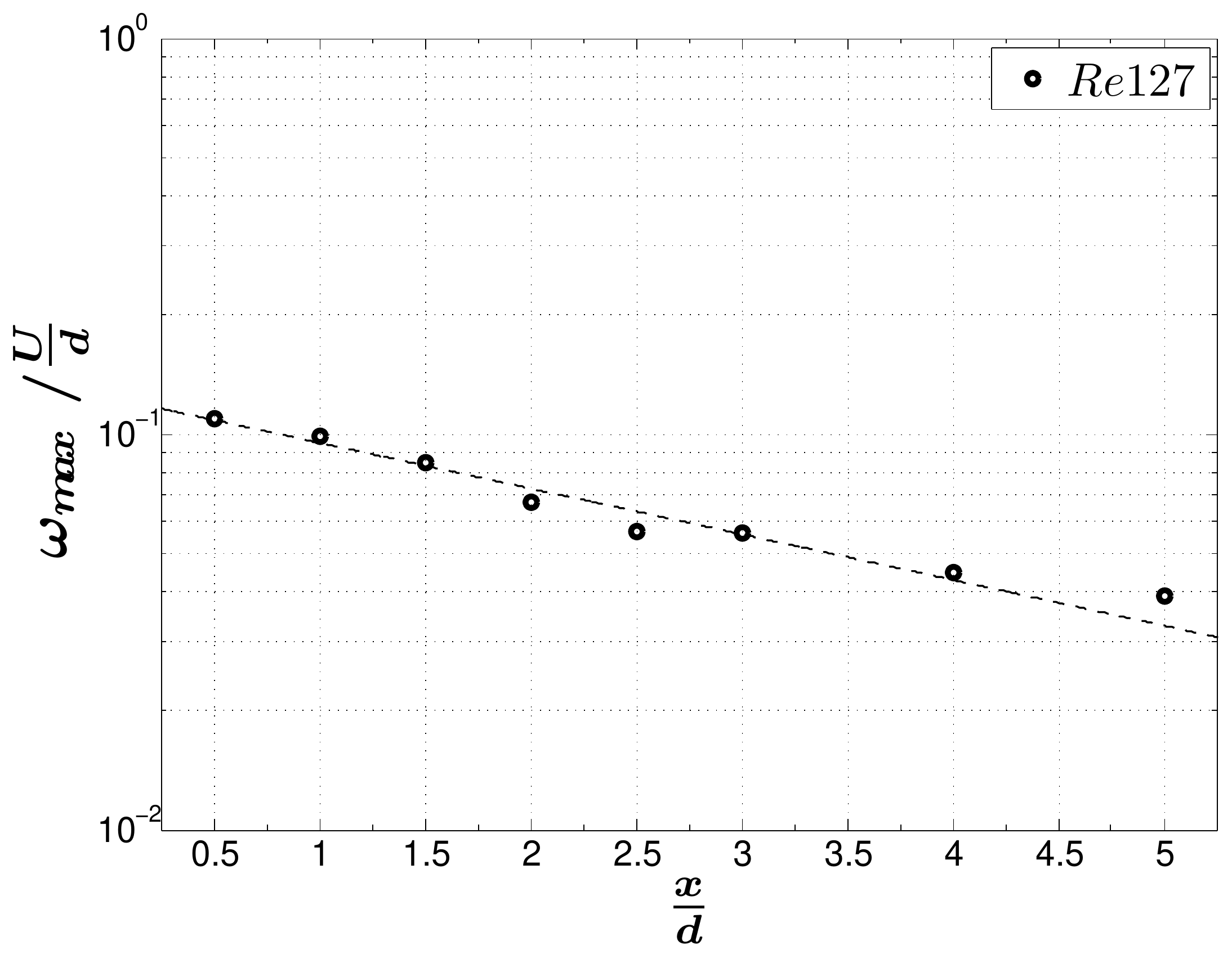}
                \caption{}
                \label{fig:Figure8a}   
        \end{subfigure}
        
        \begin{subfigure}[b]{0.8\textwidth}
                \centering
                \includegraphics[width=\textwidth]{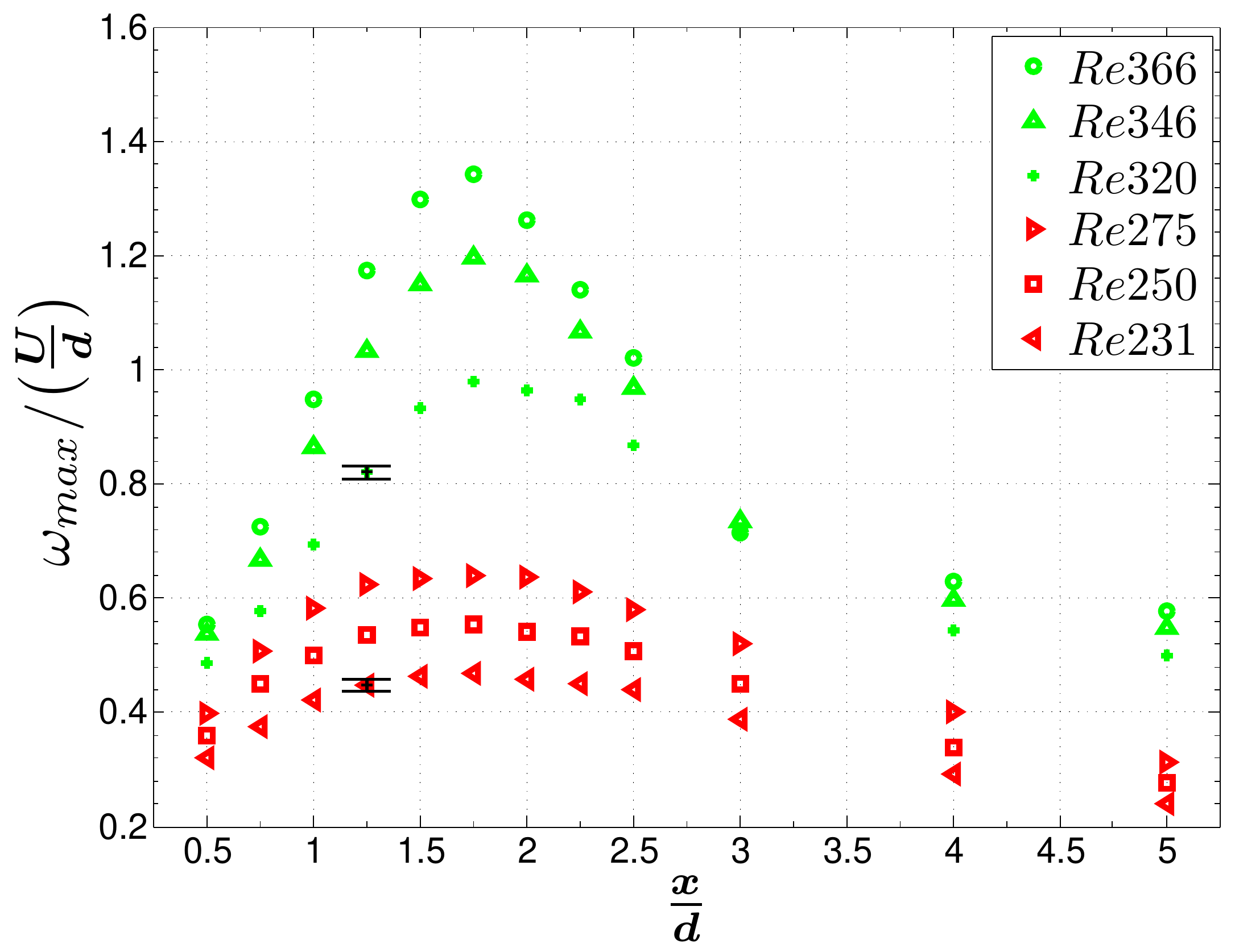}
                \caption{}
                \label{fig:Figure8b}
        \end{subfigure}
        \end{center}
	\end{minipage}	
	\caption{(Colour online) Vorticity evolution: \textit{(a)} exponential decrease for the basic flow at $Re=127$, \textit{(b)} envelope of streamwise vorticity fluctuations for the two counter-rotating vortices ($Re=231, 250, 275$) and the hairpin vortex shedding ($Re=320, 346, 366$) regimes. }
	\label{fig:Figure8}
	\end{center}
\end{figure}

\subsection{Maximal longitudinal vorticity as a function of streamwise distance}\label{sec:GLOBAL_MODE}

\indent As described earlier, for this type of geometry, three different types of flow structure can be observed for the investigated range of Reynolds numbers. To further investigate the transition process we have performed a detailed study of the $\omega_x$ evolution in the streamwise direction. For this purpose, a laser sheet was positioned normally to the undisturbed flow. In the case of the basic flow, an exponential decrease of the amplitude of $\omega_x$ was observed due to viscous relaxation. This evolution, in the form of exp$(-x/\xi)$, is presented in semi-log plot (figure~\ref{fig:Figure8a}), where $\xi$=3.8$d$ is the estimated attenuation length while $x$ denotes the distance between the measurement plane and the rear wall of the cube. We have averaged all the measured vorticity fields for each Reynolds number and subsequently extracted maximum positive and minimum negative values of vorticity. The mean of their absolute value is taken as the quantity describing the spatial evolution of the $\omega_x$. 

\indent Once the instability appears, for both the two counter-rotating vortices ($Re=231, 250, 275$) and the hairpin vortex shedding regimes ($Re=320, 346, 366$), the longitudinal vorticity $\omega_x$ grows, reaching a maximum, and then decays in the streamwise direction. Such spatial behaviour is characteristic for the envelope of a global mode instability. This spatial variation is presented in figure~\ref{fig:Figure8b}. It should to be noted that in the present study, the maximum of the longitudinal component of vorticity always occurred at $x/d=1.75$. Such a spatial evolution of the global mode is also reported for a sphere, for several quantities considered: (i) energy of streamwise velocity fluctuation (\citealt{s_ormieres}; \citealt{s_schouveiler}), (ii) root mean square (r.m.s.) of streamwise velocity (\citealt{s_tomboulides}) and (iii) longitudinal vorticity (\citealt{s_thompson}; \citealt{e_tomek}).

\indent In addition, one can observe that $\omega_x$ continues to increase after the Hopf transition (reaching values of the order of $10^{0}$ compared to the $10^{-1}$ in stationary regime). This is due to the contribution of the periodically shed hairpin vortices which can be visualised as a street of vortical loops with dominant transversal component of vorticity. However, as the length of a cube edge is finite, the flow in the recirculation zone is fully three dimensional and the shed structures are bent, adding the longitudinal component of vorticity, which is observable in the plane perpendicular to the free stream.

\subsection{Longitudinal enstrophy as a function of Reynolds number}
\begin{figure}
	\begin{center}
	\begin{subfigure}[ht!]{0.8\textwidth}
	\centerline{\includegraphics[width=1\textwidth]{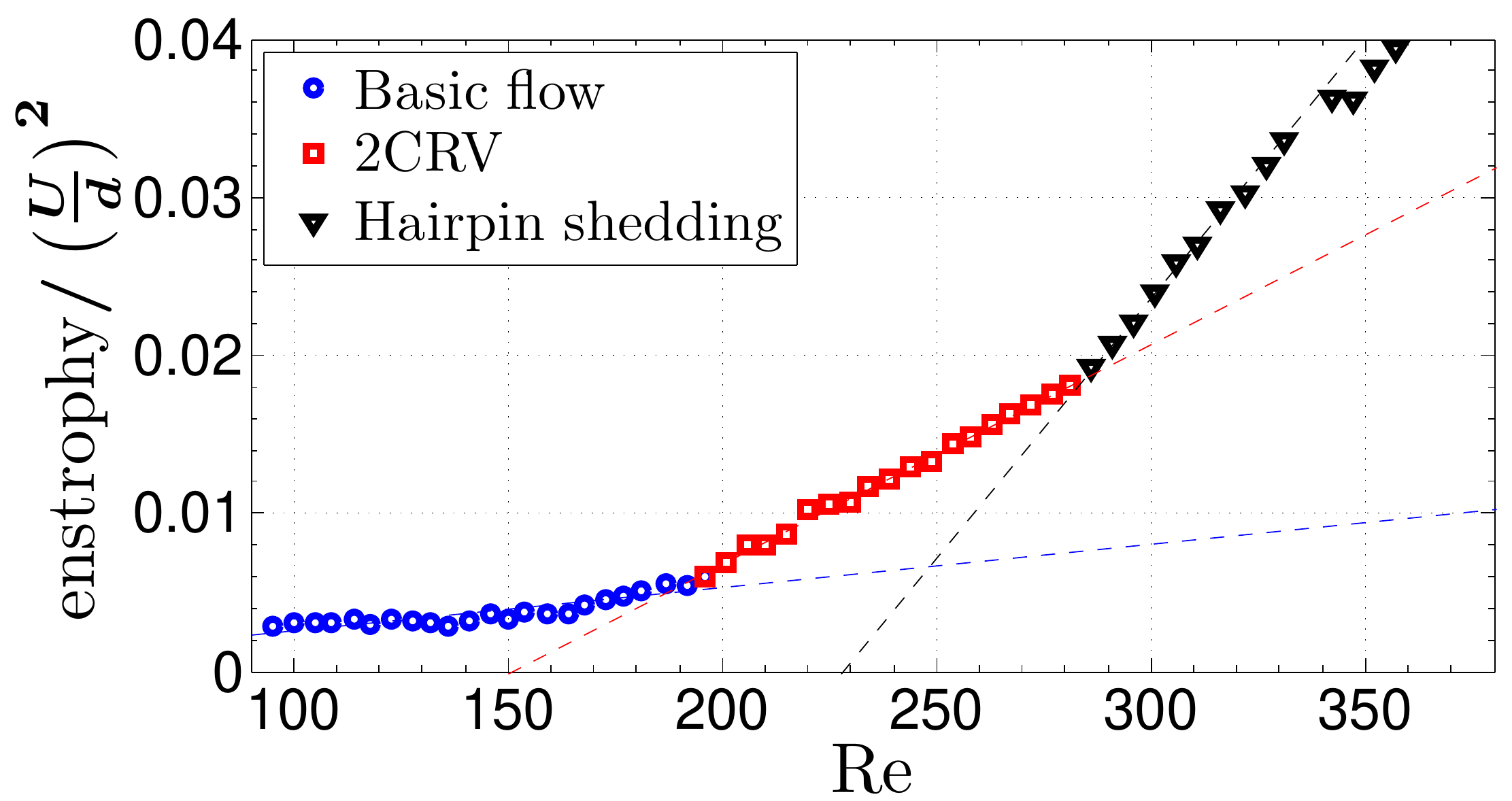}}
	\caption{}
	\label{fig:Figure9a}
	\end{subfigure}
	
	\begin{subfigure}[ht!]{0.8\textwidth}
	\centerline{\includegraphics[width=1\textwidth]{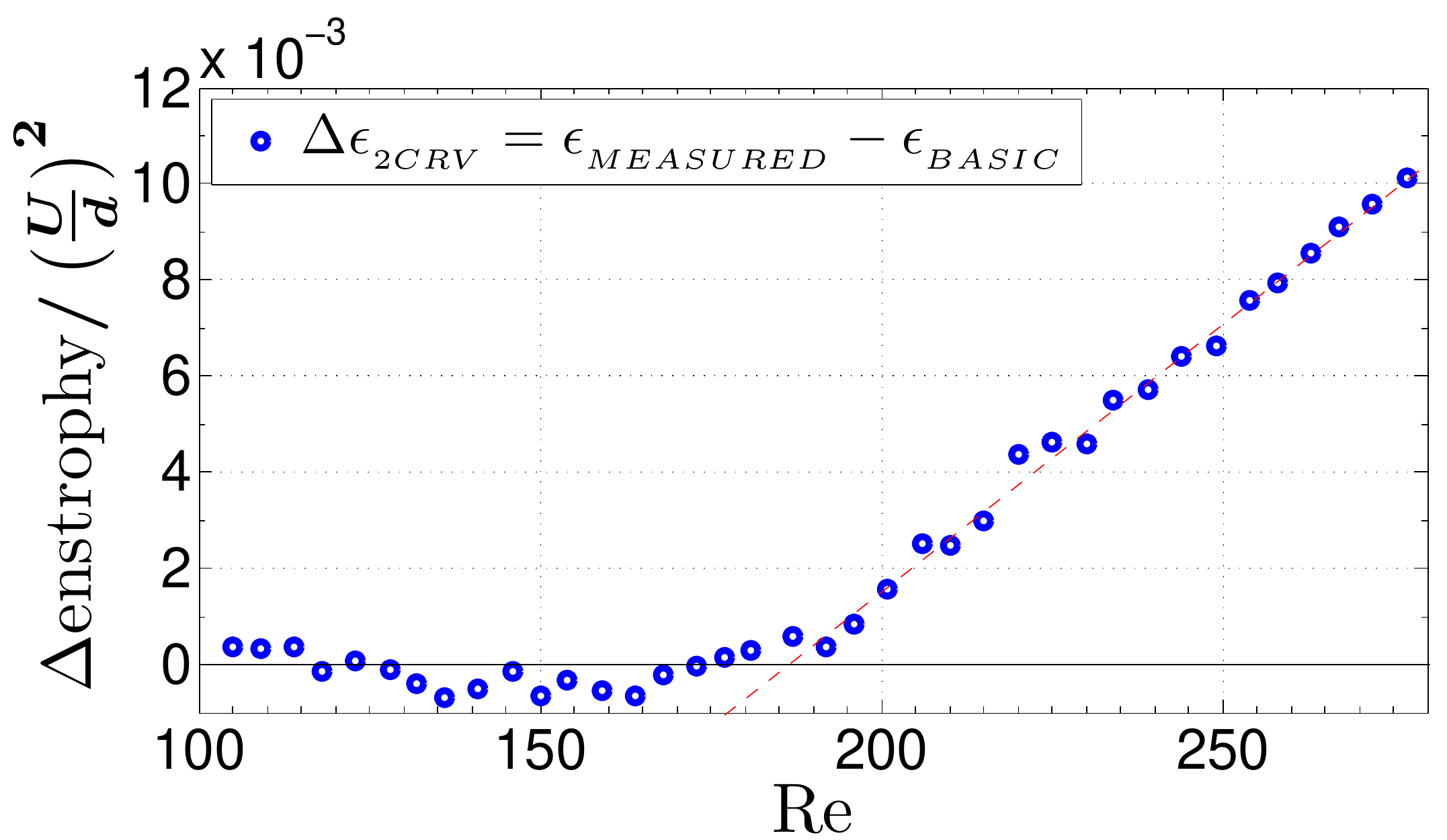}}
	\caption{}
	\label{fig:Figure9b}
	\end{subfigure}
	
	\begin{subfigure}[ht!]{0.8\textwidth}
	\centerline{\includegraphics[width=1\textwidth]{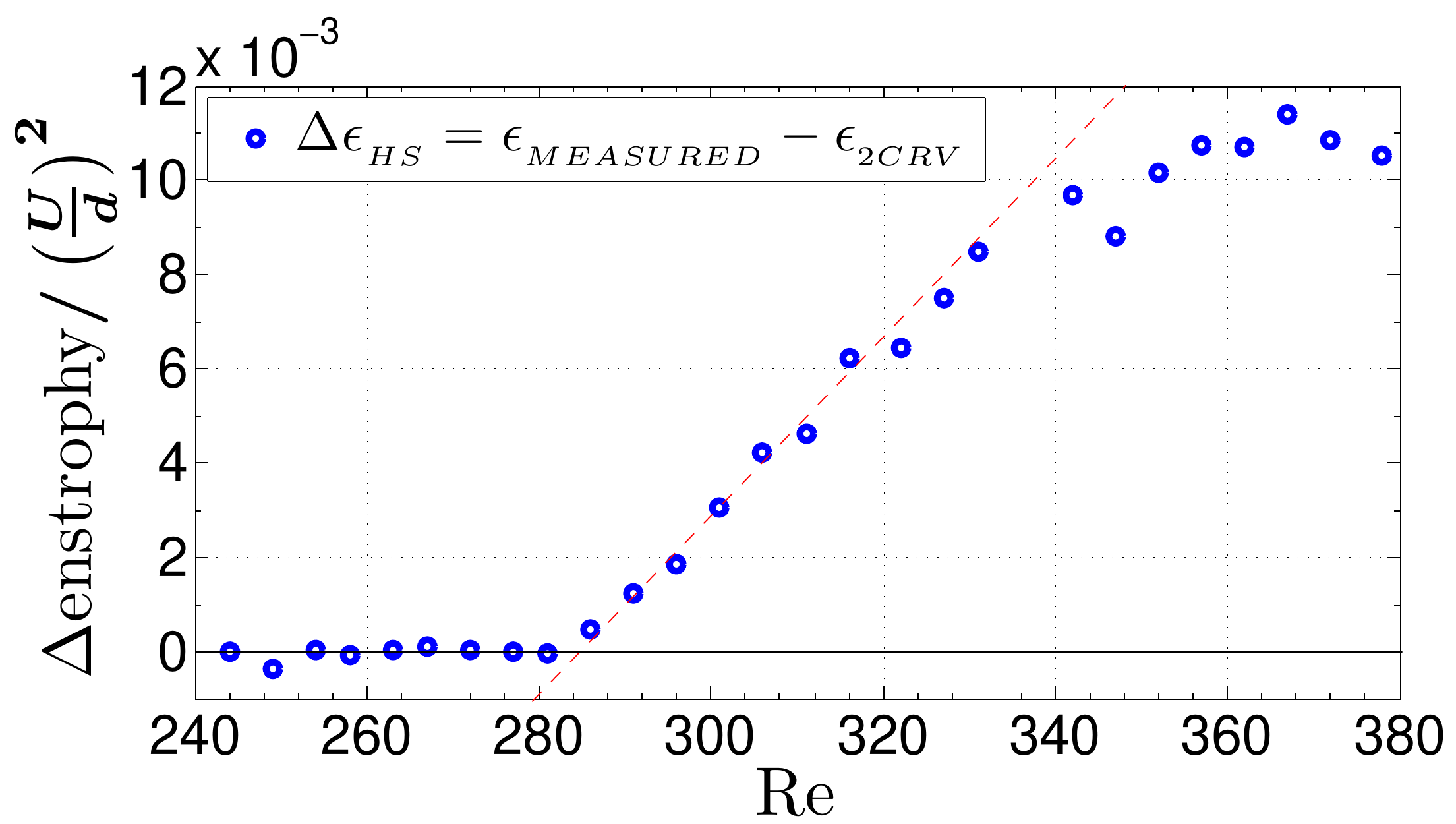}}
	\caption{}
	\label{fig:Figure9c}
	\end{subfigure}
	
	\end{center}
	\caption{(Colour online) Extraction of enstrophy components: \textit{(a)} evolution of normalised longitudinal enstrophy $\varepsilon$; \textit{(b)} evolution of stationary instability $\Delta\varepsilon_{_{2CRV}}$ associated with the two counter-rotating vortices regime; \textit{(c)} evolution of non-stationary instability $\Delta\varepsilon_{_{HS}}$ associated with the hairpin vortex shedding regime.}
	\label{fig:Figure9}
\end{figure}

\indent Using results of PIV measurements obtained with the laser sheet positioned perpendicularly to the stream, we have calculated the enstrophy $\varepsilon$ of the longitudinal vorticity component $\omega_x$, defined as:\\
\begin{equation}
    \varepsilon ={\frac{1}{|\Omega |}}\cdot{\int\limits_{\Omega}{\omega_x}^2 d\Omega},
    \label{eqn:wzor0}
\end{equation}\\
where $|\Omega |$ denotes the area of a two-dimensional domain $\Omega$ in which the measurements take place. As was mentioned in Section~\ref{sec:GLOBAL_MODE}, the vorticity distribution in the streamwise direction is different for the basic flow and for the two subsequent regimes. For this reason, the distance of the laser sheet from the rear wall of the cube ($x/d=1.5$) was selected so that the highest possibly vorticity for all regimes is obtained. For each Reynolds number, longitudinal enstrophy $\varepsilon$ was calculated by time-averaging of all captured snapshots (even for a non-stationary hairpin vortex shedding regime). The resulting evolution is presented in figure~\ref{fig:Figure9a}. In order to determine when the transition occurs, we use the method described in the next section.

\section{ANALYSIS OF EXPERIMENTAL DATA}\label{sec:ANALYSIS}
\indent Two different and complementary studies of the evolution of longitudinal enstrophy $\varepsilon$ (\ref{eqn:wzor0}) are presented in this Section.

\indent The first study concentrates on direct extraction of components corresponding to the basic flow and to stationary and non-stationary instabilities associated with the two counter-rotating vortices and the hairpin vortex shedding regimes respectively. The second one consists of an azimuthal Fourier modal decomposition carried out in order to obtain a detailed description of bifurcation branches and a more direct physical description of the wake behaviour.

\indent In addition, we have tested the validity of a supercritical or Landau`s model of instability to determine the values of both onsets. In the framework of this model, the squared amplitude of the instability parameter grows linearly with increasing Reynolds number (in the vicinity of the threshold). We have taken $\omega_x$ as the parameter describing the instability. It should be noted that the enstrophy $\varepsilon$ is proportional to the squared longitudinal vorticity and therefore its evolution should remain linear near the onset (if the supercritical transition takes place).

\begin{figure}
	\begin{center}
	\begin{minipage}[!hb]{0.49\textwidth}
		\begin{center}
		\begin{subfigure}[b]{\textwidth}
       			\centering
                \includegraphics[width=\textwidth]{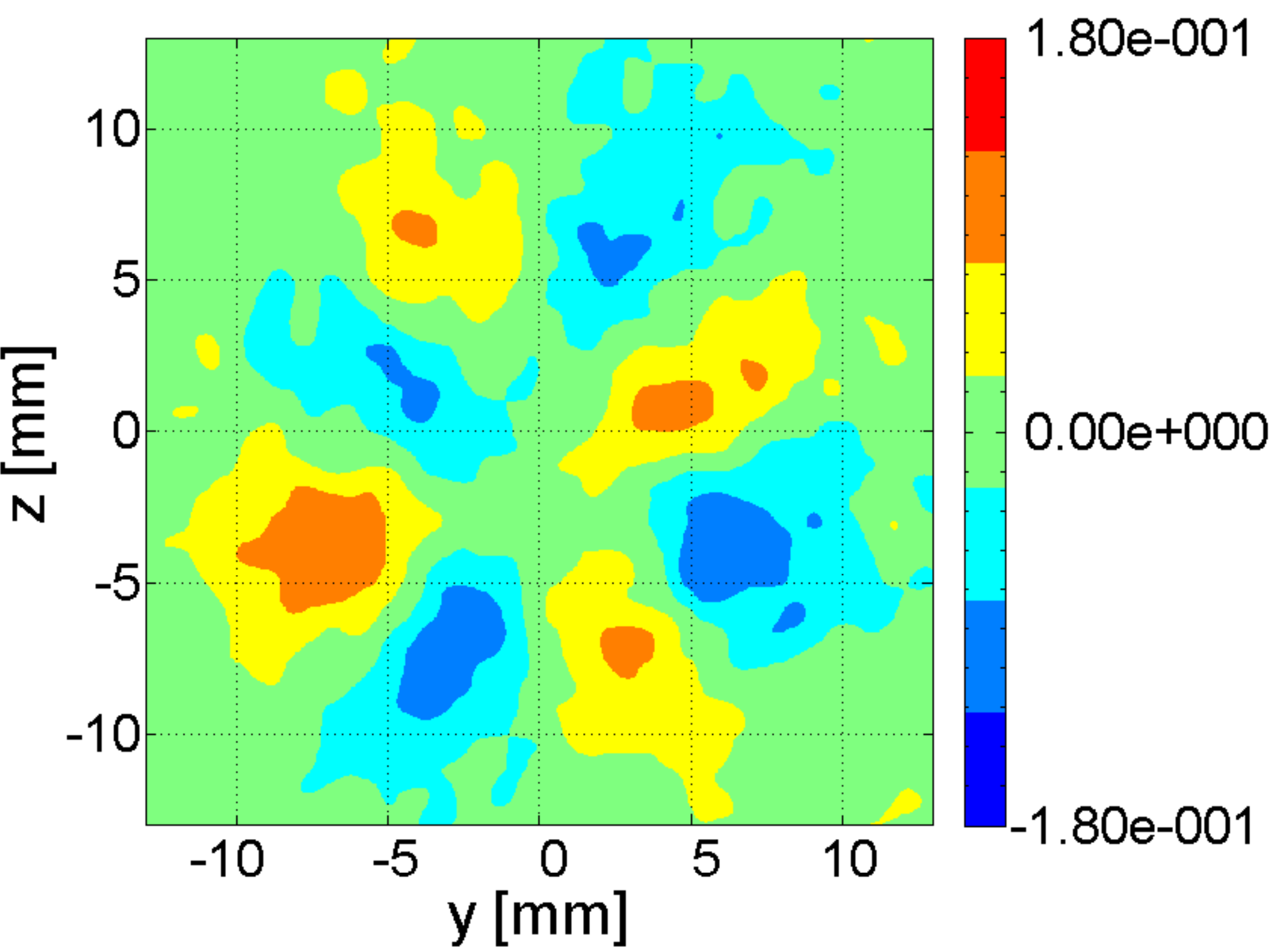}
                \caption{}
                \label{fig:Figure10a}   
        \end{subfigure}
        \end{center} 
    \end{minipage}	
    \begin{minipage}[!hb]{0.49\textwidth}
		\begin{center}
		\begin{subfigure}[b]{\textwidth}
       			\centering
                \includegraphics[width=\textwidth]{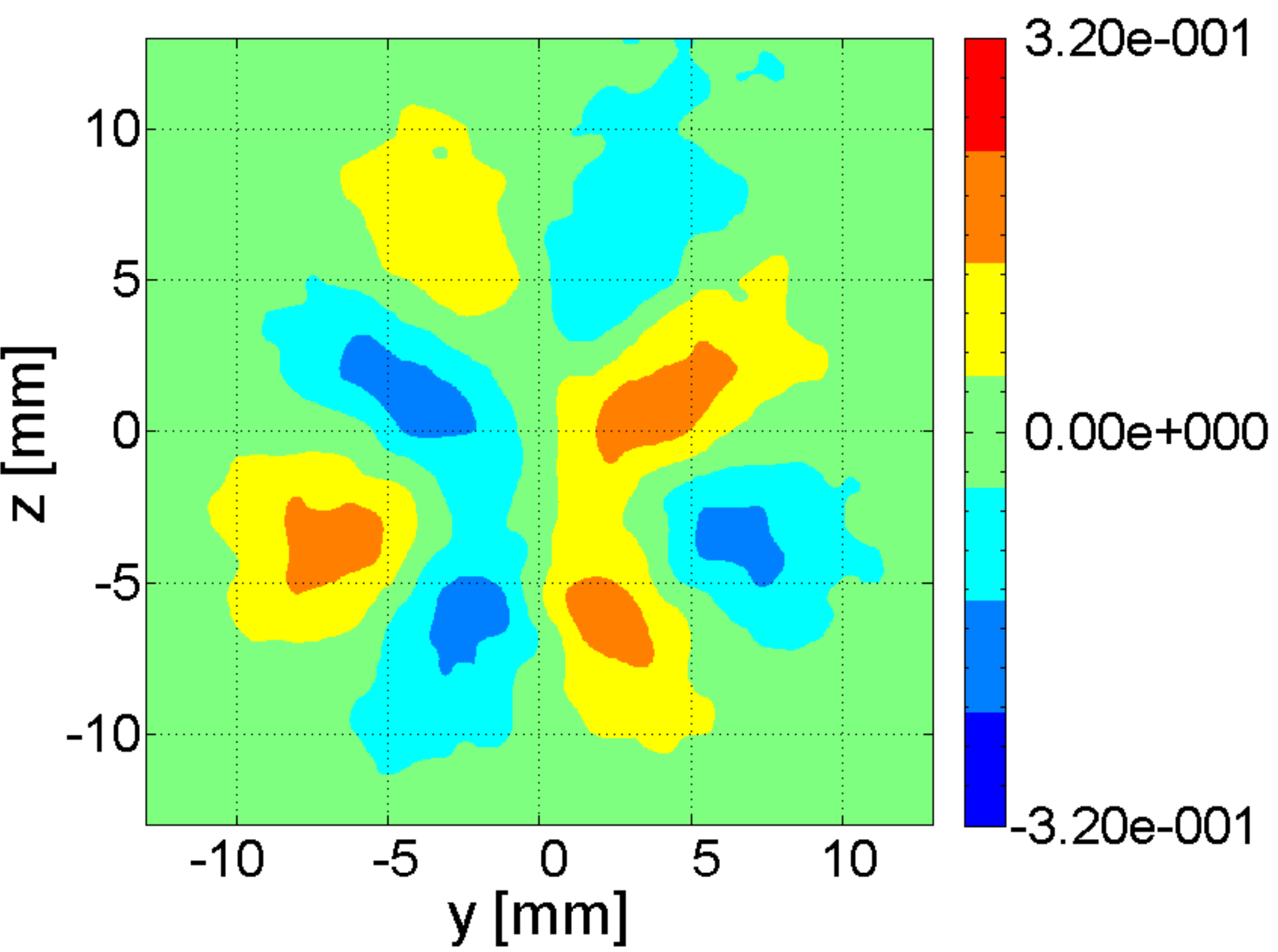}
                \caption{}
                \label{fig:Figure10b}   
        \end{subfigure}
        \end{center} 
    \end{minipage}	
    \begin{minipage}[!hb]{0.49\textwidth}
		\begin{center}
		\begin{subfigure}[b]{\textwidth}
       			\centering
                \includegraphics[width=\textwidth]{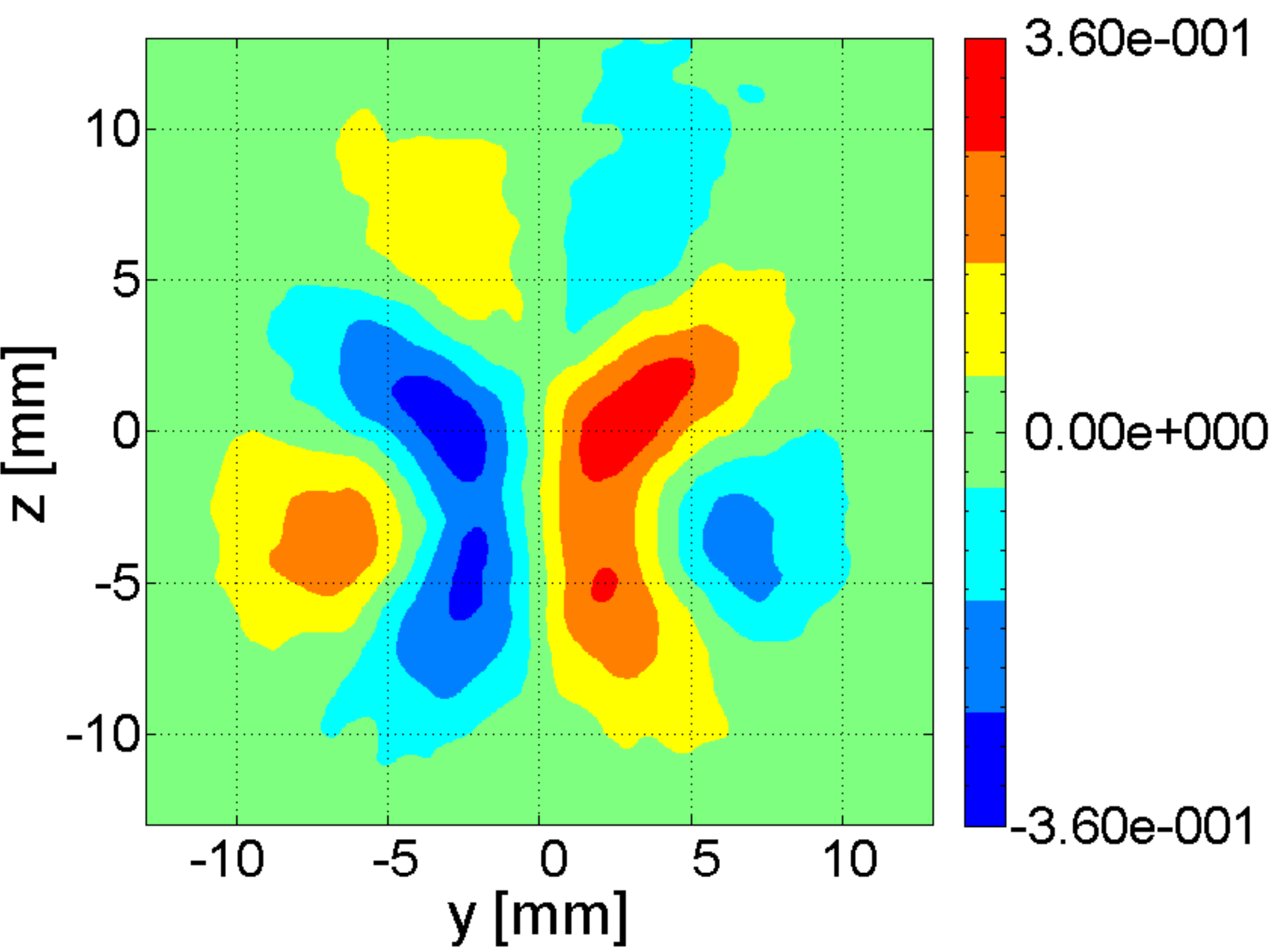}
                \caption{}
                \label{fig:Figure10c}   
        \end{subfigure}
        \end{center} 
    \end{minipage}	
    \begin{minipage}[!hb]{0.49\textwidth}
		\begin{center}
		\begin{subfigure}[b]{\textwidth}
       			\centering
                \includegraphics[width=\textwidth]{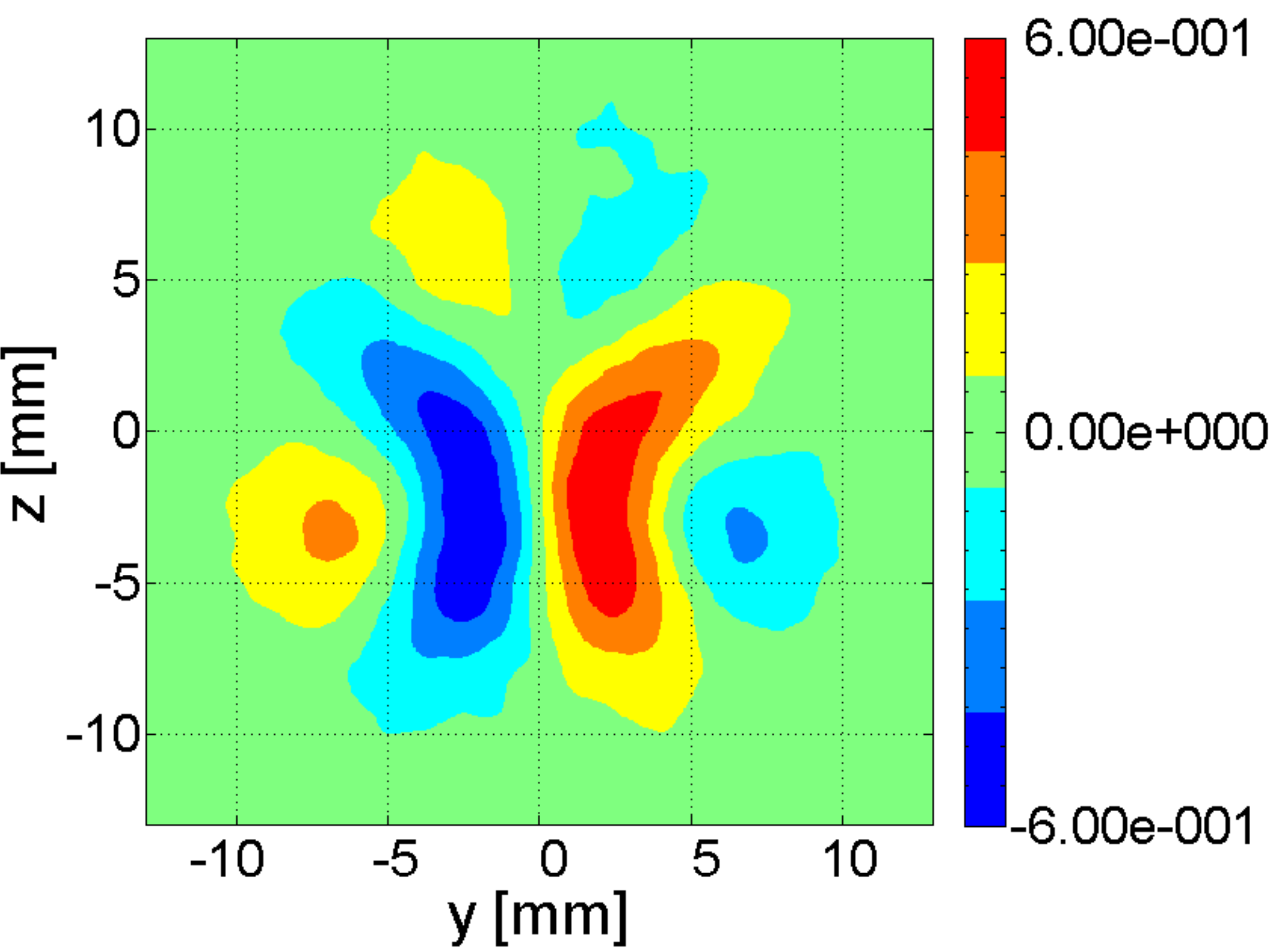}
                \caption{}
                \label{fig:Figure10d}   
        \end{subfigure}
        \end{center} 
    \end{minipage}	
	\caption{(Colour online) Flow evolution described by streamwise vorticity during the transition from the basic flow to the two counter-rotating vortices regime; gradual change of topology can be observed. \textit{(a)} $Re=132$, \textit{(b)} $Re=168$, \textit{(c)} $Re=196$, \textit{(d)} $Re=230$.}
	\label{fig:Figure10}
	\end{center}
\end{figure}
\begin{figure}
	\begin{center}
	\begin{minipage}[!hb]{0.49\textwidth}
		\begin{center}
		\begin{subfigure}[b]{\textwidth}
       			\centering
                \includegraphics[width=\textwidth]{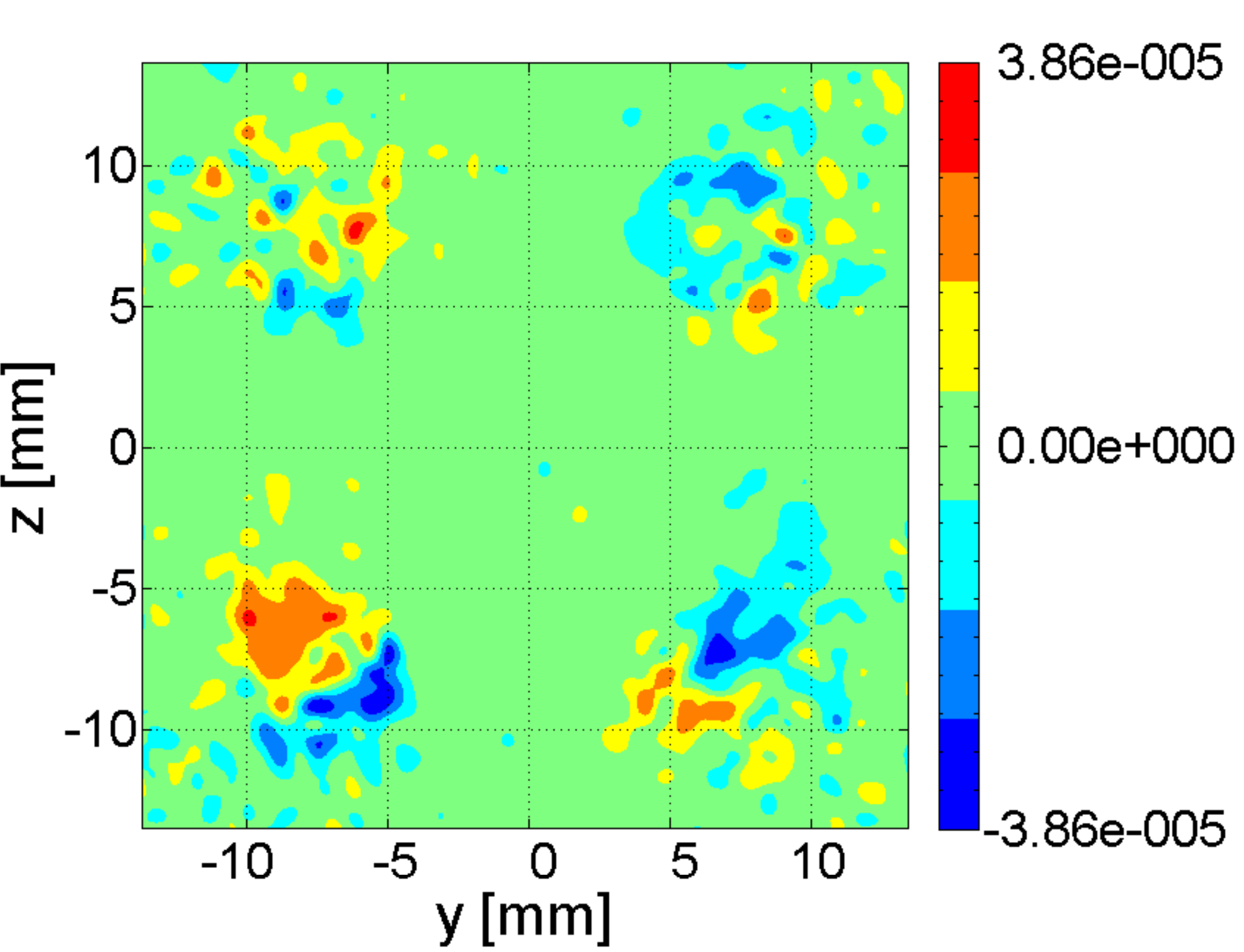}
                \caption{}
                \label{fig:Figure11a}   
        \end{subfigure}
        \end{center} 
    \end{minipage}	
    \begin{minipage}[!hb]{0.49\textwidth}
		\begin{center}
		\begin{subfigure}[b]{\textwidth}
       			\centering
                \includegraphics[width=\textwidth]{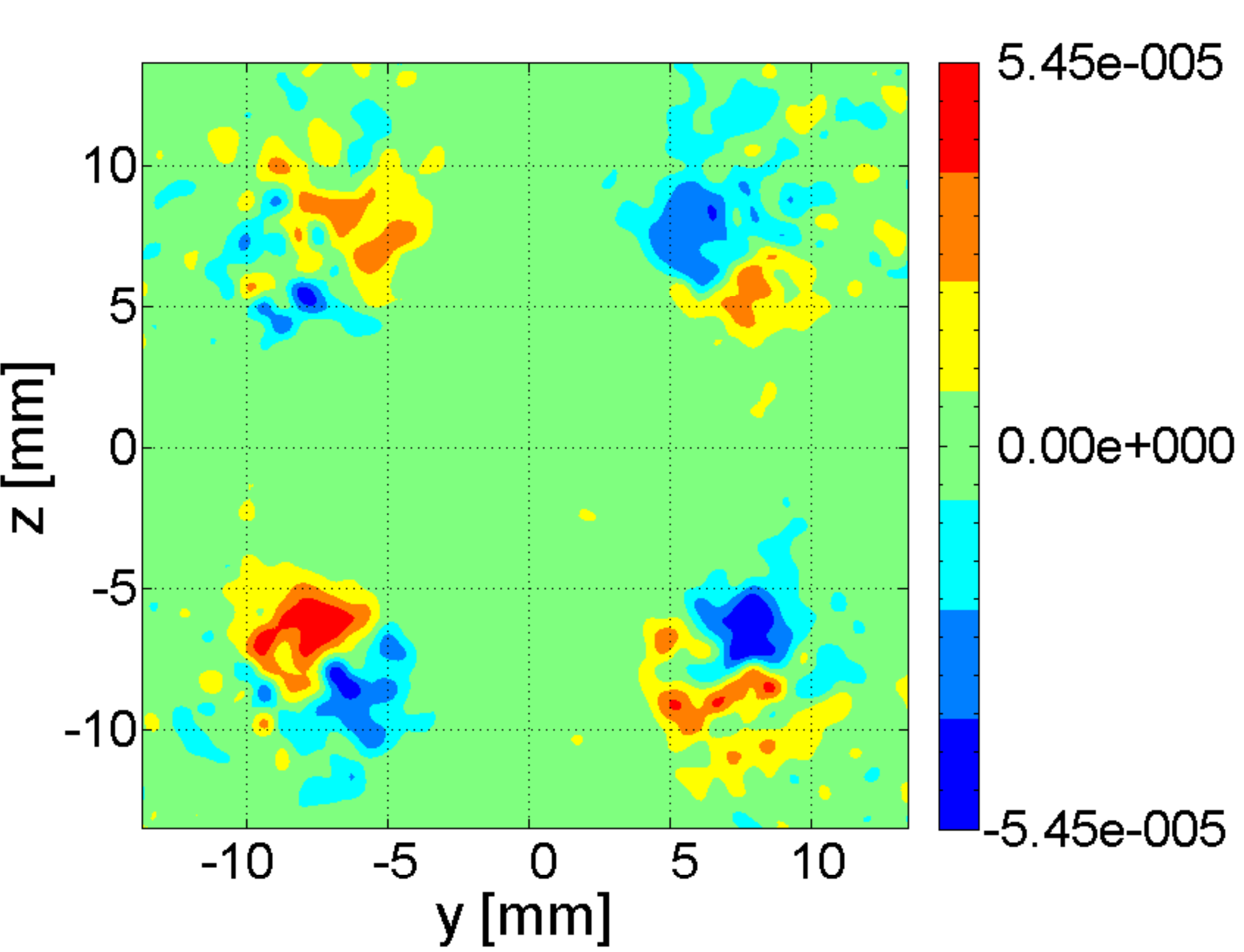}
                \caption{}
                \label{fig:Figure11b}   
        \end{subfigure}
        \end{center} 
    \end{minipage}	
    \begin{minipage}[!hb]{0.49\textwidth}
		\begin{center}
		\begin{subfigure}[b]{\textwidth}
       			\centering
                \includegraphics[width=\textwidth]{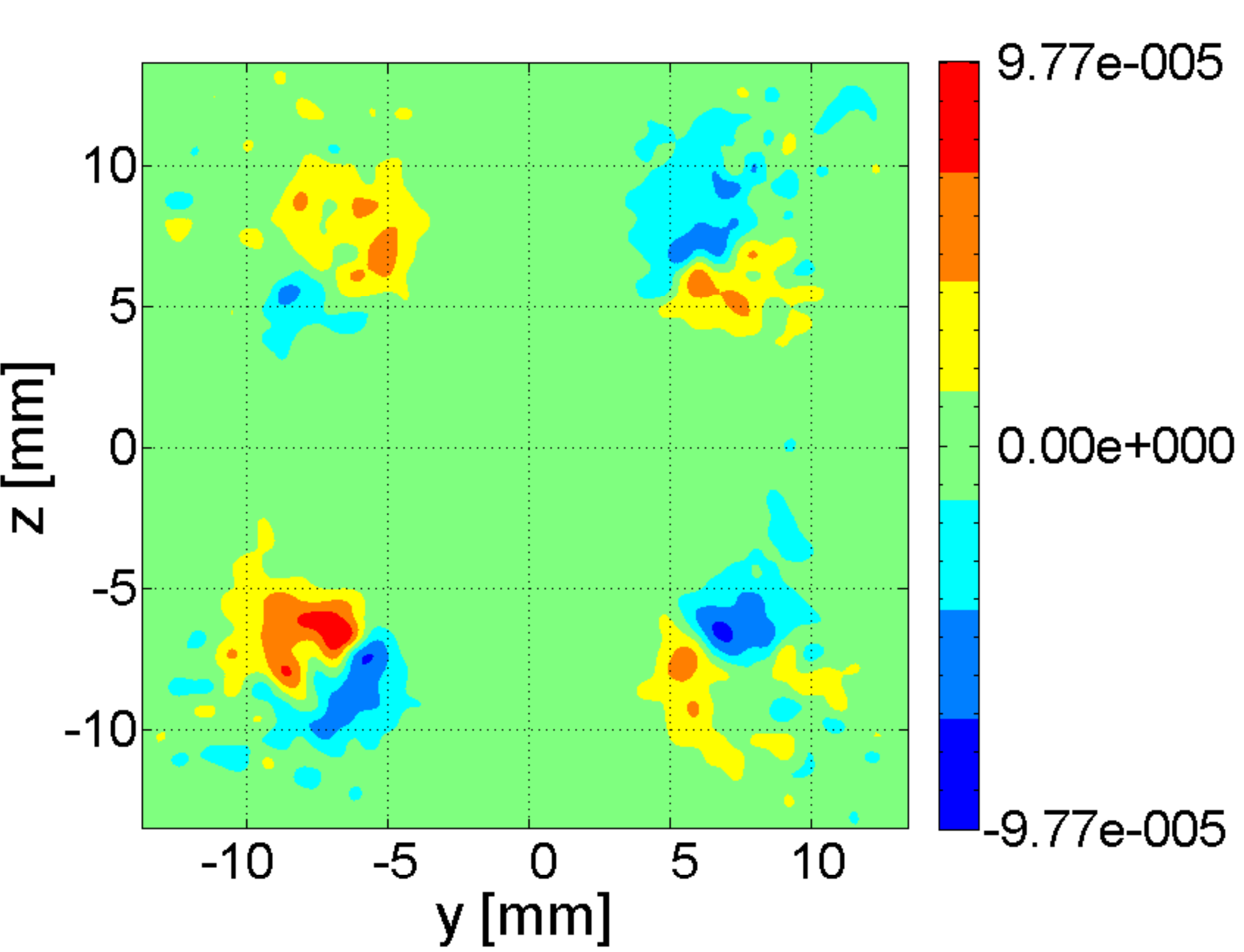}
                \caption{}
                \label{fig:Figure11c}   
        \end{subfigure}
        \end{center} 
    \end{minipage}	
    \begin{minipage}[!hb]{0.49\textwidth}
		\begin{center}
		\begin{subfigure}[b]{\textwidth}
       			\centering
                \includegraphics[width=\textwidth]{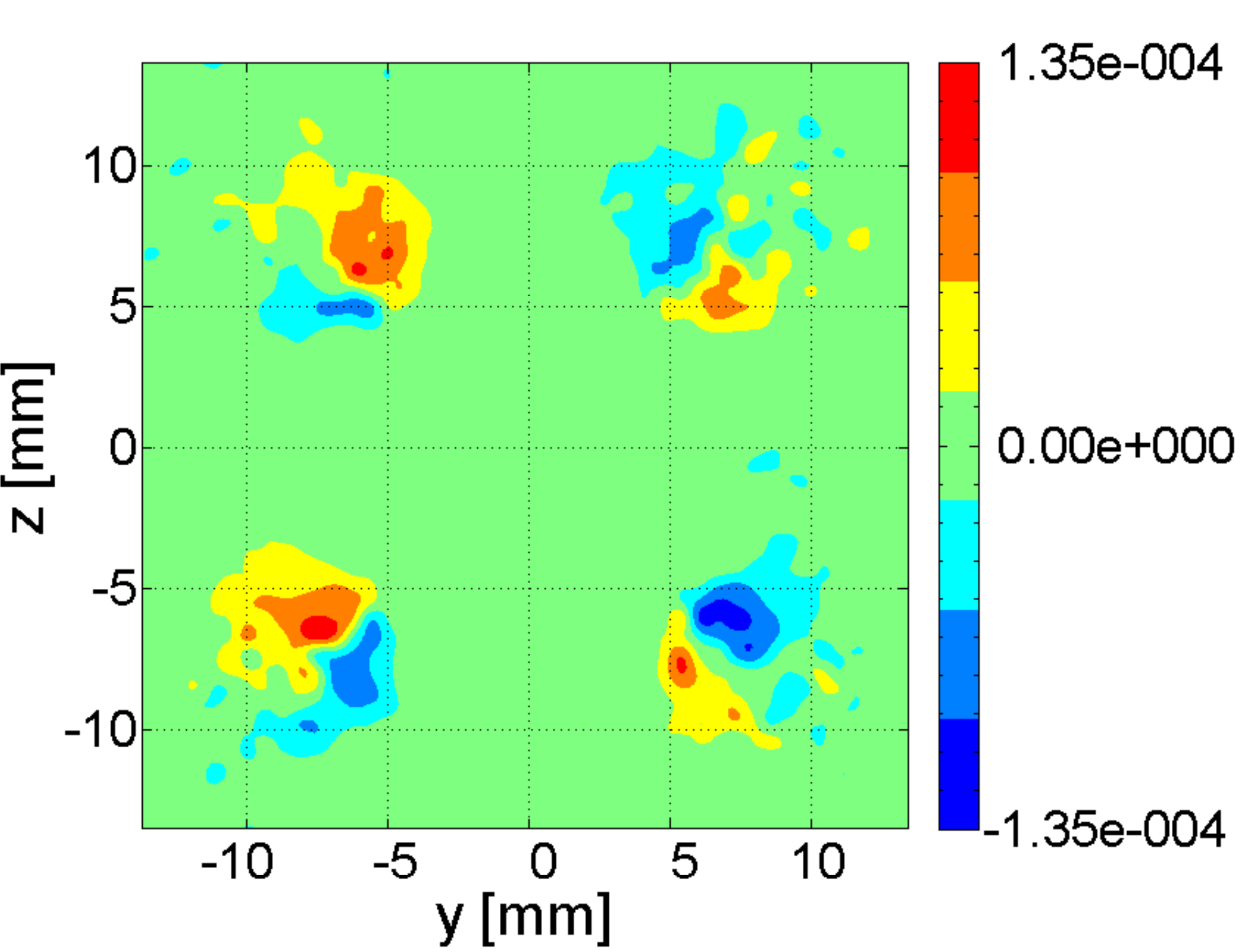}
                \caption{}
                \label{fig:Figure11d}   
        \end{subfigure}
        \end{center} 
    \end{minipage}	
	\caption{(Colour online) Evolution of the longitudinal component of $\lambda_{cx}$ as a function of the Reynolds number (to be compared with figure~\ref{fig:Figure10}). \textit{(a)} $Re=132$, \textit{(b)} $Re=168$, \textit{(c)} $Re=196$, \textit{(d)} $Re=230$.}
	\label{fig:Figure11}
	\end{center}
\end{figure}
\subsection{Extraction of vortical components}
In order to identify transition onsets, different components of enstrophy $\varepsilon$ associated with the observed regimes were extracted. Figure~\ref{fig:Figure9a} presents the evolution of the longitudinal enstrophy $\varepsilon$ (normalised with the free-stream velocity $U$ and the cube edge length $d$) in which both bifurcations are clearly apparent. 

\indent For the lowest Reynolds numbers only extrinsic vorticity is present (circles in figure~\ref{fig:Figure9a}). It is associated with the non-axisymmetric basic flow induced by the pressure gradient, characterised by four pairs of opposite-sign vortices. The linear dependence on Reynolds number can be expressed as:\\
\begin{equation}
    \varepsilon_{_{BASIC}}/\left(\frac{U}{d}\right)^2=A_{_{BASIC}}\textup{Re}+B_{_{BASIC}},
    \label{eqn:wzor1}
\end{equation}\\
where coefficients $A_{_{BASIC}}={2.76}\cdot{10^{-5}}$ and $B_{_{BASIC}}={-2.08}\cdot{10^{-4}}$ were obtained by least-squares approximation. 
\begin{figure}
	\begin{center}
	\begin{minipage}[!hb]{\textwidth}
		\begin{center}
		\begin{subfigure}[b]{0.8\textwidth}
       			\centering
                \includegraphics[trim=0 0 160 0,clip, angle=270,width=\textwidth]{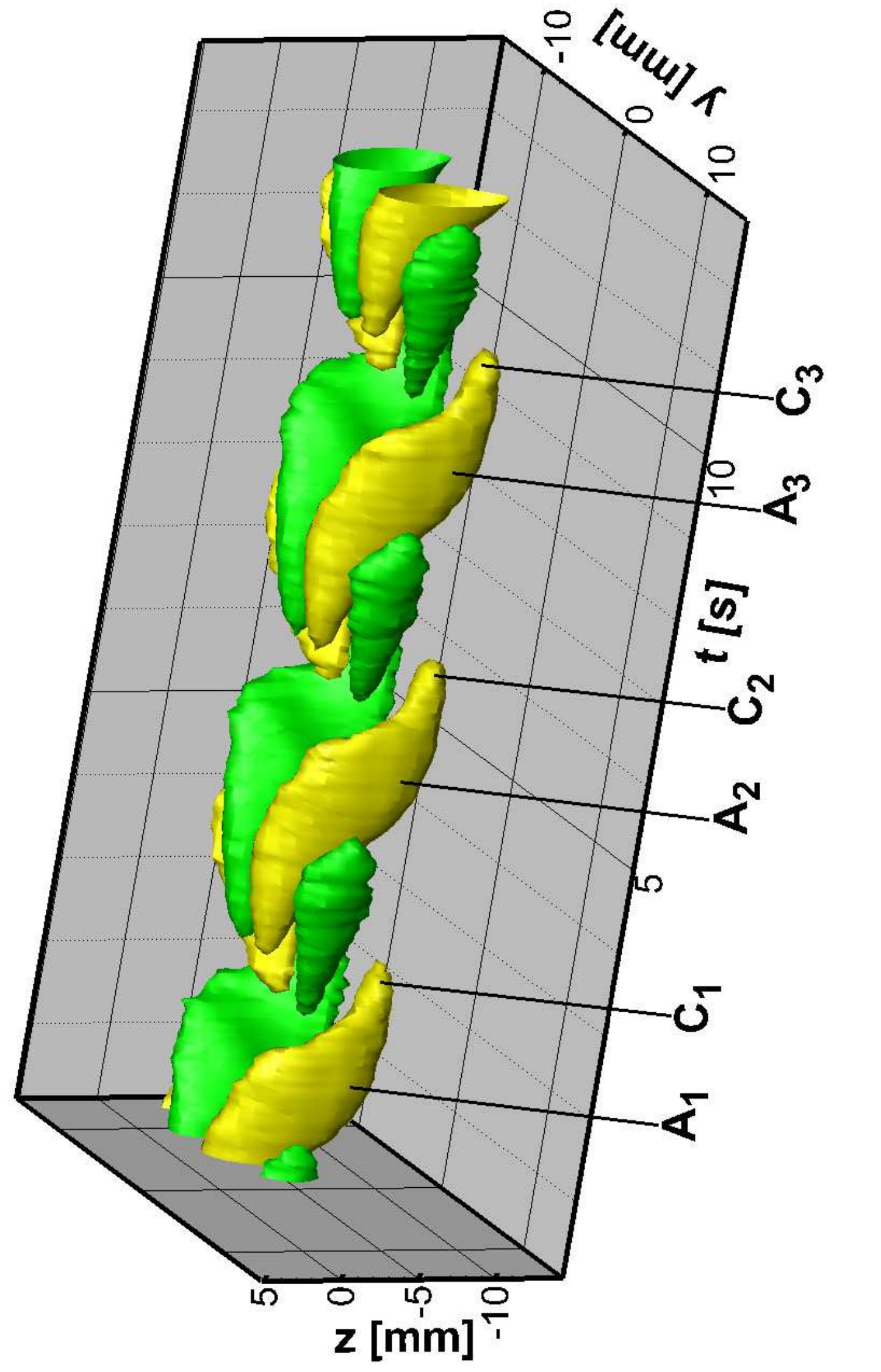}
                \caption{}
                \label{fig:Figure12a}   
        \end{subfigure}
        
        \begin{subfigure}[b]{0.8\textwidth}
                \centering
                \includegraphics[trim=0 0 160 0,clip, angle=270,width=\textwidth]{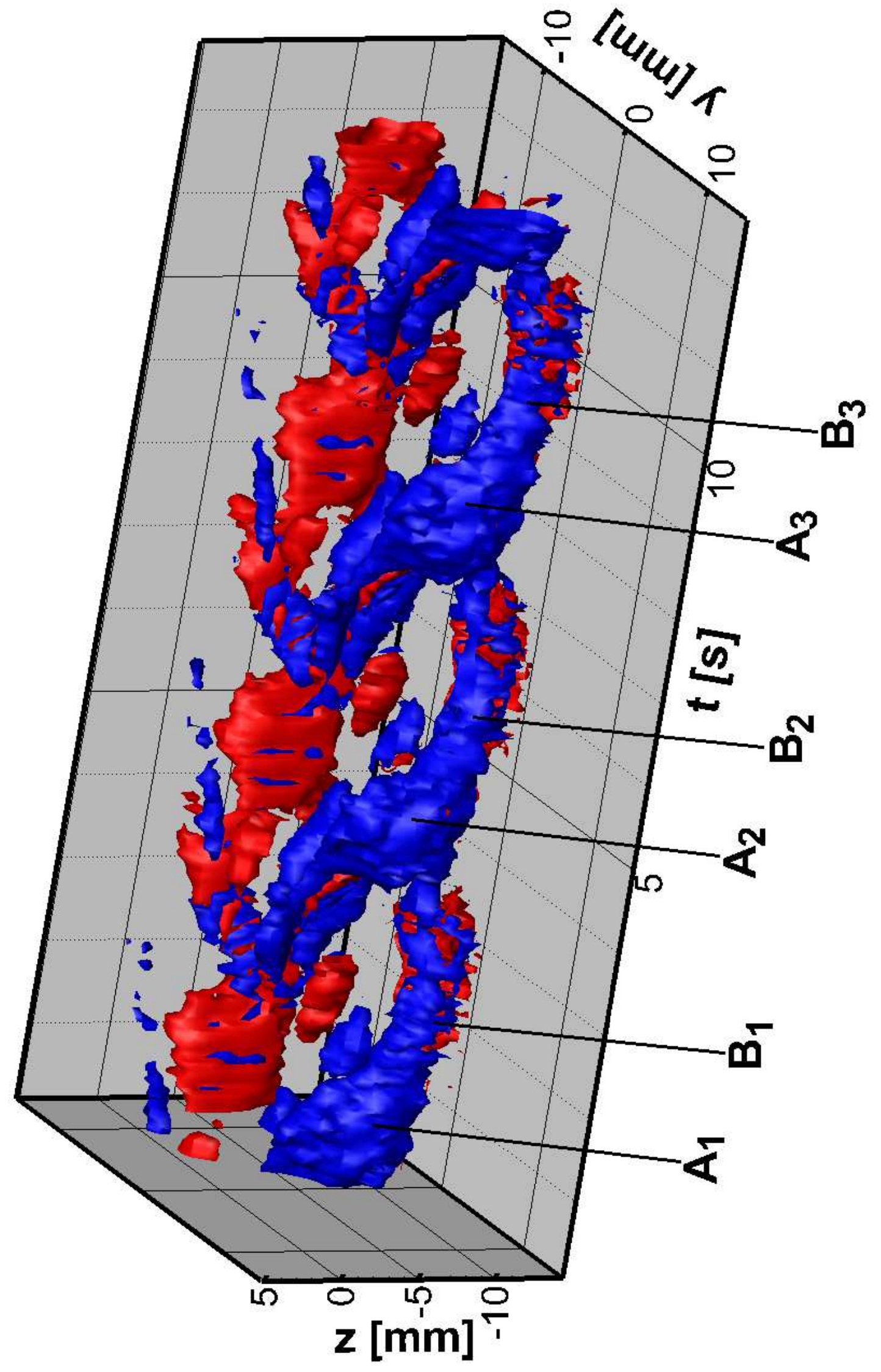}
                \caption{}
                \label{fig:Figure12b}
        \end{subfigure}
        \end{center}
	\end{minipage}	
	\caption{Temporal evolution of the longitudinal component of (a) the vorticity $\omega_x$ and (b) the complex eigenvalue of velocity gradient $\lambda_{cx}$ (for $Re=316$). Both are calculated from the in-plane velocity measured in the laser plane positioned normally to the free-stream velocity. The sequence was filtered using POD decomposition.}
	\label{fig:Figure12}
	\end{center}
\end{figure}

\indent As the Reynolds number is increased, a first bifurcation occurs. One can observe the change of slope due to the appearance of some additional enstrophy associated with the two steady counter-rotating vortices regime. This supplementary component, called stationary instability, is denoted by $\Delta\varepsilon_{_{2CRV}}$. In order to determine its amplitude, we extrapolate the value of $\varepsilon_{BASIC}$ from formula~\ref{eqn:wzor1} for higher Reynolds numbers and subtract it from the total original $\varepsilon$ (see figure~\ref{fig:Figure9a}). The result describes the growth of stationary instability related to the appearance of the major pair of bean-shaped vortices after the regular transition from the basic flow to the two counter-rotating vortices regime. The total observed longitudinal enstrophy is a superposition of the basic flow enstrophy $\varepsilon_{BASIC}$ (four pairs of vortices) and the contribution from the stationary instability (one pair of vortices) $\Delta\varepsilon_{_{2CRV}}$. This evolution of corresponding enstrophy, obtained as the difference of the measured enstrophy and the extrapolated value of the base state, is presented in figure~\ref{fig:Figure9b}. For Reynolds numbers higher than 190, the linear growth is quite apparent and agrees well with Landau`s model of instability. The function describing this evolution can be estimated as:\\
\begin{equation}
    \Delta\varepsilon_{_{2CRV}}/\left(\frac{U}{d}\right)^2=A_{_{2CRV}}\textup{Re}+B_{_{2CRV}},
    \label{eqn:wzor2}
\end{equation}\\
where $A_{_{2CRV}}={1.11}\cdot{10^{-4}}$ and $B_{_{2CRV}}={-2.07}\cdot{10^{-2}}$. After performing the linear extrapolation, the value of first onset is determined as $Re_1=186$.

\indent In figures~\ref{fig:Figure9a} and~\ref{fig:Figure9b} a continuous change of slope in the vicinity of the regular transition from the basic flow to the two counter-rotating vortices can be observed. This relatively smooth transition is due to an imperfect bifurcation and figure~\ref{fig:Figure10} illustrates this process. The change of topology from the basic flow (orthogonal symmetry, with 4 pairs of counter-rotating vortices) to the subsequent regime (planar symmetry, one major pair of counter-rotating vortices) occurs gradually as the transition is rounded. Two additional problems contribute to the difficulty with the localisation of the first threshold value. First of all, the basic flow consists of four pairs of extrinsic vortices produced by the corners of the bluff body and this vortical structure is still present after the regular transition, when the two intrinsic counter-rotating vortices appear due to the stationary instability. The second reason is the influence of the support tube, due to which there exists one preferred symmetry plane.

\indent In addition, we have calculated the imaginary part of the complex eigenvalue of the velocity gradient tensor (hereinafter denoted $\lambda_{cx}$, for details see~\citealt{lam_zhou}) for the two in-plane velocity components, measured on the plane normal to the free-stream velocity. figure~\ref{fig:Figure11} displays the evolution of $\lambda_{cx}$ as the Reynolds number is increased. Basically, this criterion does not distinguish the direction of the rotation. To add this supplementary information $\lambda_{cx}$ has been coloured in accordance with the sign of vorticity.

\indent One should notice that, contrary to the case of vorticity, $\lambda_{cx}$ is not the linear operator of a velocity gradient. For this reason, it was not possible to calculate the mean of $\lambda_{cx}$ directly from time-averages of measured velocity fields. Instead, it was obtained by averaging instantaneous $\lambda_{cx}$ calculated for each velocity field filtered with a proper orthogonal decomposition (POD) (using the first six modes which represent about 80\% of the total energy of the flow).

\indent In figure~\ref{fig:Figure11}, one may clearly observe the pattern associated with the basic flow (four pairs of longitudinal counter-rotating vortices). However, even for higher Reynolds numbers (Re=$230$, figure~\ref{fig:Figure11d}) no trace of the counter-rotating vortices can be found with this method. This is in contrast to the spatial distribution of vorticity (figure~\ref{fig:Figure10d}) with a clear presence of a major pair of longitudinal counter-rotating vortices. It seems therefore that the $\lambda_c$ criterion, even if the term $\lambda_c^2$ is analogous to the enstrophy (see~\citealt{lam_zhou}), remains less sensitive than the method based on the an analysis of the enstrophy of longitudinal vorticity.

\indent One should also note that \citet{k_saha1}, in his numerical study, claims that the first transition occurs for higher Reynolds number (in the range of 216 to 218, compared with the present 186). Perhaps this difference is due to the fact that he has used relatively high values of the imaginary part of $\lambda_{c}$ as the threshold to capture the appearance of the main pair of counter-rotating vortices and first bifurcation. It should also be noted, that Saha refers to his threshold values "as an upper limit of transition" (see p.1633 in \citealt{k_saha1}) rather than as an exact result, recognising apparent limitations of his numerical method.

\indent Finally, after further increasing the Reynolds number, a Hopf transition takes place. Figures~\ref{fig:Figure12a} and~\ref{fig:Figure12b} reveal the time sequences of the longitudinal components of $\omega_x$ and $\lambda_{cx}$ respectively. They were calculated at subsequent time intervals in the hairpin vortex shedding regime for Re=$316$. The instantaneous snapshots of velocity  were filtered using the POD, with the first six modes representing about 80\% of the energy of the flow. Assuming the Taylor hypothesis of frozen turbulence, one may consider elapsed time as the spatial coordinate in the free-stream direction.

\indent The regular, periodic structure of the flow is clearly observed. Letters $A_1$, $A_2$ and $A_3$ in figure~\ref{fig:Figure12} indicate regions of high values of $\omega_x$ and $\lambda_{cx}$ due to the influence of the hairpin legs. They move toward the lower part of the wake as they are advected downstream (letters $C_1$, $C_2$ and $C_3$ in figure~\ref{fig:Figure12a}). One may expect that, in these regions, hairpin heads are formed and, as a result, the transversal components of $\omega$ and $\lambda_{c}$ are more important than the longitudinal components. Note that we have only measured $\omega_x$ and $\lambda_{cx}$, therefore we were unable to visualise the full loop of the hairpin head (which would require determination of all components of $\omega$ and $\lambda_{c}$).

\indent As is the case for the stationary instability, the non-stationary one can also be characterised by supplementary enstrophy related to the hairpin vortex shedding contribution. It can be observed as the second change of slope in figure~\ref{fig:Figure9a}. From the longitudinal enstrophy $\varepsilon$ in the hairpin vortex shedding regime, we subtract the extrapolated component $\varepsilon_{_{2CRV}}$, as we have proceeded in the case of the first instability. The remainder grows linearly with the Reynolds number (figure~\ref{fig:Figure9c}) and can be described by the function:\\
\begin{equation}
    \Delta\varepsilon_{_{HS}}/\left(\frac{U}{d}\right)^2=A_{_{HS}}\textup{Re}+B_{_{HS}},
    \label{eqn:wzor3}
\end{equation}\\
where $A_{_{HS}}={1.89}\cdot{10^{-4}}$ and $B_{_{HS}}={-5.39}\cdot{10^{-2}}$ are again obtained by the least-squares procedure. The value of the onset of the Hopf bifurcation estimated by the linear interpolation equals Re$_2=285$, compared with 265-270 from \citet{k_saha1}.

\indent It should also be noted that the analysis presented above confirms the contribution of the hairpin vortices to the time-averaged longitudinal vorticity field measured on the plane perpendicular to the free-stream velocity. As described earlier, in the hairpin vortex shedding regime the loops of vorticity formed in the recirculation zone are periodically shed into a far wake. These structures are initially oriented mainly in the transversal direction. However, as they convect downstream, they are partially bent and reoriented towards the streamwise direction with a transversally oriented head and longitudinally oriented legs. Due to the non-stationary additional contribution of the hairpins legs, the mean value of the maximum of the absolute value of vorticity is larger than that which would result from the separate stationary instability. This can be clearly observed in figure~\ref{fig:Figure12b}, where letters $B_1$, $B_2$ and $B_3$ refer to the two counter-rotating vortices, associated with a stationary instability, while $A_1$, $A_2$ and $A_3$ characterise the non-stationary instability, indicating the additional influence of the hairpin legs.

\begin{figure}
	\begin{minipage}[hb!]{\textwidth}
	\begin{center}
	\begin{subfigure}[ht!]{0.3\textwidth}
	\centerline{\includegraphics[width=\textwidth]{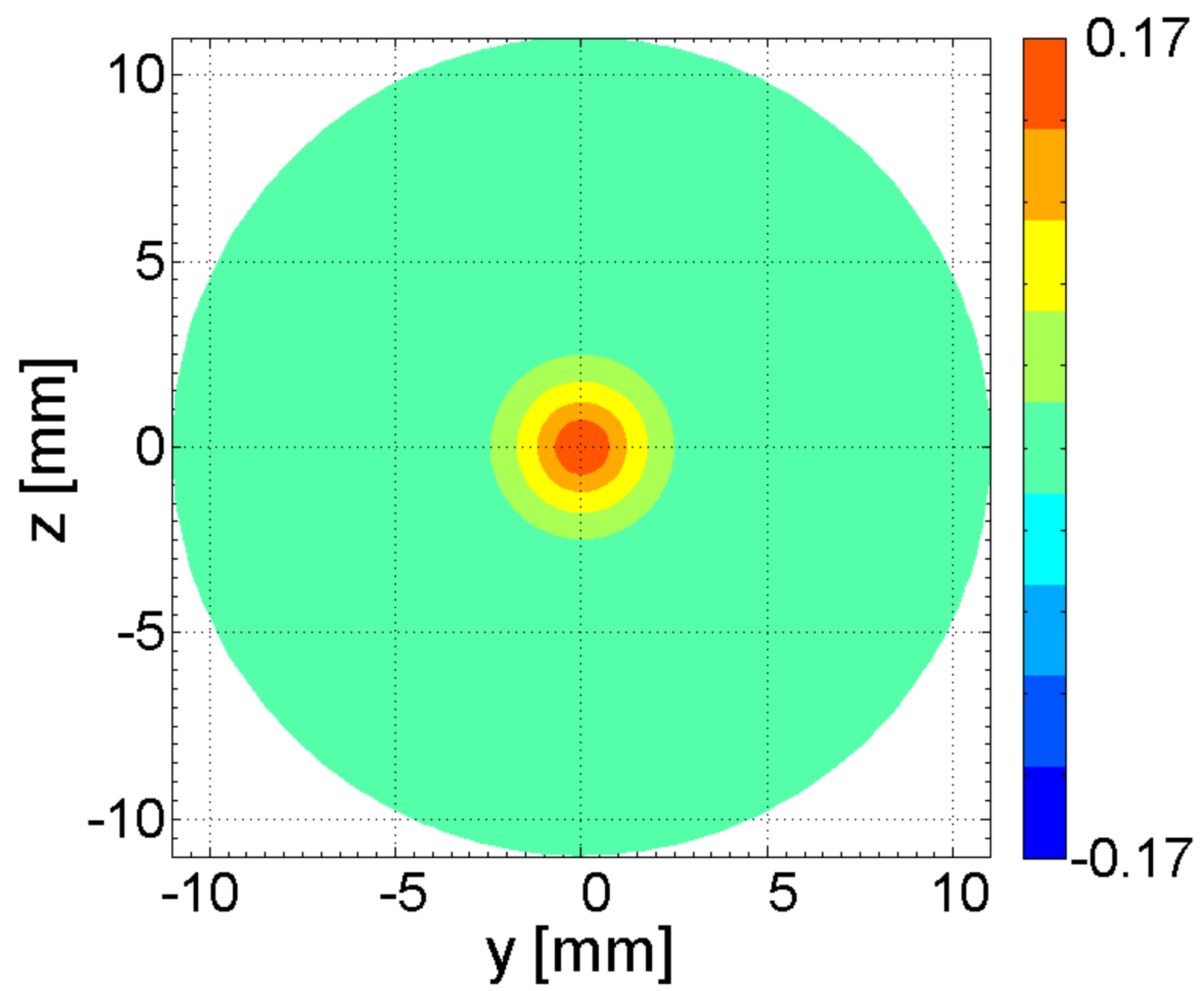}}
	\caption{}
	\label{fig:Figure13a}
	\end{subfigure}
	\begin{subfigure}[ht!]{0.3\textwidth}
	\centerline{\includegraphics[width=\textwidth]{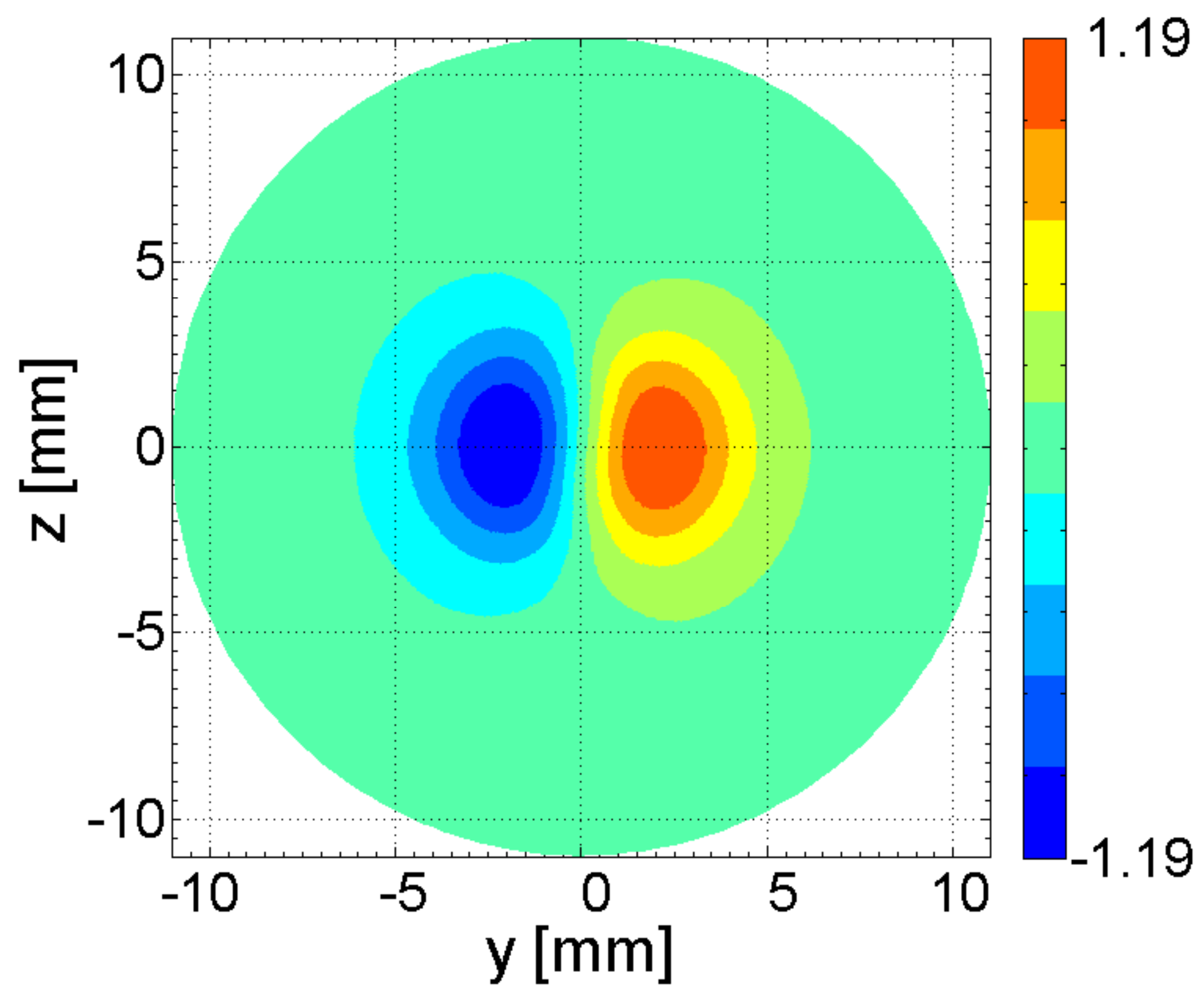}}
	\caption{}
	\label{fig:Figure13b}
	\end{subfigure}
	\begin{subfigure}[ht!]{0.3\textwidth}
	\centerline{\includegraphics[width=\textwidth]{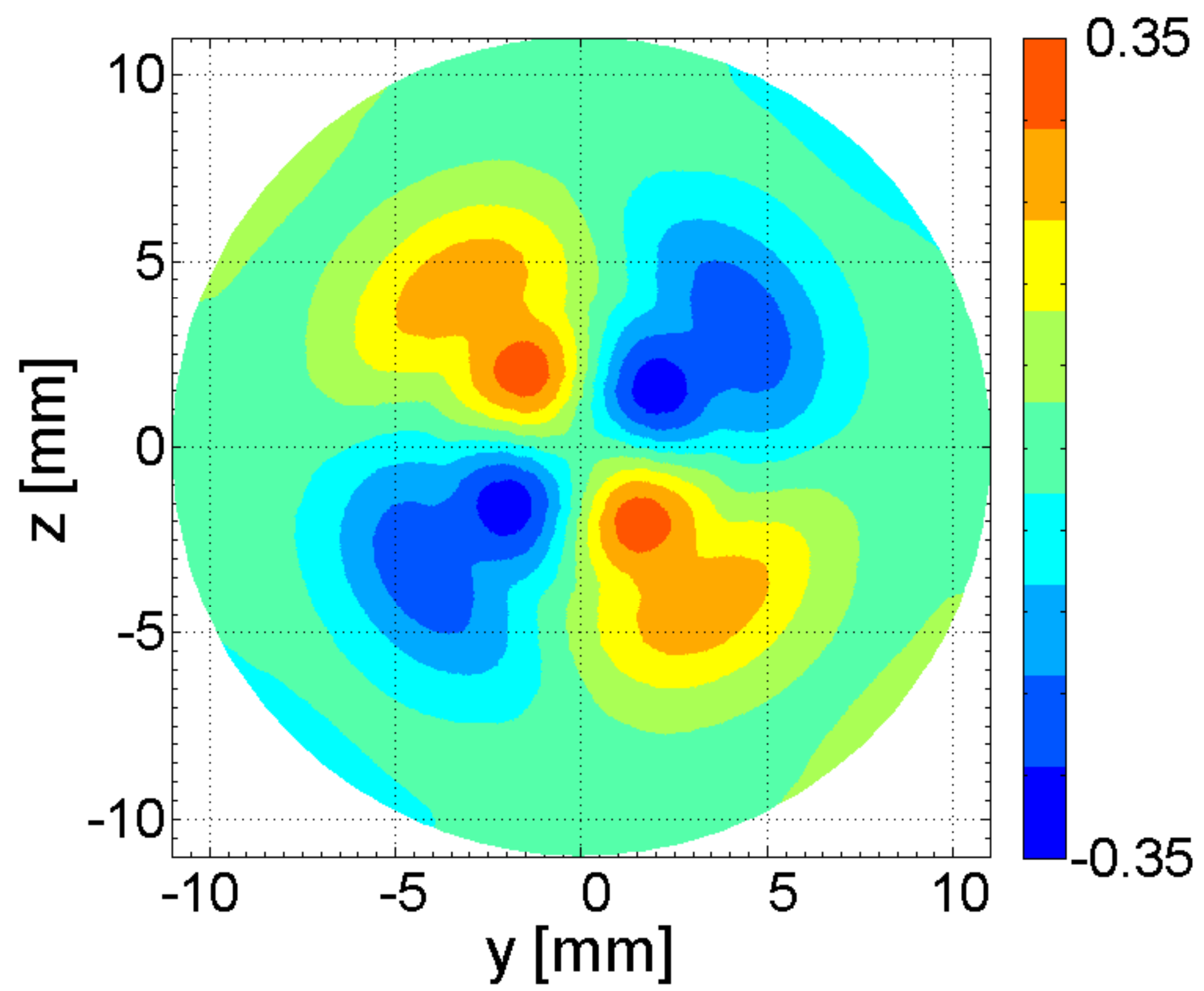}}
	\caption{}
	\label{fig:Figure13c}
	\end{subfigure}
	\end{center}
	\end{minipage}
	
	\begin{minipage}[hb!]{\textwidth}
	\begin{center}
	\begin{subfigure}[ht!]{0.3\textwidth}
	\centerline{\includegraphics[width=\textwidth]{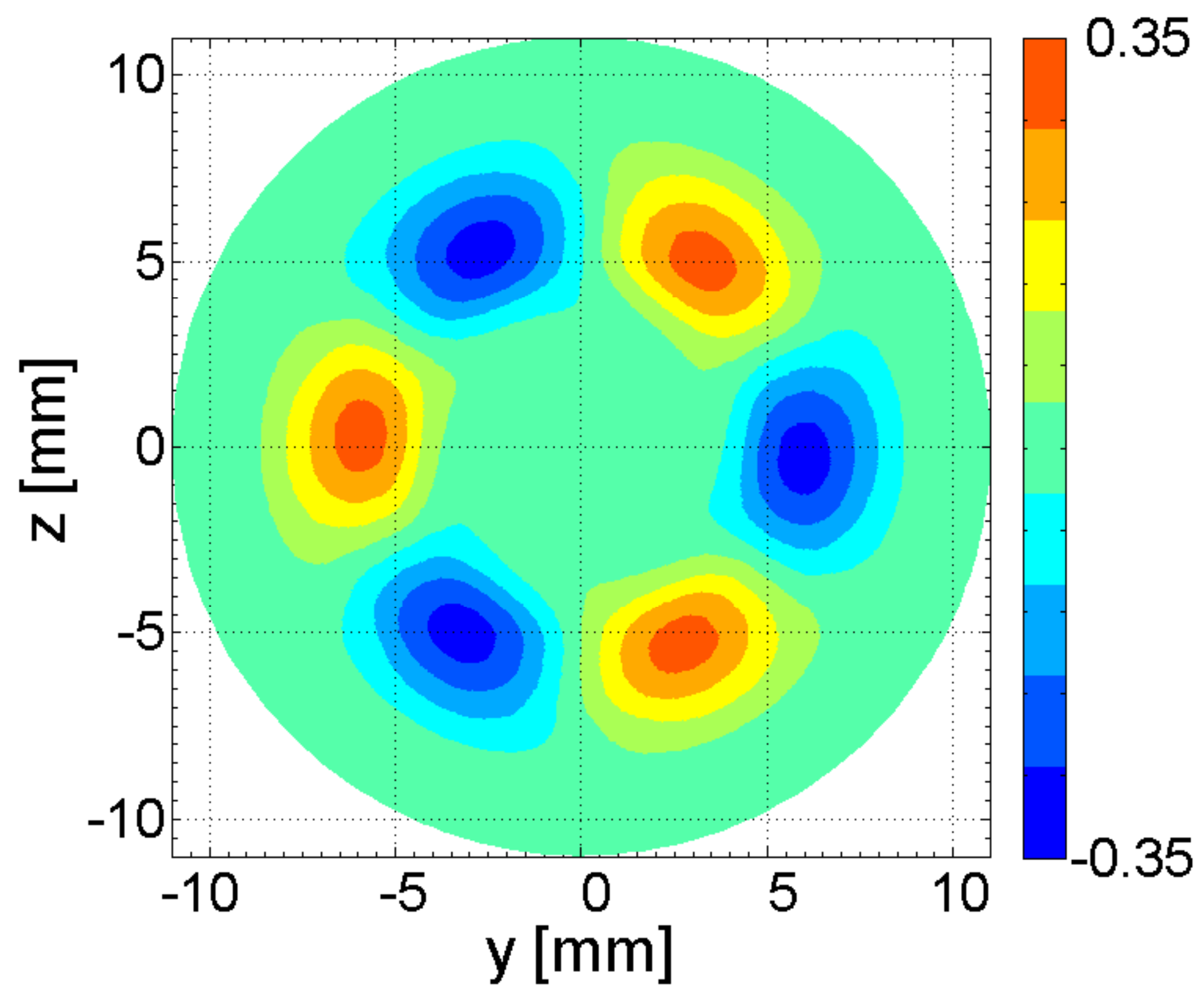}}
	\caption{}
	\label{fig:Figure13d}
	\end{subfigure}
	\begin{subfigure}[ht!]{0.3\textwidth}
	\centerline{\includegraphics[width=\textwidth]{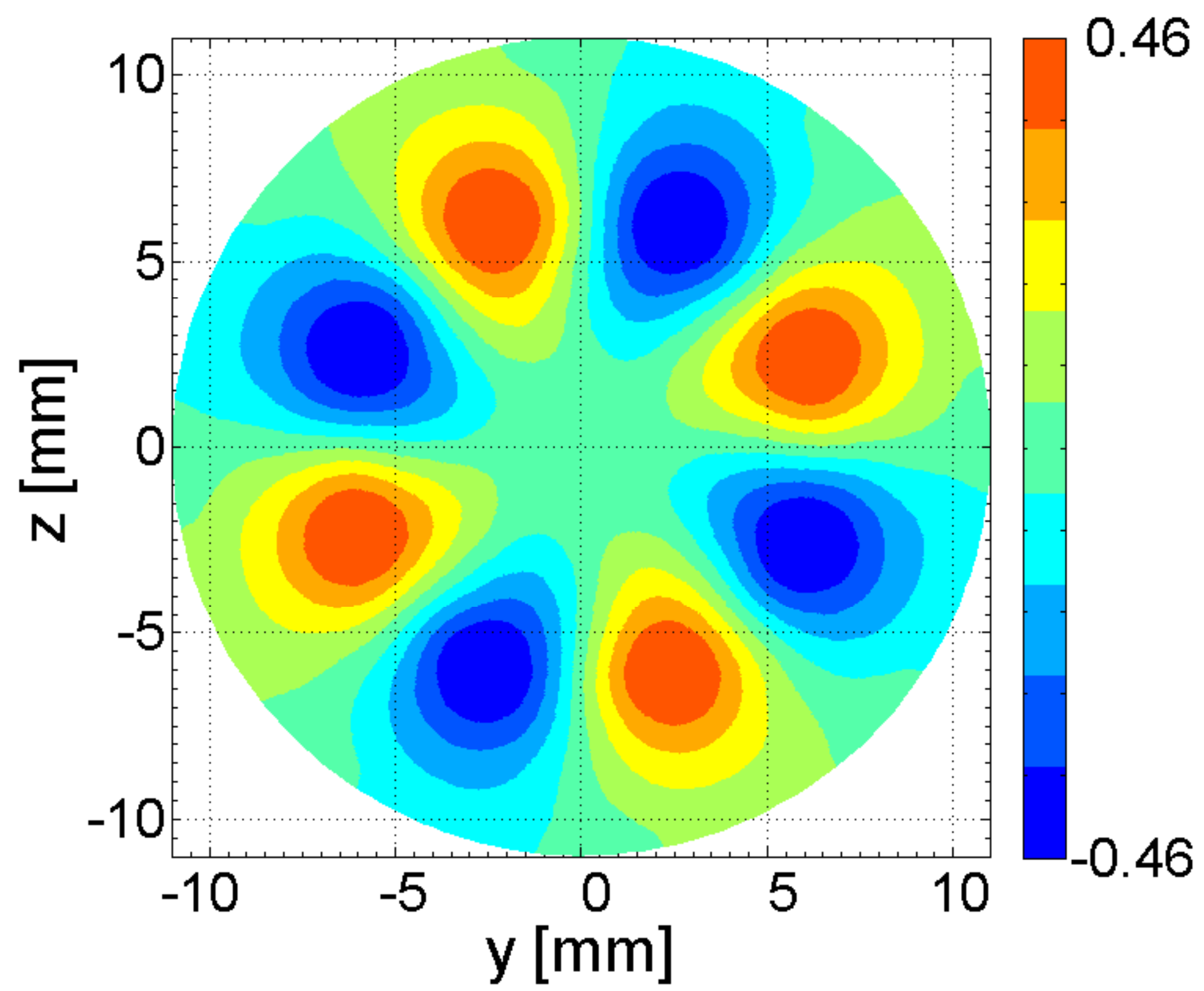}}
	\caption{}
	\label{fig:Figure13e}
	\end{subfigure}
	\end{center}
	\end{minipage}
	
	\caption{(Colour online) Mode patterns of Fourier decomposition for time-averaged vorticity field at $Re=296$. For each mode the range of color bar is adjusted to its amplitude. \textit{(a)} Mode $m=0$, \textit{(b)} Mode $m=1$, \textit{(c)} Mode $m=2$, \textit{(d)} Mode $m=3$, \textit{(e)} Mode $m=4$.}
	\label{fig:Figure13}
	
\end{figure}
\begin{figure}
	\begin{center}
	\begin{minipage}[!hb]{0.49\textwidth}
		\begin{center}
		\begin{subfigure}[b]{\textwidth}
       			\centering
                \includegraphics[width=\textwidth]{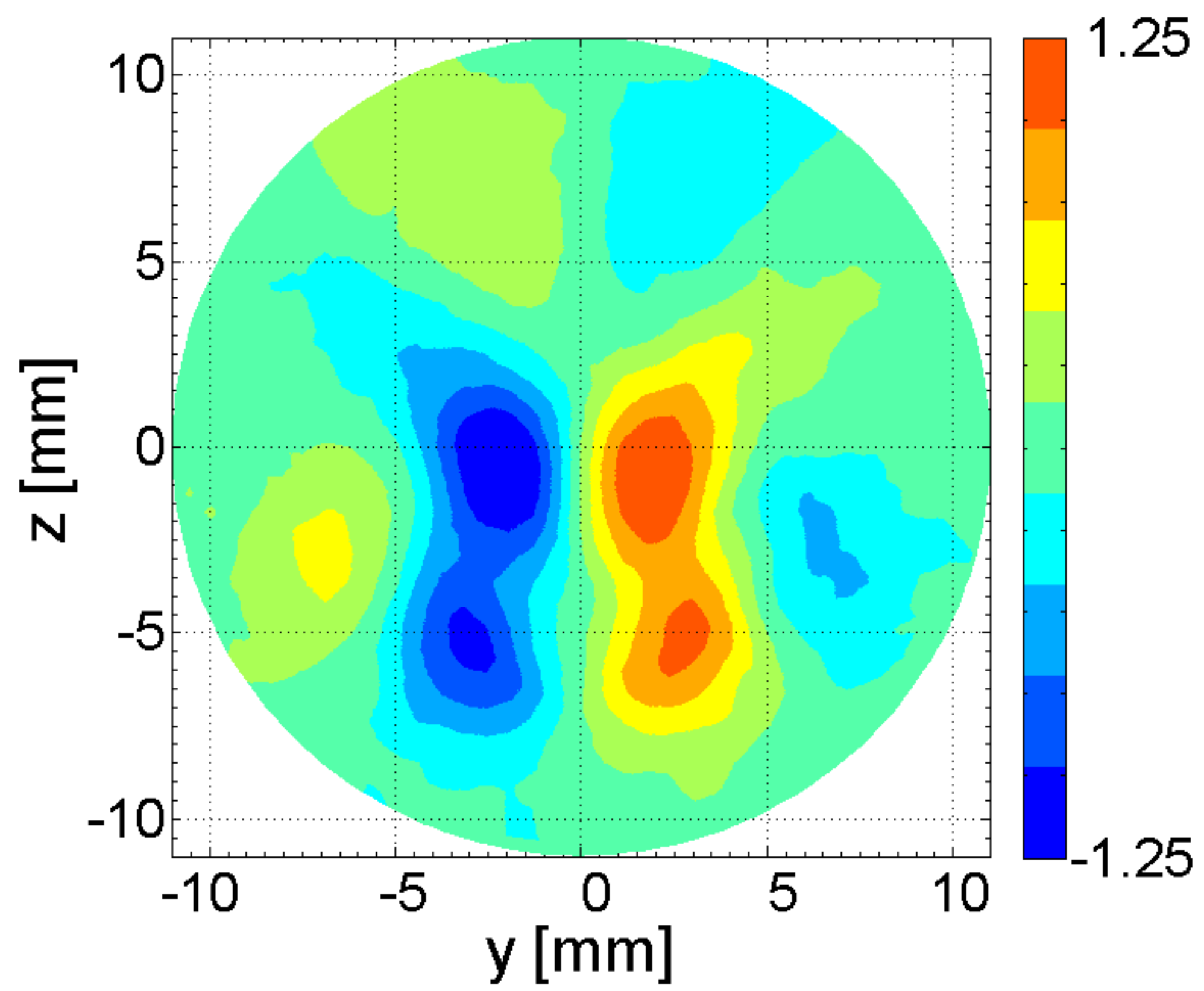}
                \caption{}
                \label{fig:Figure14a}   
        \end{subfigure}
        \end{center} 
    \end{minipage}	
    \begin{minipage}[!hb]{0.49\textwidth}
		\begin{center}
		\begin{subfigure}[b]{\textwidth}
       			\centering
                \includegraphics[width=\textwidth]{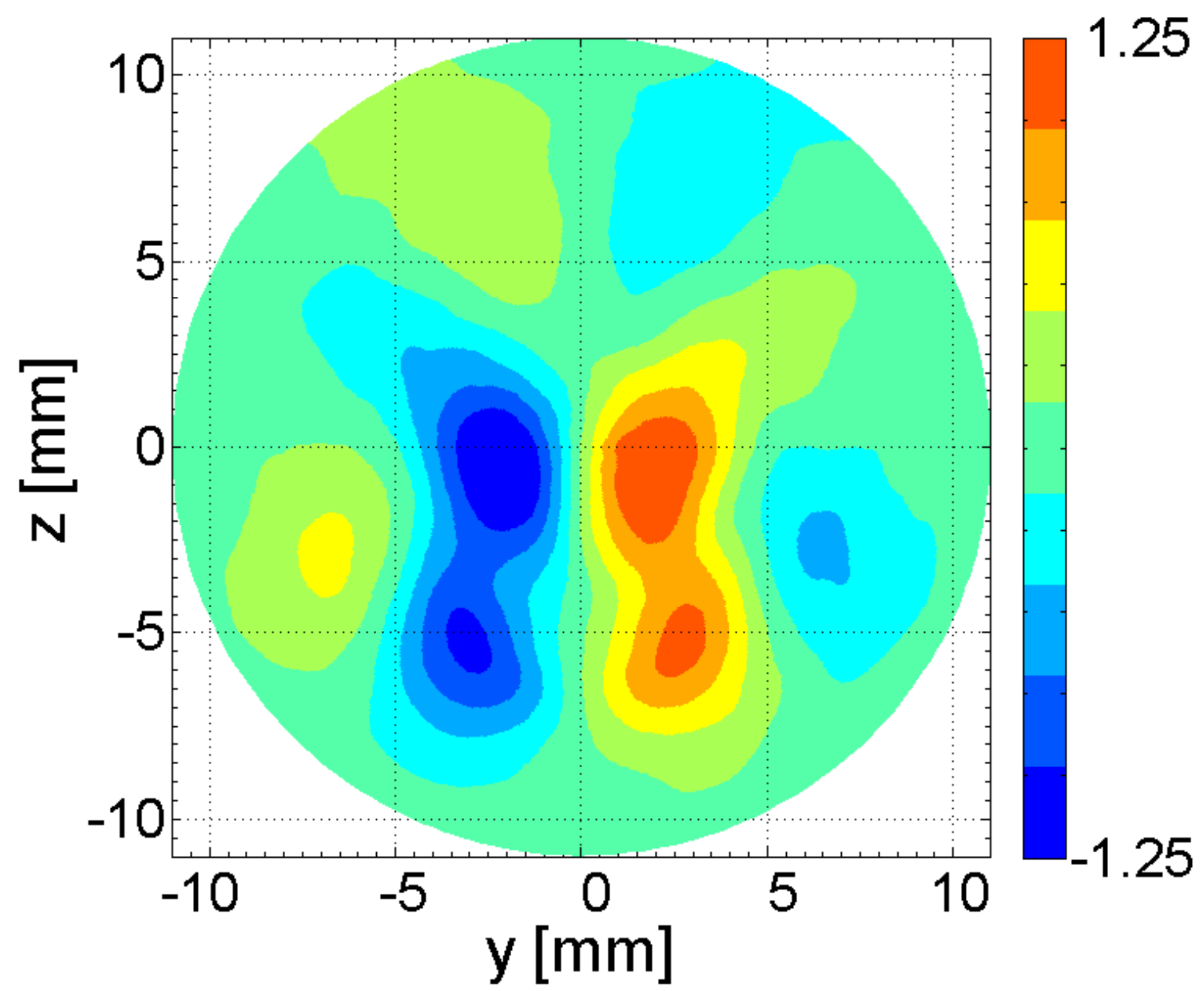}
                \caption{}
                \label{fig:Figure14b}   
        \end{subfigure}
        \end{center} 
    \end{minipage}	
	\caption{(Colour online) Time-averaged vorticity field: (a) PIV measurements and (b) Fourier reconstruction with the first five azimuthal modes at $Re=296$.}
	\label{fig:Figure14}
	\end{center}
\end{figure}

\subsection{Fourier decomposition of the longitudinal vorticity field}
To obtain more detailed information about the evolution of the flow field as the Reynolds number is increased, we have performed azimuthal Fourier decomposition of the streamwise vorticity. This is justified by the fact that the investigated instabilities are triggered by azimuthal breaking of symmetries.

\indent The longitudinal vorticity $\omega_x(y_i,z_j)$ obtained by measurements at a cross-section of the channel is available on a Cartesian grid $(y_i,z_j)\in\Re^2$, where $y$ and $z$ are axes of the cross-section. These data were reinterpolated over a polar grid $(r_k,\theta_n)\in\Re^2$ using an algorithm implemented in MATLAB$^{\textregistered}$ software. The grid consisted of 140 nodes in the radial and 240 nodes in the angular directions respectively. Subsequently a sequence of one-dimensional polar Fourier transforms were performed:\\
\begin{equation}
    \Omega(r_k,m)=\sum_{n}\omega_x(r_k,\theta_n)e^{im\theta_n}.
    \label{eqn:wzor4}
\end{equation}\\
Finally the amplitudes of azimuthal modes were obtained by numerical integration in the radial direction:\\
\begin{equation}
    \Omega(m)={\frac{1}{\sum_{k}r_{k}\Delta r}}\cdot{\sum_{k}\Omega(r_k,m) r_{k}\Delta r}.
    \label{eqn:wzor5}
\end{equation}\\

 \begin{figure}
	\begin{center}
	\begin{subfigure}[ht!]{0.8\textwidth}
	\centerline{\includegraphics[width=1\textwidth]{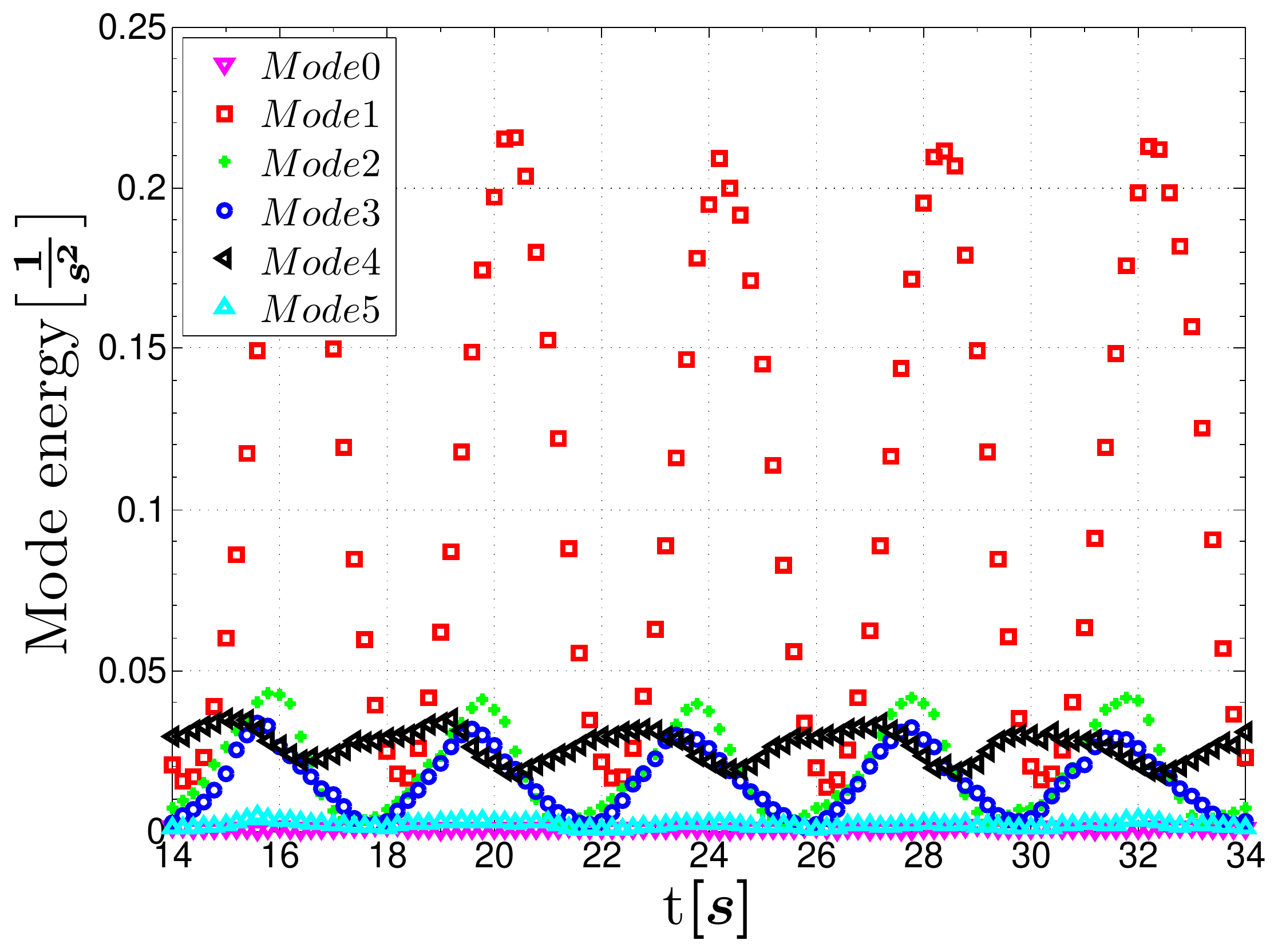}}
	\end{subfigure}
	\caption{(Colour online) Mode energy evolution in time ($Re=301$).}
	\label{fig:Figure15}
	\end{center}
\end{figure}
\begin{figure}
	\begin{center}
	\begin{minipage}[!hb]{\textwidth}
		\begin{center}
		\begin{subfigure}[b]{0.8\textwidth}
       			\centering
                \includegraphics[width=\textwidth]{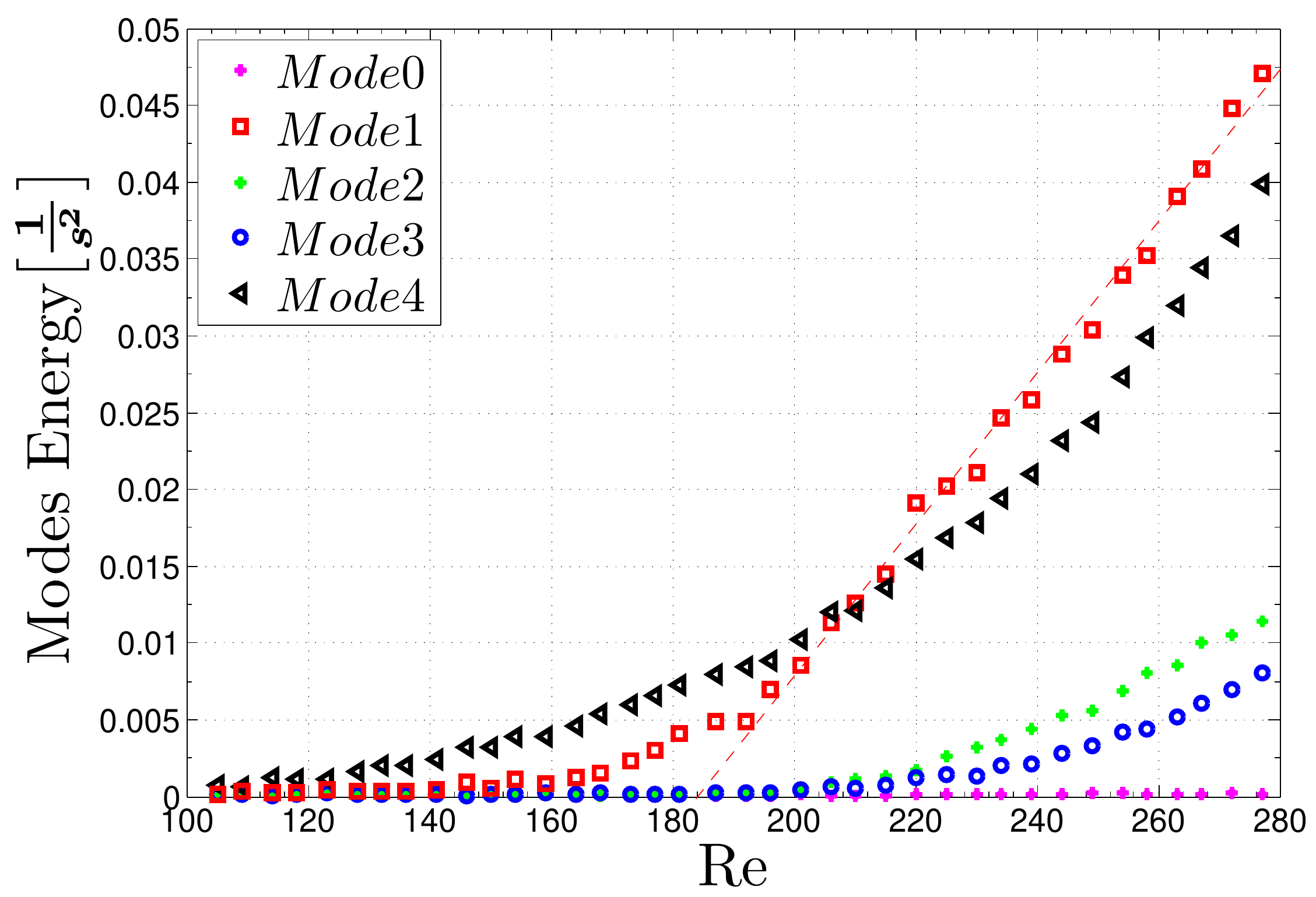}
                \caption{}
                \label{fig:Figure16a}   
        \end{subfigure}
        
        \begin{subfigure}[b]{0.8\textwidth}
                \centering
                \includegraphics[width=\textwidth]{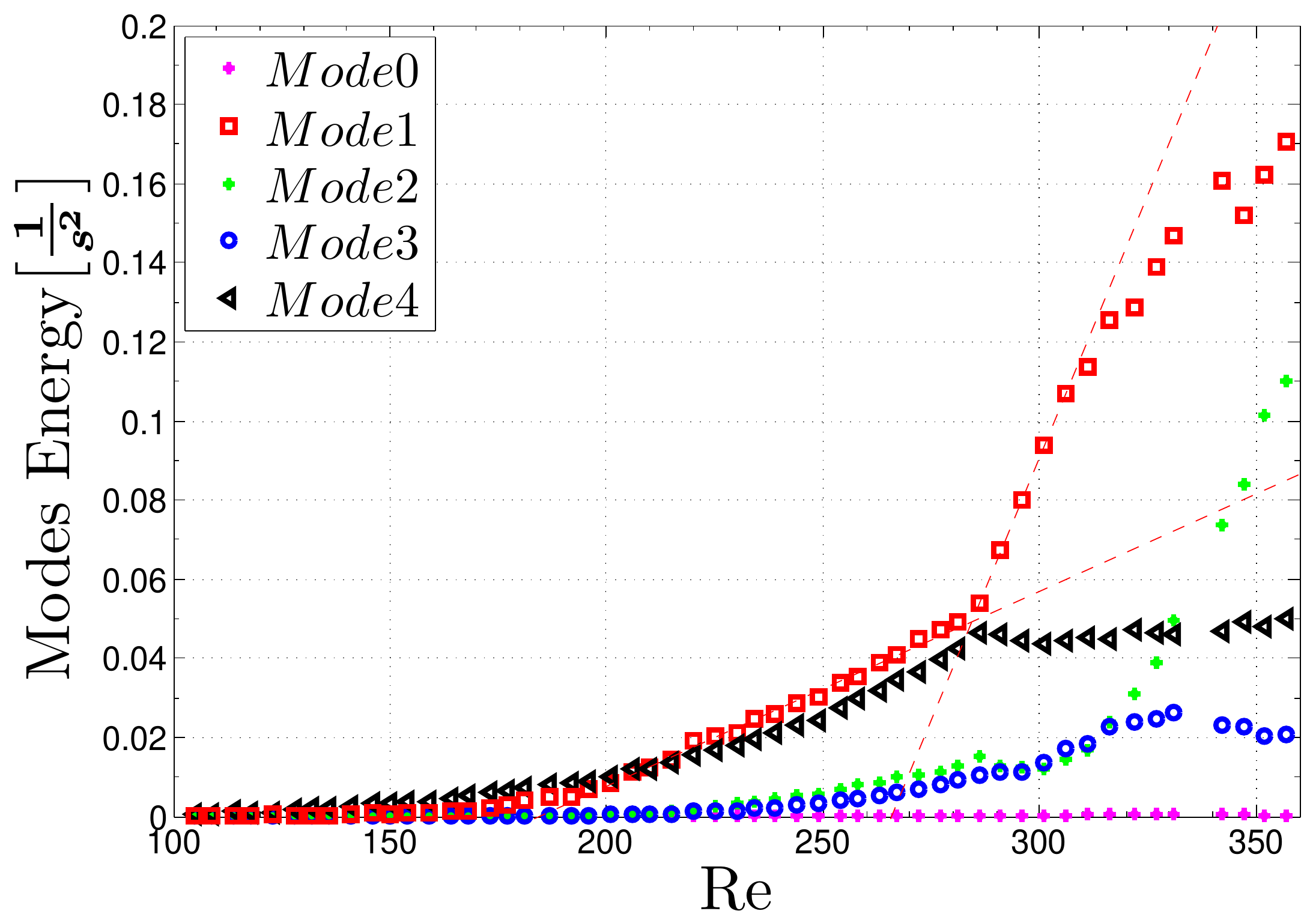}
                \caption{}
                \label{fig:Figure16b}
        \end{subfigure}
        \end{center}
	\end{minipage}	
	\caption{(Colour online) The first five azimuthal modes of enstrophy as a function of Reynolds number: \textit{(a)} detailed view of the bifurcation from the basic flow to the two stationary counter-rotating vortices regime and determination of the onset of transition ($Re_1=184$), \textit{(b)} an estimate of transition threshold from the two counter-rotating vortices to the hairpin vortex shedding regime ($Re_2=284$). }
	\label{fig:Figure16}
	\end{center}
\end{figure}

\begin{figure}
	\begin{center}
	\begin{subfigure}[ht!]{0.8\textwidth}
	\centerline{\includegraphics[width=1\textwidth]{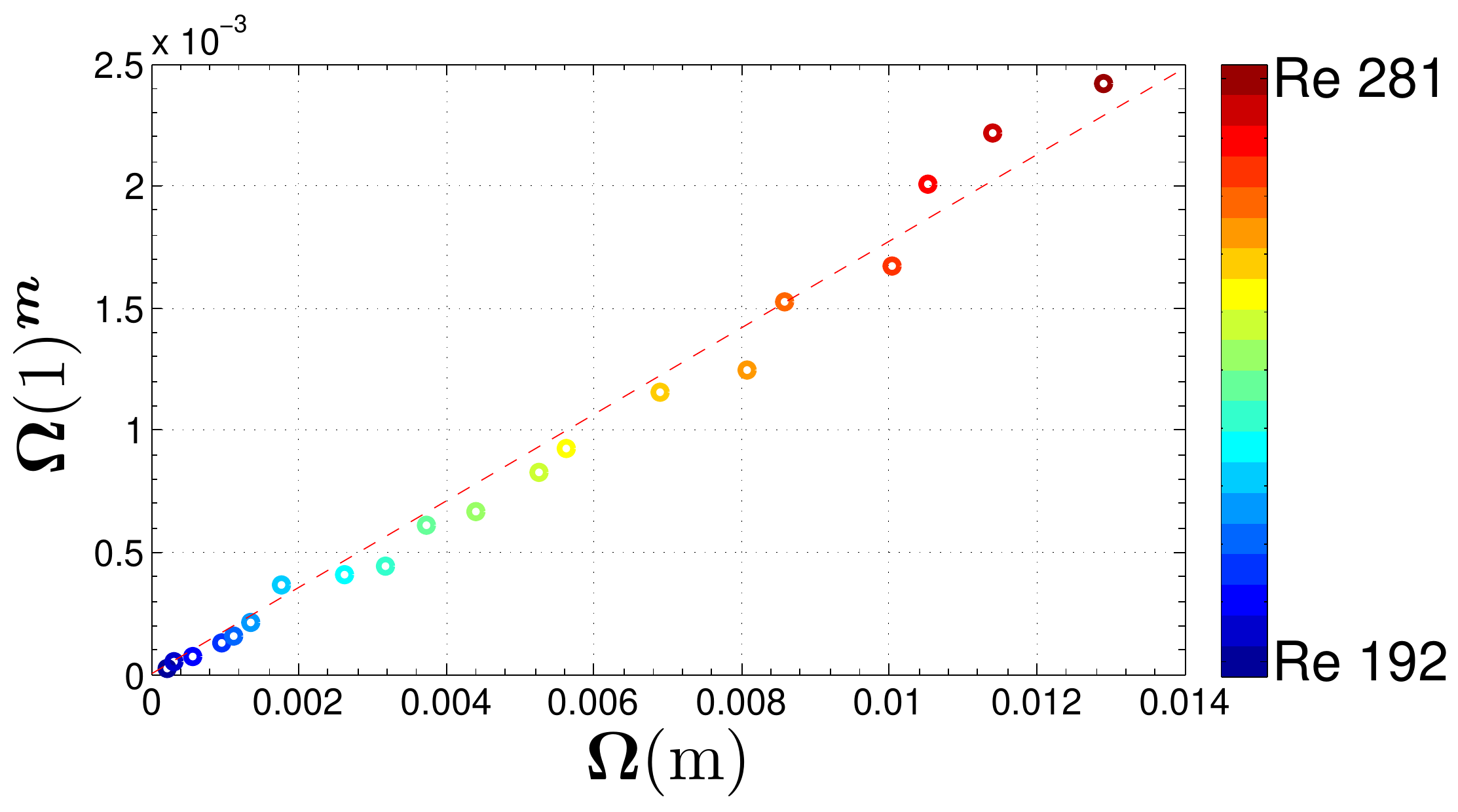}}
	\caption{}
	\label{fig:Figure17a}
	\end{subfigure}
	
	\begin{subfigure}[ht!]{0.8\textwidth}
	\centerline{\includegraphics[width=1\textwidth]{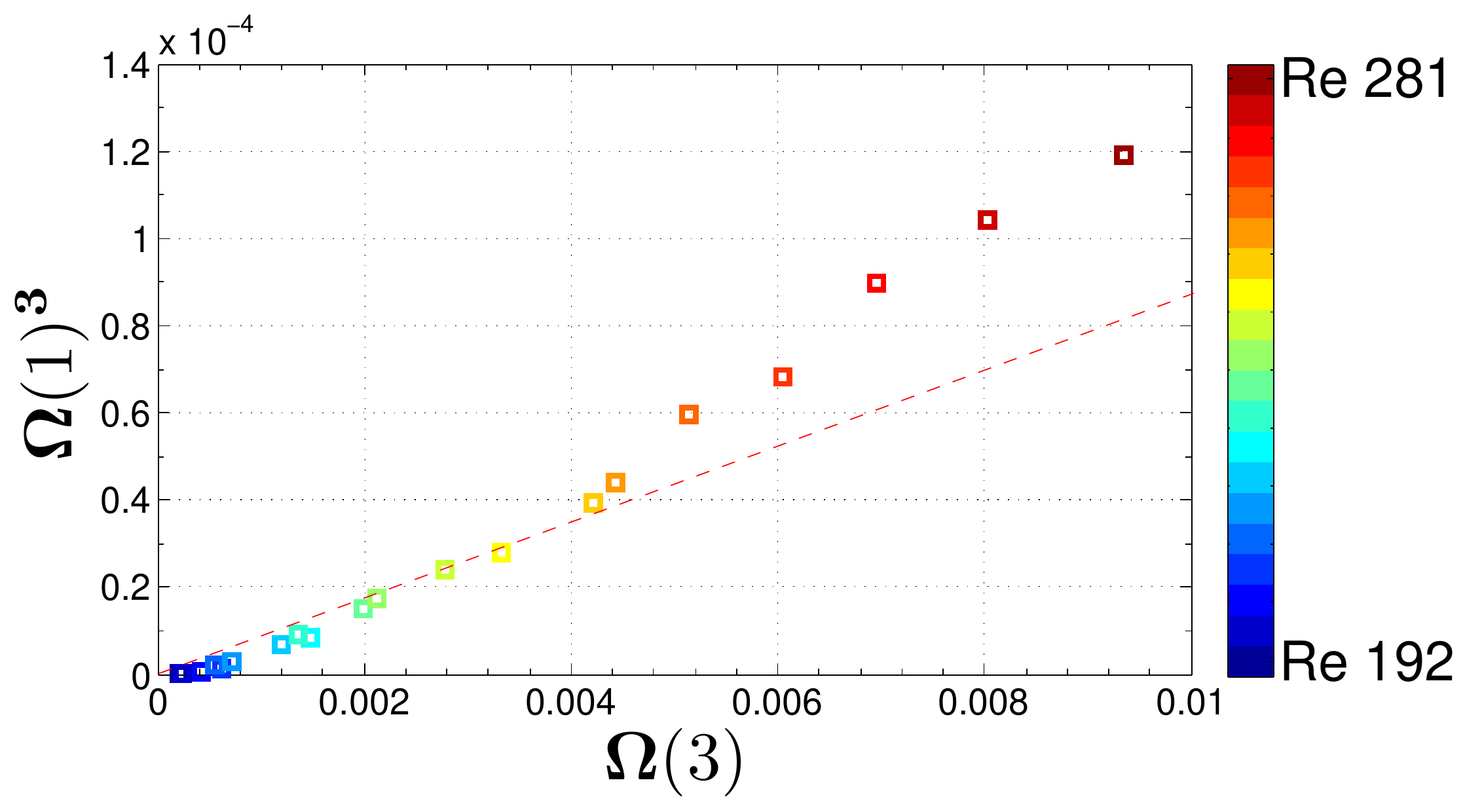}}
	\caption{}
	\label{fig:Figure17b}
	\end{subfigure}
	
	\begin{subfigure}[ht!]{0.8\textwidth}
	\centerline{\includegraphics[width=1\textwidth]{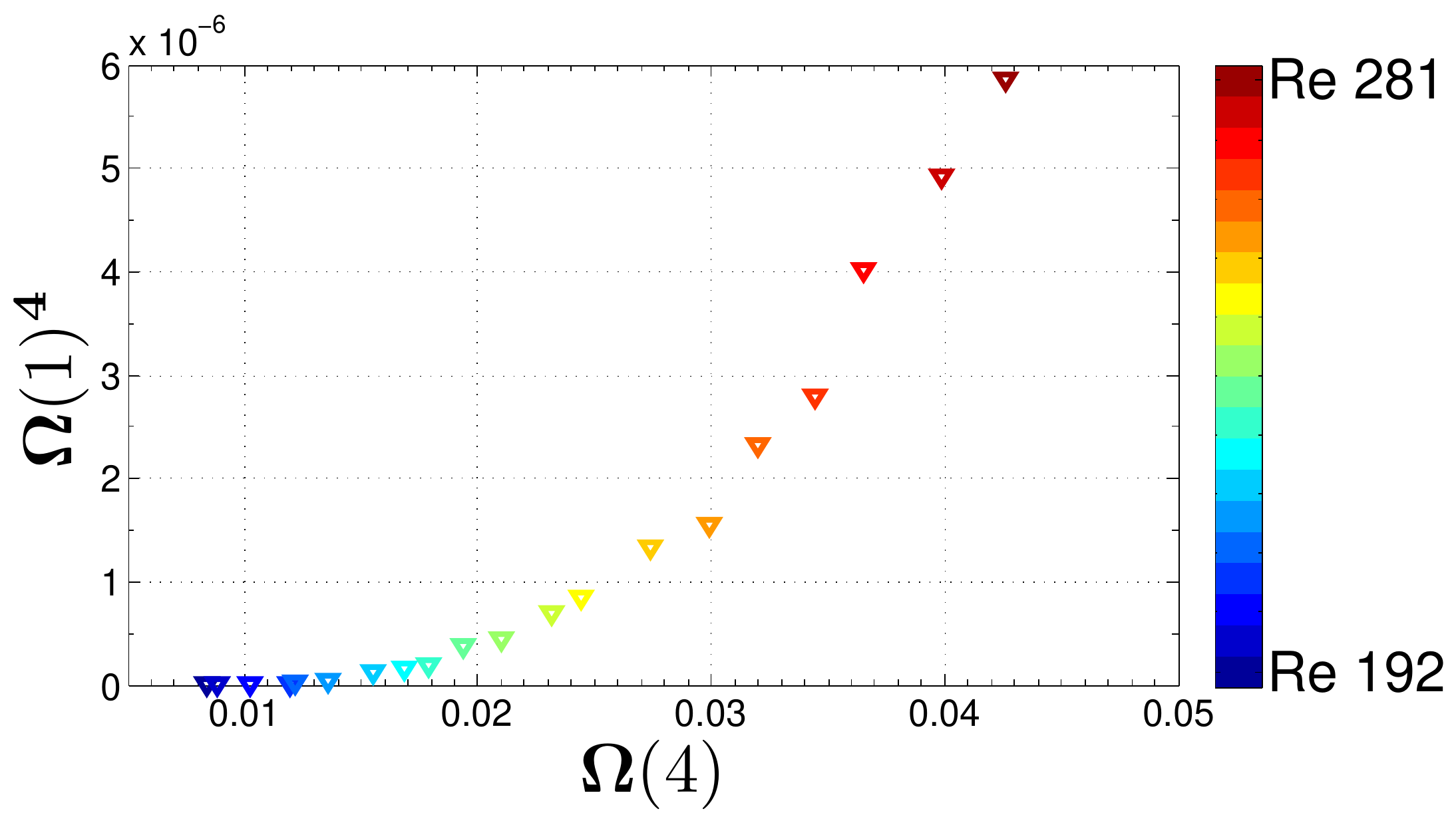}}
	\caption{}
	\label{fig:Figure17c}
	\end{subfigure}
	
	\end{center}
	\caption{(Colour online) Ratio of nonlinear dependency of amplitude of the azimuthal modes $m=2,3,4$ on the amplitude of the first mode, at different Reynolds numbers.}
	\label{fig:Figure17}
\end{figure}

\indent The main mode patterns (from  $m=0$ to $m=4$) obtained by the Fourier decomposition of a time-averaged vorticity field are presented in figure~\ref{fig:Figure13} (the unit of $\omega_x$ is $s^{-2}$). Figure~\ref{fig:Figure14} shows an example of a longitudinal vorticity field and the field obtained by reconstruction of the first five azimuthal modes. Similarly to the previously presented extraction of vortical components, we have also performed (for each Reynolds number) the decomposition of vorticity fields resulting from the time-average of all captured snapshots.

\indent Figure~\ref{fig:Figure15} presents the energy of the lower modes of enstrophy $\varepsilon_m$ as a function of time. The higher-order modes $m\geq5$ can be neglected due to their low amplitude. For this reason, the current analysis concentrates on the first five modes ($m=0,...,4$). The leading role of mode $m=1$ should be noted, as it contributes the most to the energy of the flow instability. This mode is strongly associated with both bifurcations, due to the fact that longitudinal vorticity patterns in both regimes contain only one symmetry plane and mode $m=1$ represents this planar symmetry.

\indent From the study of the temporal behaviour of the streamwise vorticity one can observe that the amplitudes $\Omega(m)$ of modes $m=1,2,3$ have almost sinusoidal behaviour, while the amplitude of mode $m=4$ is significantly less regular. This is due to their different origins. Modes $m=1,2,3$ are related to instability effects. In contrast, mode $m=4$ includes the extrinsic vorticity of the basic flow which is orthogonally symmetric (4 symmetry planes inclined at $45^\circ$ one to another) and mode $m=4$ also represents this type of symmetry (see figures~\ref{fig:Figure11} and~\ref{fig:Figure13e}), which is already present in the basic flow state. This is opposite to the situation for axisymmetric bluff bodies, such as spheres or disks, where the base flow is well represented by mode $m=0$. However, in the case of a cube, there is almost no contribution of this mode to the flow energy, as the base flow corresponds to the mode $m=4$.

\indent In addition, we have investigated the dependence of the modal decomposition on the Reynolds number in order to compare the values at onsets obtained with this method and with the previously described extraction of stationary and non-stationary instabilities as the manifestation of symmetry breaking. This time, we have taken the enstrophy of the mode $m=1$ as a quantity describing the instabilities, as an order parameter. We test whether this parameter (proportional to the square of the vorticity) has a linear variation with the Reynolds number, as follows from 
Landau`s model. From the results presented in figures~\ref{fig:Figure16a} and~\ref{fig:Figure16b} for the first and the second transitions respectively, we found the onset values to be equal to Re$_1=184$ and Re$_2=284$ respectively. The threshold of the first instability was estimated by extrapolating the amplitude of the mode $m=1$ to zero. These values are in a close agreement with the previous analysis of longitudinal enstrophy which confirms the validity of the results. Note that stationary instability follows the planar symmetry given by mode $m=1$, as the regular bifurcation is represented by the break of orthogonal symmetry. For this reason we consider the value of first onset obtained with azimuthal Fourier decomposition to be more accurate than that obtained by the separation of stationary and non-stationary contributions to the enstrophy of the longitudinal vorticity, presented in the previous subsection.

\indent Finally, as the first mode plays the major role in distributing the energy of longitudinal vorticity for the hairpin vortex shedding regime, we have also analysed the ratio of nonlinear dependence of the azimuthal modes $m=2,3,4$ on the mode $m=1$ for the flow prior to the Hopf bifurcation in a two counter-rotating vortices regime. In figure~\ref{fig:Figure17} three data sets are plotted. Each point represents one particular realisation at a given Reynolds number. The abscissa and ordinate correspond to the mode amplitudes $\Omega(m)$ and $\Omega(1)^m$ respectively. The change of Reynolds number is represented by the variation of colour.

\indent The circles in figure~\ref{fig:Figure17a} denote the evolution of $\Omega(1)^2$ as a function of $\Omega(2)$. It illustrates essentially a linear relation, which proves that both these modes are associated as slave modes.  

\indent The squares in figure~\ref{fig:Figure17b} display the dependence of $\Omega(1)^3$ as a function of $\Omega(3)$. It differs from a strictly linear evolution due to the higher-order nonlinear couplings. However, one can see that this deviation is quite weak.

\indent In contrast, one may observe in figure~\ref{fig:Figure17c} (data set represented by triangles) that $\Omega(1)^4$ as a function of $\Omega(4)$ significantly differs from linear behaviour. For small mode amplitude corresponding to low Reynolds numbers, the mode m=4 is uncoupled and independent from $\Omega(1)$,representing only the extrinsic vorticity. As Reynolds number is increased the fully nonlinear coupling appears, represented by continuous growth of the ratio $\Omega(1)^4$/$\Omega(4)$.

\indent This confirms our earlier hypothesis, that the mode $m=4$ corresponds to the vorticity structure induced by the geometry of the cube and it is not related to the instability effect connected with modes $m=1,2,3$ (intrinsic vorticity). It is consistent  with previously obtained results associated with the evolution in time of the energy of modes for unsteady flow in a hairpin vortex shedding regime.

\section{CONCLUSIONS}\label{sec:CONC}
\indent We have presented experimental results for the wake flow behind a cube and analysis of corresponding flow instabilities in the Reynolds number range of 100-400. Existence of three well-defined regimes was confirmed, namely the non-axisymmetric basic flow, the regular state with two counter-rotating vortices and the hairpin vortex shedding regime. In the wake of a cube, unlike axisymmetric bluff bodies, one observes longitudinal vorticity in the basic flow originating from a transversal gradient of pressure induced by corners of the obstacle. The wake consists of four pairs of counter-rotating vortices with orthogonal symmetry. 

\indent Subsequently, after a regular bifurcation, two threaded counter-rotating vortices appear while the flow preserves only one symmetry plane. The observed vorticity field is a result of interaction between the basic flow vorticity and a dipole of vorticity related to a regular, stationary instability. Finally, due to a Hopf bifurcation, the flow becomes periodically time-dependent with regular oscillations of the recirculation zone manifested by one-sided hairpin vortex shedding. The symmetry plane and its orientation remain the same as in the previous regime. The scenario of transitions agrees with a previous numerical simulation of the wake behind a cube carried out by \citet{k_saha1}.

\indent In order to estimate the values for the bifurcation onsets, we have extracted from the longitudinal vorticity the components associated with each investigated regime, namely the basic flow component, the stationary instability related to the two counter-rotating vortices and the non-stationary instability corresponding to the hairpin vortex shedding regime. We have shown that the squared amplitude of both instability fluctuations depends linearly on $Re-Re_{cr}$ which is consistent with a supercritical instability (Re$_{cr}$ is the appropriate onset of either regular or Hopf bifurcation). We observe a rounding in the first bifurcation or an imperfect bifurcation, originating essentially from the initial existence of longitudinal vorticity of the basic flow and the influence of the cube support.

\indent We have also shown a different spatial evolution in the streamwise direction of maximum vorticity between the basic flow and regimes related to both consecutive instabilities. The former decreases exponentially as it is an extrinsic phenomenon, while the two latter, being intrinsic, have a global mode envelope.

\indent Another very important result of this investigation is the accurate measurement of the change of $\omega_x$ caused by the Hopf bifurcation (from the two counter-rotating vortices to the hairpin vortex shedding regime). In the stationary flow the counter-rotating vortices are well defined but their magnitude is weak. When the Hopf bifurcation occurs and the flow becomes periodic, the mean amplitude of maximum vorticity extracted from the time-averaged vorticity field continues to increase. This growth is larger than predicted by the a contribution of separate stationary instability. It is due to the reorientation from the transversal to streamwise direction of the periodically shed hairpin vortex heads.

\section{ACKNOWLEDGMENTS}
The authors thank Laurette Tuckerman for the help with modal analysis, as well as Tomasz Bobinski for fruitful discussion concerning the subject of this paper. We would like also to acknowledge Xavier Benoit Gonin for technical assistance. \\

\bibliographystyle{jfm}

\end{document}